\RequirePackage{fix-cm}
\RequirePackage{amsmath}

\documentclass[smallextended]{svjour3}       
\smartqed  

\usepackage{graphicx}
\usepackage{amsmath}

\usepackage{amssymb}

\usepackage{mathptmx}  
\journalname{Journal of Statistical Physics}
\begin{document}

\title{Dynamical large deviations for plasmas below the Debye length and
the Landau equation}


\author{Ouassim Feliachi         \and
        Freddy Bouchet 
}


\institute{O. Feliachi \at
              Institut Denis Poisson, Université d’Orléans, CNRS,
Université de Tours, France\\
Univ Lyon, Ens de Lyon, Univ Claude Bernard, CNRS, Laboratoire de Physique, Lyon, France \\      
              \email{ouassim.feliachi@univ-orleans.fr}           
           \and
           F. Bouchet \at
              Univ Lyon, Ens de Lyon, Univ Claude Bernard, CNRS, Laboratoire de Physique, Lyon, France \\
              \email{freddy.bouchet@ens-lyon.fr}           
}

\date{Received: date / Accepted: date}

\maketitle

\begin{abstract}
We consider a homogeneous plasma composed of $N$ particles of the
same electric charge which interact through a Coulomb potential. In
the large plasma parameter limit, classical kinetic theories justify
that the empirical density is the solution of the Balescu--Guernsey--Lenard
equation, at leading order. This is a law of large numbers. The Balescu--Guernsey--Lenard equation
is approximated by the Landau equation for scales much smaller than
the Debye length. In order to describe typical and rare fluctuations,
we compute for the first time a large deviation principle for dynamical
paths of the empirical density, within the Landau approximation. We
obtain a large deviation Hamiltonian that describes fluctuations and
rare excursions of the empirical density, in the large plasma parameter
limit. We obtain this large deviation Hamiltonian either from the
Boltzmann large deviation Hamiltonian in the grazing collision limit,
or directly from the dynamics, extending the classical kinetic theory
for plasmas within the Landau approximation. We also derive the large
deviation Hamiltonian for the empirical density of $N$ particles,
each of which is governed by a Markov process, and coupled in a mean
field way. We explain that the plasma large deviation Hamiltonian
is not the one of $N$ particles coupled in a mean-field way. 
\end{abstract}

\section{Introduction: kinetic theories, dynamical large deviations and equilibrium
statistical mechanics }

In the field of statistical physics, the literature that describes
the static fluctuations of a system around equilibrium and its relaxation
to equilibrium is very rich. For instance, working in the appropriate
thermodynamic ensemble, we can express the probability of observing
a given state of a system as a function of the corresponding thermodynamic
potential. Beyond equilibrium, classical kinetic theories describe
the relaxation to equilibrium in some asymptotic regimes. For instance
the Boltzmann equation describes the relaxation to equilibrium of
a dilute gas in the Boltzmann-Grad limit, and the Balescu-Guernsey-Lenard
equation in the opposite limit of particles with long range interactions,
for instance plasma in the weak coupling limit or self-gravitating
systems. The Landau equation is either an approximation of the Balescu-Guernsey-Lenard
equation that describes the relaxation of plasma at a scale much smaller
than the Debye length, or an approximation of the Boltzmann equation
in the weak scattering limit. All those classical kinetic equations
describe the relaxation of the empirical distribution $g_{N}(\mathbf{r},\mathbf{v},t)\equiv\frac{1}{N}\sum_{n=1}^{N}\delta(\mathbf{v}-\mathbf{v}_{n}(t))\delta\left(\mathbf{r}-\mathbf{r}_{n}(t)\right)$,
where $\delta$ are Dirac delta functions, $t$ is time, $\left(\mathbf{r}_{n}(t),\mathbf{v}_{n}(t)\right)_{1\leq n\leq N}$
are the $N$ particle positions and velocities. The six-dimensional
space of one-particle position-velocity, with points $\left(\mathbf{r},\mathbf{v}\right)$,
is called the $\mu$-space. $g_{N}$ is a distribution over the $\mu$-space
that evolves with time.

The probability $\mathbf{P}_{eq}\left(g_{N}=g^{0}\right)$ to observe
$g_{N}$ close to a given distribution $g_{0}$ of the $\mu$-space,
at some fixed arbitrary time, in the microcanonical ensemble, satisfies
\begin{equation}
\mathbf{P}_{eq}\left(g_{N}=g^{0}\right)\propto\text{e}^{N\frac{\mathcal{S}[g^{0}]}{k_{B}}}.\label{eq:Einstein}
\end{equation}
This is the classical Einstein formula relating the specific entropy
$\mathcal{S}[g^{0}]$ of the macrostate $g^{0}$ with its equilibrium
probability. $k_{B}$ is the Boltzmann constant. This can be seen
as a definition of the Boltzmann entropy $\mathcal{S}[g^{0}]$ of
the macrostate $g^{0}$. For a dilute gas, because the particles are
independent at leading order, of for systems with long range interactions,
because the two-body interactions are weak, it is known that $\mathcal{S}$
is the negative of the Boltzmann $\mathcal{H}$ function ($\mathcal{S}[g^{0}]=-k_{B}\int\text{d}\mathbf{r}\text{d}\mathbf{v}\,g^{0}\log g^{0}$) if the macrostate $g^{0}$satisfies the conservation
laws (mass, momentum and energy), and $\mathcal{S}\left[g^{0}\right]=-\infty$
otherwise. 

However all those classical works and results in equilibrium statistical
mechanics and kinetic theory do not describe the probability of paths
that may lead to any macrostate $g^{0}$. More generally, the macroscopic
or mesoscopic stochastic process for $g_{N}$ is not described by
classical theories, and dynamical description is restricted to relaxation
to equilibrium. In principle, very rarely, the microscopic dynamics
can lead the distribution function to follow other paths than the
relaxation paths described by the kinetic equation. What is the probability
of such rare excursions? How do these probabilities depend on the
paths? Those are key questions. Answering them are the starting point
for solving many other non-equilibrium problems. Moreover, if the
microscopic dynamics is time-reversible (in the sense of dynamical
systems), for instance if the microscopic dynamics is Hamiltonian,
then we expect the stochastic process for $g_{N}$ to be also time-reversible
(in the sense of stochastic processes). It is a fundamental question
to describe this stochastic process for the empirical distribution
$g_{N}$.

More precisely we need to estimate the probability $\mathbf{P}\left(\left\{ g_{N}(t)\right\} _{0\leq t\leq T}=\left\{ g(t)\right\} _{0\leq t\leq T}\right)$
to observe the evolution of $\left\{ g_{N}(t)\right\} $ to be in a neighborhood of 
any prescribed path $\left\{ g(t)\right\} $, for times $0\leq t\leq T$,
in some asymptotic limit when the kinetic description is valid, with the prescription that $g_{N}(t=0)$ is in the neighborhood of $g(t=0)$. The
mathematical and theoretical formalism adapted to this problem is
large deviation theory. We need to prove the large deviation result
\begin{equation}
\mathbf{P}\left(\left\{ g_{N}(t)\right\} _{0\leq t\leq T}=\left\{ g(t)\right\} _{0\leq t\leq T}\right)\underset{\epsilon\rightarrow0}{\asymp}\text{e}^{-\frac{1}{\epsilon}\int_{0}^{T}\text{d}t\,\text{Sup}_{p}\left\{ \int\dot{g}p\,\text{d}\mathbf{r}\text{d}\mathbf{v}-H[g,p]\right\} },\label{eq:Grandes_Deviations_Theorie_Cinetique}
\end{equation}
where $\dot{g}$ is the time derivative of $g$, $p$ is a function
over the $\mu$-space and is called the conjugated momentum of $\dot{g}$,
the Hamiltonian $H$ is a functional of $g$ and $p$ that characterizes
the dynamical fluctuations, and where the symbol $\underset{\epsilon\rightarrow0}{\asymp}$
roughly means a logarithmic equivalence ($g_{\epsilon}\underset{\epsilon\downarrow0}{\asymp}\exp(\varphi/\epsilon)\iff\lim_{\epsilon\downarrow0}\epsilon\log g_{\epsilon}=\varphi$). A mathematical definition of a large deviation principle is found in classical textbooks \cite{varadhan1984large}. We
note that $H$ is not the Hamiltonian of the microscopic dynamics
but $H$ rather defines a statistical field theory that quantifies
the probabilities of paths of the empirical distribution. $H$ is
associated with a Lagrangian $L\left[g,\dot{g}\right]=\text{Sup}_{p}\left\{ \int\dot{g}p\,\text{d}\mathbf{r}\text{d}\mathbf{v}-H[g,p]\right\} $
and an action $\int_{0}^{T}\text{d}t\,L\left(g,\dot{g}\right)$. The
large deviation speed $\epsilon$ is a small parameter associated
to the kinetic limit. $\epsilon$ could be $1/N$, but more generally
it will depend on the physical system under consideration.

In the paper \cite{Bouchet_Boltzmann_JSP}, we explained why deriving
a dynamical large deviation principle like (\ref{eq:Grandes_Deviations_Theorie_Cinetique})
shed an illuminating perspective on the irreversibility paradox. In
a nutshell, if the microscopic dynamics is time-reversible, then $H$
will automatically verify a time-reversal symmetry, relating the microscopic
time-reversibility to the time-reversibility of the stochastic process
of the empirical distribution. The entropy will be automatically related
to the quasipotential, quantifying precisely the relation between
the dynamical properties of the field theory determined by $H$, to
the interpretation of the entropy as characterizing the static properties
through the Einstein formula (\ref{eq:Einstein}). The increase of
the entropy for relaxation paths will immediately follow as a general
property of the quasipotential, as a mere consequence of the convexity
of $H$ with respect to the variable $p$, a property which is always
true for a large deviation Hamiltonian. Then (\ref{eq:Grandes_Deviations_Theorie_Cinetique})
characterizes the large deviations of a time-reversible process, and
thus does not break the time reversibility. The most probable evolution
of this time-reversible process will break time-reversal symmetry
because we consider a specific path, and will be the solution of the
kinetic equation. This explains why the kinetic equation increases
$S$ although the microscopic dynamics is time-reversible. Moreover,
(\ref{eq:Grandes_Deviations_Theorie_Cinetique}) characterizes the
probability of any paths at the large deviation level, and quantifies
very precisely the exponential concentration close to the solution
of the kinetic equation. 

Several works recently computed the dynamical large deviations for
particle systems. One of the firsts was a work by Derrida, Lebowitz
and Speer \cite{Derrida_Lebowitz_Speer_2002_PhRvL} for systems of
particles that have a Markovian dynamics, for instance the SEP (Simple
Exclusion Process). Following this work, Rome's group derived a consistent
general formalism to describe phenomenologically macroscopic fluctuation
theories \cite{bertini2015macroscopic} of systems which mesoscopic
dynamics is diffusive. Those two complementary approaches nicely describe
the dynamical large deviations for a large class of particle systems.
However, it would be interesting to deal with large deviation principles
for particle systems with a more physical dynamics than the one considered
so far, starting from the Hamiltonian dynamics of atoms or molecules. 

This paper is the second of a series of three in which we address
the computation of the large deviation Hamiltonian $H$, and of the
large deviation parameter $\epsilon$, for the three classical kinetic
theories associated respectively to the dilute gases (the Boltzmann
equation), mean field interactions, plasma and self-gravitating stars
(the Balescu--Guernsey--Lenard equation), and
plasma at a scale much smaller than the Debye length and in a weak
coupling limit (the Landau equation). In our first paper \cite{Bouchet_Boltzmann_JSP},
we explained that for dilute gases, $\epsilon$ is the inverse of
the number of particles in a volume of the size of the mean free path.
In this first paper, we also derived the Boltzmann large deviation
Hamiltonian (see formulas (\ref{eq:LDP_Boltzmann-1}-\ref{eq:LDP_Boltzmann_HT-1})
in section \ref{sec:Large-dev-from-boltzmann} of the present paper)
from the natural Boltzmann hypothesis of molecular chaos. Long before
our work \cite{Bouchet_Boltzmann_JSP}, Rezakhanlou has proven \cite{rezakhanlou1998}
a large deviation result for 1D stochastic dynamics mimicking the
hard sphere dynamics. The functional form of the large deviation Hamiltonian
we deduced from Boltzmann's molecular chaos hypothesis is actually
the same as Rezakhanlou's one. Moreover, for the specific case of
hard spheres and in the Boltzmann-Grad limit, Bodineau, Gallagher,
Saint-Raymond and Simonella \cite{bodineau2020fluctuation} have rigorously
proven large deviation asymptotics that give an information equivalent
to the large deviation formulas (\ref{eq:LDP_Boltzmann-1}-\ref{eq:LDP_Boltzmann_HT-1}),
and which is valid for times of order of one collision time, as Lanford
result for the kinetic equation. 

The aim of the present paper, is to derive the large deviation Hamiltonian,
and the formula for $\epsilon$, for plasma in the weak coupling limit,
and scales much smaller than the Debye length, whose kinetic equation
is the Landau equation. The aim of our third paper, in preparation,
is to derive the large deviation Hamiltonian, and the formula for
$\epsilon$, associated to plasma in the weak coupling limit and systems
with long range interactions, independently on the hypothesis that
perturbations are at scales much smaller than the Debye length. The
kinetic equation for this third case is the Balescu--Guernsey--Lenard
\cite{Lifshitz_Pitaevskii_1981_Physical_Kinetics,Nicholson_1991}.
In both the second and third paper, we consider first the case of
homogeneous dynamics, for simplicity.\\

In this paper, we deal with the case of the kinetic theory that leads
to the Landau equation \cite{Lifshitz_Pitaevskii_1981_Physical_Kinetics,Nicholson_1991}.
The Landau equation is the law of large numbers for the relaxation
to equilibrium of a homogeneous plasma, in the weak coupling limit
and for perturbations at scales much smaller than the Debye length.
We consider more generally any system with long range interactions
at a scale much smaller than the Debye length scale (the scale at
which inertia and interaction effects do balance each others). For
these systems, we consider the rescaled empirical density $g_{\Lambda}(\mathbf{r},\mathbf{v},t)\equiv\Lambda^{-1}\sum_{n=1}^{N}\delta\left(\mathbf{v}-\mathbf{v}_{n}(t)\right))\delta\left(\mathbf{r}-\mathbf{r}_{n}\left(t\right)\right)$,
where $\Lambda$ is plasma parameter, e.g. the number of particles
in a box of the size of the Debye length. The main result of this
paper is the derivation of the Landau Hamiltonian $H_{\text{Landau}}$
that describes the dynamical large deviations for the probability
of any homogeneous evolution paths $\left\{ f(t)\right\}_{0\leq t\leq T}$
for the empirical density $\left\{ g_{\Lambda}(t)\right\}_{0\leq t\leq T}$. The natural evolution of $g_{\Lambda}$ occurs on time scales of order $\Lambda$ (except in dimension d=1 \cite{Yamaguchi_Barre_Bouchet_DR:2004_PhysicaA}). After time rescaling $\tau=t/ \Lambda$, we study the probability of $g_{\Lambda}^{s}\left(\mathbf{v},\tau\right)=g_{\Lambda}\left(\mathbf{v},\Lambda\tau\right)$ (by abuse of notation and for convenience, we still denote $g_{\Lambda}^{s}=g_{\Lambda}$). We justify that the probability that a path $\left\{ g_{\Lambda}(\tau)\right\} _{0\leq\tau\leq T}$ remains in the neighborhood of a prescribed path $\left\{ f(\tau)\right\} _{0\leq t\leq T}$ satisfies the large deviation principle
\begin{equation}
\mathbf{P}\left(\left\{ g_{\Lambda}(\tau)\right\} _{0\leq \tau \leq T}=\left\{ f(\tau)\right\} _{0\leq \tau \leq T}\right)\underset{\Lambda\rightarrow\infty}{\asymp}\text{e}^{-\Lambda\int_{0}^{T}\text{d}\tau\,\text{Sup}_{p}\left\{ \int\text{d}\mathbf{r}\text{d}\mathbf{v}\,\dot{f}p-H_{\text{Landau}}[f,p]\right\} },\label{eq:Large_Deviations_Landau}
\end{equation}
where $p\left(\mathbf{v},t\right)$ is a homogenous function over
the $\mu$-space, and where the large deviation Hamiltonian $H_{\text{Landau}}[f,p]$
is
\begin{equation}
H_{\text{Landau}}[f,p]=H_{MF}\left[f,p\right]+H_{I}\left[f,p\right],\label{eq:ham_from_boltz-2}
\end{equation}
with 
\begin{equation}
H_{MF}\left[f,p\right]=\int\text{d}\mathbf{r}\text{d}\mathbf{v}f\left\{ \mathbf{b}\left[f\right].\frac{\partial p}{\partial\mathbf{v}}+\frac{\partial}{\partial\mathbf{v}}.\left(\mathbf{D}\left[f\right].\frac{\partial p}{\partial\mathbf{v}}\right)+\mathbf{D}\left[f\right]:\frac{\partial p}{\partial\mathbf{v}}\frac{\partial p}{\partial\mathbf{v}}\right\} ,\label{eq:Hamiltonien_Mean_Field}
\end{equation}
and 
\begin{equation}
H_{I}\left[f,p\right]=-\int\text{d}\mathbf{r}\text{d\ensuremath{\mathbf{v}_{1}}}\text{d}\mathbf{v}_{2}f(\mathbf{v}_{1})f(\mathbf{v}_{2})\frac{\partial p}{\partial\mathbf{v}_{1}}\frac{\partial p}{\partial\mathbf{v}_{2}}:\mathbf{B}\left(\mathbf{v}_{1},\mathbf{v}_{2}\right).\label{eq:Hamiltonien_Interaction}
\end{equation}
The drift $\mathbf{b}$, diffusion tensor $\mathbf{D}$, and interaction
tensor \textbf{$\mathbf{B}$ }will\textbf{ }be defined in the following
sections. In particular, in this paper we show that whenever the size
of the domain is larger than the Debye length $\lambda_d$, the relevant large
deviation parameter is the plasma parameter $\Lambda$, and $H_{\text{Landau}}$ describes correctly the large deviations for any fluctuations with wave numbers $k$ with $k\lambda_{D}\gg1$ (those are the same as the validity conditions for the Landau equation). Whenever the size of the domain is smaller than the Debye length, the relevant
large deviation parameter is the number of particles, and $H_{\text{Landau}}$ describes correctly the large deviation for all fluctuations.

We give two derivations of this Hamiltonian $H_{\text{Landau}}$.
The first derivation starts from the large deviation Hamiltonian $H_{B}$
\cite{Bouchet_Boltzmann_JSP} of a dilute gas in the Boltzmann--Grad
limit (the large deviation for the Boltzmann kinetic theory) and considers
the weak scattering limit. Both the Landau equation and the large
deviation Hamiltonian $H_{\text{Landau}}$ are obtained in the weak
scattering limit from the large deviations of the Boltzmann kinetic
theory. As a second derivation, we compute the large deviation Hamiltonian
$H_{\text{Landau}}$ directly from the plasma dynamics. 

Independently from these two derivations, we also derive another new
and important result: the large deviation Hamiltonian for the empirical
density of $N$ particles driven by $N$ independent Markov processes
(equation (\ref{eq:Hamiltonien_N_Markov_Processes})). In the case
of $N$ diffusions with mean field interactions we obtain the Hamiltonian
(\ref{eq:Hamiltonien_Mean_Field}). One of the conclusions of this
paper is that, while the Landau equation can be understood as a diffusion
equation for $N$ independent particles (Fokker-Planck interpretation),
the large deviation Hamiltonian associated to the Landau equation
is not the large deviation Hamiltonian of $N$ independent particles.
The weak physical interactions impose a new interaction term (\ref{eq:Hamiltonien_Interaction})
which is essential for describing the large deviations. We prove that
this interaction term (\ref{eq:Hamiltonien_Interaction}) is also
crucial for the energy conservation properties of the statistical
field theory. Finally, all along the paper we prove the expected properties
of the obtained Hamiltonian: conservation law symmetries, time-reversal
symmetry, and we prove that the entropy is the negative of the quasipotential
up to conservation laws. 

We also explain that the path large deviation principle for the empirical
distribution implies a gradient structure for the Landau equation.
This gradient structure does not involve the Wasserstein distance
as in many kinetic theories, but another more intricate distance that
takes into account of the effect of weak interaction between particles
in the kinetic limit.\\

The subject of plasma fluctuations is a classical one, see for instance  \S 51 of \cite{Lifshitz_Pitaevskii_1981_Physical_Kinetics}, or chapter 11 of \cite{akhiezer1975plasma}, among hundreds of other publications. For instance, the space-time two-point correlations for the fluctuations of the distribution function and potential of a plasma with a non-equilibrium distribution function which is stable for Vlasov dynamics, for times much smaller than the evolution time of the distribution function itself, can be computed either from a Klimontovich approach \cite{Lifshitz_Pitaevskii_1981_Physical_Kinetics}, a truncation of the BBGKY hierarchy \cite{Nicholson_1991}, or using equipartition of local van Kampen modes \cite{morrison2008fluctuation}. One may wonder how the present work connects to those classical results. First, as will be clear in section \ref{subsec:The-quasi-stationary-Gaussian}, our derivation starts from the classical formulas for the local in time fluctuations of non-equilibrium stable distributions. Then our approach is fully consistent with the classical results of fluctuations in plasma. However, we address a question of a nature that has never been considered so far: the probability that those local fluctuations lead to a large deviation in the long term evolution of the distribution function. Our main result, the large deviation Hamiltonian that describes the long term path probability for the distribution function, is thus entirely new, as far as we know. It is fully compatible with the classical theories of local fluctuations in plasmas.\\

The kinetic theory of plasmas and systems with long-range interactions is also a very active subject in mathematics, currently, with the proof of the validity of the Balescu--Guernsey--Lenard equation up to time scales of order $N^r$ with $r<1$ \cite{duerinckx2021lenard}, the study of fluctuations \cite{lancellotti2016time} and correlation functions \cite{paul2019size}, the proof of a central limit theorem for fluctuations for short times \cite{duerinckx2021size}, and the study of two-point correlation functions \cite{velazquez2018two}.

In section \ref{sec:Dynamical-large-deviations}, we present the expected
general properties for the dynamical large deviations of a kinetic
theory. In section \ref{sec:Dynamical-large-deviations}, we also
present heuristically two important and classical frameworks for dynamical
large deviation theory: large deviations due to $N$ independent small
increments leading to an effect of order $1$, and large deviations
for slow-fast systems. In section \ref{sec:Dynamics-of-plasma}, we
present the dynamics of $N$ particles with Coulomb interactions and
the related kinetic equations: the Vlasov, the Balescu--Guernsey--Lenard
and the Landau equations. Inspired by the structure of the Landau
and Balescu--Guernsey--Lenard equation, which
can be seen as non-linear Fokker--Planck equations, we compute
in section \ref{sec:N independent diffusions and N diffusions with mean field coupling}
the large deviation Hamiltonian for the empirical density of $N$
particles with diffusions coupled in a mean field way. We show that
it cannot be the large deviation Hamiltonian for neither the Balescu--Guernsey--Lenard
nor the Landau equation. In section \ref{sec:Large-dev-from-boltzmann},
we derive the large deviation Hamiltonian for the kinetic theory associated to the Landau equation,
from the one previously obtained for the Boltzmann equation. This
Hamiltonian is quadratic in $p$ the conjugated variable to $\dot{f}$,
showing that for the Landau equation Gaussian fluctuations properly
describe path large deviations. It is natural to use this Hamiltonian
large deviation principle for the Landau equation kinetic theory, to conjecture a
Hamiltonian large deviation principle for the kinetic theory leading to the Balescu--Guernsey--Lenard
equation, by replacing the Landau collision kernel by the Balescu--Guernsey--Lenard
one. We call this Hamiltonian the dressed Landau Hamiltonian. However,
we show in section \ref{sec:Large-dev-from-boltzmann} that this dressed
Landau Hamiltonian is not the large deviation Hamiltonian associated to the
kinetic theory leading to the Balescu--Guernsey--Lenard equation. We argue that
the large deviation Hamiltonian for the Balescu--Guernsey--Lenard kinetic theory 
is not quadratic in the conjugated momentum (the large deviations
are driven by non-Gaussian fluctuations). Finally, in section  \ref{sec:LD-from-microscopic-dynamics}, we compute the large deviation Hamiltonian directly from the $N$ particle dynamics. We show that a cumulant
expansion coincides with the dressed Landau Hamiltonian, up to a certain
truncation in terms of the power of the interaction potential. We explain that this justifies
that the large deviation Hamiltonian for the kinetic theory associated to the Landau equation is quadratic in the conjugated momentum, 
because of the limit of small scales compared to the Debye length. This result is fully consistent with the one obtained in section \ref{sec:Large-dev-from-boltzmann}.

\section{Dynamical large deviations and kinetic theories \label{sec:Dynamical-large-deviations}}

The aim of many works in statistical mechanics is to describe the
evolution of the empirical density of particle dynamics. For instance,
in this work, we will consider the rescaled empirical distribution
$g_{\epsilon}(\mathbf{r},\mathbf{v},t)=\epsilon\sum_{n=1}^{N}\delta(\mathbf{r}-\mathbf{r}_{n}(t))\delta(\mathbf{v}-\mathbf{v}_{n}(t))$.
A large deviation principle for the dynamics of the empirical distribution
is a result that reads
\begin{equation}
\mathbf{P}\left(\left\{ g_{\epsilon}(t)\right\} _{0\leq t\leq T}=\left\{ g(t)\right\} _{0\leq t\leq T}\right)\underset{\epsilon\rightarrow0}{\asymp}\text{e}^{-\frac{1}{\epsilon}\int_{0}^{T}\text{d}t\,\text{Sup}_{p}\left\{ \int\dot{g}p\text{d}\mathbf{r}\text{d}\mathbf{v}-H[g,p]\right\} },\label{eq:Large_Deviations_Landau-1}
\end{equation}
with the prescription that $g_{\epsilon}(t=0)$ is in the neighborhood of $g(t=0)$, where $\epsilon$ is a small parameter that can be related to $N$.
This section present a set of known results about large deviation
theory which are essential for the following discussion. In section
\ref{subsec:LD_Kinetic_theories} we describe the expected properties
of any such large deviation principle for the kinetic theory of the
empirical distribution. A more detailed account of a similar discussion
can be found in \cite{Bouchet_Boltzmann_JSP}. In section \ref{subsec:Dynamical-large-deviations},
we present two important frameworks that allow to compute dynamical
large deviations: on one hand, large deviations due to $N$ independent
small increments leading to an effect of order $1$, and on the other
hand, large deviations for slow-fast systems.

\subsection{Large deviation for kinetic theories\label{subsec:LD_Kinetic_theories}}

\subsubsection{General properties of path large deviations and expected properties
for large deviations for kinetic theories}

\paragraph{Most probable evolution}

We consider the properties of a stochastic process whose rare fluctuations
are described, at the level of large deviations, by the action 
\begin{equation}
\mathcal{A}\left[g\right]=\int_{0}^{T}\mbox{d}t\,L\left[g,\dot{g}\right]=\int_{0}^{T}\mbox{d}t\,\text{Sup}_{p}\left[\int p\dot{g}-H\left[g,p\right]\right].\label{eq:Action}
\end{equation}
(see equation (\ref{eq:Large_Deviations_Landau-1})). The kinetic
equation is expected to be the most probable evolution corresponding
to the action (\ref{eq:Action}), and with initial condition $g_{r}(t=0)=g$.
It is also called a \textbf{\emph{relaxation path issued from $g$}}.
It solves $\frac{\partial g_{r}}{\partial t}=R\left[g_{r}\right]$,
with initial condition $g_{r}(t=0)=g$, where $R\left[g\right]=\arg\inf_{\dot{g}}L\left[g,\dot{g}\right]$.
Then one easily proves that 
\begin{equation}
\dot{g}=\frac{\delta H}{\delta p}\left[g,p=0\right],\label{eq:Most-Probab_Evo}
\end{equation}
is the kinetic equation. 

\paragraph{Quasipotential and macrostate entropy}

We assume that the stochastic process $g_{\epsilon}$ has a stationary
distribution $P_{s}$ whose dynamics follows the large deviation principle
\begin{equation}
P_{s}(g)\equiv\mathbb{E}\left[\delta\left(g_{\epsilon}-g\right)\right]\underset{\epsilon\downarrow0}{\asymp}\exp\left(-\frac{U\left[g\right]}{\epsilon}\right),\label{eq:Quasipotential_Definition-1}
\end{equation}
where $U$ is called the \textbf{\emph{quasipotential}}. In order
to simplify the following discussion, we also assume that the relaxation
equation has a single fixed point $g_{0}$ and that any solution to
the relaxation equation converges to $g_{0}$ . Then the quasipotential
satisfies 
\[
U\left[g\right]=\inf_{\left\{ \left\{ \tilde{g}(t)\right\} {}_{-\infty\leq t\leq0}\left|\tilde{g}(-\infty)=g_{0}\,\,\,\mbox{and}\,\,\,\tilde{g}(0)=g\right.\right\} }\int_{-\infty}^{0}\mbox{d}t\,L\left[\tilde{g},\dot{\tilde{g}}\right].
\]
The minimizer of this variational problem, that is the most probable
path starting from $g_{0}$ and ending at $g$, is denoted $g_{f}(t,g)$
and is called the \textbf{\emph{fluctuation path ending at $g$.}}

For many kinetic theory, we expect from equilibrium statistical mechanics
that the quasipotential $U\left[g\right]$ is the opposite of the
entropy $S\left[g\right]=-\int\text{d}\mathbf{v}\text{d}\mathbf{r}\,g\log g$
constrained by the conserved quantities 
\[
U\left[g\right]=\left\{ \begin{array}{l}
-S\left[g\right]\,\,\,\text{if}\,\,\,M\left[g\right]=1,\,\,\,\mathbf{P}\left[g\right]=0,\,\,\,\mbox{and}\,\,\,E\left[g\right]=E_{0}\\
-\infty\,\,\,\mbox{otherwise}.
\end{array}\right.
\]
\\

We have the following properties which are direct consequences of
the definitions of $H$ and $L$, and whose proofs are classical and
given for example in sections 7.2 to 7.4 of \cite{Bouchet_Boltzmann_JSP}:
\begin{enumerate}
\item $H$ is a convex function of the variable $p$ and $H\left[g,p=0\right]=0$,
see sec. 7.2.1 of \cite{Bouchet_Boltzmann_JSP}.
\item The relaxation paths solve the equation $\frac{\partial g}{\partial t}=R\left[g\right]$
with $\inf_{\dot{g}}L[g,\dot{g}]=0=L[g,R\left[g\right]]$, and $R\left[g\right]=\frac{\delta H}{\delta p}\left[g,0\right]$,
see sec. 7.2.2 of \cite{Bouchet_Boltzmann_JSP}.
\item The quasipotential solves the stationary \textbf{Hamilton--Jacobi
equation} 
\begin{equation}
H\left[g,\frac{\delta U}{\delta g}\right]=0,\label{eq:Stationary_Hamilton_Jacobi}
\end{equation}
see sec. 7.2.3 of \cite{Bouchet_Boltzmann_JSP}.
\item \textbf{The fluctuation paths} solve 
\[
\dot{g}=F\left[g\right]\equiv\frac{\delta H}{\delta p}\left[g,\frac{\delta U}{\delta g}\right],
\]
see sec. 7.2.4 of \cite{Bouchet_Boltzmann_JSP}.
\item As $H$ is convex, the quasipotential decreases along the relaxation
paths 
\begin{equation}
\frac{\mbox{d}U}{\mbox{d}t}\left[g_{r}\right]=H[g_{r},0]-H\left[g_{r},\frac{\delta U}{\delta g}\left[g_{r}\right]\right]+\int\text{d}\mathbf{r}\text{d}\mathbf{v}\,\frac{\delta H}{\delta p}\left[g_{r},0\right]\frac{\delta U}{\delta g}\left[g_{r}\right]\leq0,\label{decrease}
\end{equation}
see sec. 7.2.5 of \cite{Bouchet_Boltzmann_JSP}. For kinetic theories,
because the quasipotential is the entropy whenever the conservation
laws are verified, we can immediately conclude that the entropy will
increase along the solution of the kinetic equation.
\item As $H$ is convex, the quasipotential increases along the fluctuation
paths 
\begin{equation}
\frac{\mbox{d}U}{\mbox{d}t}\left[g_{f}\right]=H[g_{f},0]-H\left[g_{f},\frac{\delta U}{\delta g}\left[g_{f}\right]\right]+\int\text{d}\mathbf{r}\text{d}\mathbf{v}\,\frac{\delta H}{\delta p}\left[g_{f},\frac{\delta U}{\delta g}\left[g_{f}\right]\right]\frac{\delta U}{\delta g}\left[g_{f}\right]\geq0,\label{increase}
\end{equation}
see sec. 7.2.5 of \cite{Bouchet_Boltzmann_JSP}. For kinetic theories,
because the quasipotential is the entropy whenever the conservation
laws are verified, we can immediately conclude that the entropy will
decrease along the fluctuation paths.
\item \textbf{\emph{Generalized detailed balance }}(see sec. 7.3.2 of \cite{Bouchet_Boltzmann_JSP}).
Let $I$ be an involution that characterizes time-reversal symmetry
(for instance the map that correspond to velocity or momentum inversion
in many systems). We assume that $I$ is self adjoint for the $L^{2}$
scalar product, that is $\int\text{d}\mathbf{r}\text{d}\mathbf{v}\,I\left[g\right]p=\int\text{d}\mathbf{r}\text{d}\mathbf{v}\,gI\left[p\right]$.
The detailed balance conditions for the quasipotential $U$ combined
with the involution $I$ are $U\left[g\right]=U\left[I\left[g\right]\right]$
is 
\begin{equation}
H\left[I\left[g\right],-I\left[p\right]\right]=H\left[g,p+\frac{\delta U}{\delta g}\right].\label{eq:detailed balance}
\end{equation}
For any systems for which the microscopic dynamics is time reversible,
we can infer that the stochastic process of the empirical distribution
has to be time-reversal symmetric. As a consequence the large deviation
principle should verify detailed balance and the symmetry relation
has to be verified.
\item As can be easily checked, if either the detailed balance or the generalized
detailed balance conditions are verified, then $U$ satisfies the
stationary Hamilton-Jacobi equation (\ref{eq:Stationary_Hamilton_Jacobi}). 
\item If the detailed balance condition is verified, and if $U$ is the
quasipotential, then for a path $\left\{ g(t)\right\} _{0\leq t\leq T}$
and its time reversed one $\left\{ I\left[g(T-t)\right]\right\} _{0\leq t\leq T}$
we have the symmetry for the path probability 
\small
\[
P\left[\left\{ g_{\epsilon}(t)\right\} _{0\leq t\leq T}=\left\{ g(t)\right\} _{0\leq t\leq T}\right]\mbox{e}^{-\frac{U\left[g\left(t=0\right)\right])}{\epsilon}}=P\left[\left\{ g_{\epsilon}(t)\right\} _{0\leq t\leq T}=\left\{ I\left[g(T-t)\right]\right\} _{0\leq t\leq T}\right]\mbox{e}^{-\frac{U\left[I\left[g(t=T)\right]\right]}{\epsilon}},
\]
\normalsize
see sec. 7.3.1 of \cite{Bouchet_Boltzmann_JSP}.
\item \textbf{\emph{Conserved quantities }}(see sec\textbf{\emph{. }}7.2.6
of \cite{Bouchet_Boltzmann_JSP})\emph{.} At the level of the large
deviations, the condition for $C\left[g\right]$ to be a conserved
quantity is either 
\[
\mbox{for any}\,\,g\,\,\mbox{and}\,\,p,\,\,\,L\left[g,\dot{g}\right]=+\infty\,\,\,\mbox{if}\,\,\,\int\text{d}\mathbf{r}\text{d}\mathbf{v}\,\frac{\partial g}{\partial t}\frac{\delta C}{\delta g}\neq0,
\]
or 
\begin{equation}
\mbox{for any}\,\,g\,\,\mbox{and}\,\,p,\,\,\,\int\text{d}\mathbf{r}\text{d}\mathbf{v}\,\frac{\delta H}{\delta p}\left[g,p\right]\frac{\delta C}{\delta g}=0.\label{eq:Conservation_Law_H-1}
\end{equation}
In general, kinetic theories conserve at least mass, momentum and
energy. 
\item \textbf{\emph{A sufficient condition for $U$ to be the quasipotential
}}(see sec. 7.4 of \cite{Bouchet_Boltzmann_JSP}). If $U$ solves
the Hamilton--Jacobi equation, if $U$ has a single minimum
$g_{0}$ with $U\left[g_{0}\right]=0$, and if for any $g$ the solution
of the reverse fluctuation path dynamics $\frac{\partial\tilde{g}}{\partial t}=-F\left[\tilde{g}\right]=-\frac{\delta H}{\delta p}\left[\tilde{g},\frac{\delta U}{\delta\tilde{g}}\right]$
with $\tilde{g}(0)=g$ converges to $g_{0}$ for large times, then
$U$ is the quasipotential.
\end{enumerate}

\subsection{Dynamical large deviations\label{subsec:Dynamical-large-deviations}}

When the evolution of a stochastic process is the consequence of the
effect of a large number of small amplitude and statistically independent
moves, in the limit of a large number of moves, a law of large number
naturally follows. It is often very important to understand the large
deviations with respect to this law of large number. For continuous
time Markov processes, for instance diffusions with small noises,
or more generally locally infinitely divisible processes, a general
framework can be developed in order to estimate the probability of
large deviations. In section \ref{subsec:Dynamical-large-deviations},
taken from \cite{Bouchet_Boltzmann_JSP} and initially inspired by
\cite{feng2006large,FW2012}, we present this framework briefly and
the main result: the formula \eqref{eq:H-Generator} for computing
the large deviation Hamiltonian in this case. 

Another classical framework for large deviations are large deviations
for the effective dynamics of the slow variable in a slow-fast dynamics
(time averaging of the fast degrees of freedom). This classical framework
is discussed in the case of stochastic processes in \cite{FW2012,Veretennikov}.
When the slow dynamics is deterministic similar results have been
proven for instance by Kifer \cite{kifer2004averaging}. A simple heuristic account is given
in \cite{Bouchet_Grafke_Tangarife_Vanden-Eijnden_2015_largedeviations}.
\cite{Bouchet_Grafke_Tangarife_Vanden-Eijnden_2015_largedeviations}
discuss also at length the case when the fast variable is an Ornstein-Uhlenbeck
and the coupling with the slow variable is through a quadratic form.
In this specific case the Hamiltonian can be computed by solving a
matrix Riccati equation.

\subsubsection{Large deviation rate functions from the infinitesimal generator of
a continuous time Markov process \label{sec:Large_Deviations_Generator-1}}

We consider $\left\{ g_{\epsilon}(t)\right\} _{0\leq t\leq T}$, where for any $t$, $g_\epsilon(t) \in X$, a
family of continuous time Markov processes parametrized by a real
number $\epsilon$. We denote $G_{\epsilon}$ the infinitesimal generator
of the process $g_\epsilon$. $G_{\epsilon}$ acts on the space of test functions $\phi:X\rightarrow\mathbb{R}$. It is defined by
\begin{equation}
G_\epsilon\left[\phi\right](g)=\lim_{t\downarrow0}\frac{\mathbb{E}_{g}\left[\phi(g_\epsilon(t))\right]-\mathbb{\phi}(g)}{t},\label{eq:Infinitetisimal_Generator}
\end{equation}
where $\mathbb{E}_{g}$ is the average over the stochastic process $\left\{ g_{\epsilon}(t)\right\} _{0\leq t\leq T}$ conditioned on the initial condition $g_{\epsilon}(t=0)=g$.
We assume that for all $p\in L^{2}\left(\mathbb{T}^{3}\times\mathbb{R}^{3}\right)$
the limit 
\begin{equation}
H[g,p]=\lim_{\epsilon\downarrow0}\epsilon G_{\epsilon}\left[\mbox{e}^{\frac{1}{\epsilon}\int\text{d}\mathbf{r}\text{d}\mathbf{v}\,p\left(\mathbf{r},\mathbf{v}\right)g\left(\mathbf{r},\mathbf{v}\right)}\right]\mbox{e}^{-\frac{1}{\epsilon}\int\text{d}\mathbf{r}\text{d}\mathbf{v}\,p\left(\mathbf{r},\mathbf{v}\right)g\left(\mathbf{r},\mathbf{v}\right)}\label{eq:H-Generator}
\end{equation}
exists. Then the family $g_{\epsilon}$ satisfies a large deviation
principle with rate $\epsilon$ and rate function 
\begin{equation}
L\left[g,\dot{g}\right]=\sup_{p}\left\{ p\dot{g}-H\left[g,p\right]\right\} .\label{eq:Lagrangian}
\end{equation}
This means that the probability that the path $\left\{ g_{\epsilon}(t)\right\} _{0\leq t<T}$
be in a neighborhood of $\left\{ g(t)\right\} _{0\leq t<T}$, with the prescription that $g_{\epsilon}(t=0)$ is in the neighborhood of $g(t=0)$, satisfies
\begin{equation}
P\left[\left\{ g_{\epsilon}(t)\right\} _{0\leq t<T}=\left\{ g(t)\right\} _{0\leq t<T}\right]\underset{\epsilon\downarrow0}{\asymp}\exp\left(-\frac{\int_{0}^{T}\mbox{d}t\,L\left[g,\dot{g}\right]}{\epsilon}\right),\label{eq:Large deviations path}
\end{equation}
where the symbol $\underset{\epsilon\downarrow0}{\asymp}$ is a logarithm
equivalence ($g_{\epsilon}\underset{\epsilon\downarrow0}{\asymp}\exp(\varphi/\epsilon)\iff\lim_{\epsilon\downarrow0}\epsilon\log g_{\epsilon}=\varphi$).

This result is proven for specific cases (diffusions, locally infinitely
divisible processes) in the Theorem 2.1, page 127, of the third edition
of Freidlin-Wentzell textbook \cite{FW2012}. A general heuristic
derivation is given in section 7.1.2 of \cite{Bouchet_Boltzmann_JSP}.
Equation (\ref{eq:H-Generator}) will be the key starting point for
several results of this paper. For instance, we apply this framework
to the fluctuations of $N$ independent diffusions and $N$ diffusions
with mean field coupling in section \ref{sec:N independent diffusions and N diffusions with mean field coupling}.

In formula (\ref{eq:H-Generator}) the infinitesimal generator is
tested through the function $\mbox{e}^{\frac{1}{\epsilon}\int\text{d}\mathbf{r}\text{d}\mathbf{v}\,pg}$. In the small $\epsilon$ limit, this tests changes of the
observable which are of order of $\epsilon$. The $\epsilon$ prefactor
in the right hand side of equation (\ref{eq:H-Generator}) means that
the overall effect of these small changes of order $\epsilon$ is
expected to be of order $1/\epsilon$. $H$ in formula (\ref{eq:H-Generator})
thus accounts for the effects of a large number (of order $1/\epsilon$)
of small amplitude statistically independent moves (each one of order
$\epsilon$).

\subsubsection{Large deviation for slow-fast systems \label{sec:Large_Deviations_Slow_Fast}}

We consider the slow-fast dynamics
\begin{equation}
\left\{ \begin{array}{rcl}
\frac{\text{d}X_{\epsilon}}{\text{d}\tau} & = & \alpha(X_{\epsilon},Y_{\epsilon})\\
\frac{\text{d}Y_{\epsilon}}{\text{d}\tau} & = & \frac{1}{\epsilon}\beta(X_{\epsilon},Y_{\epsilon})+\frac{1}{\sqrt{\epsilon}}\gamma(X_{\epsilon},Y_{\epsilon})\frac{\mbox{d}W}{\mbox{d}\tau}
\end{array}\right.,\label{eq:slow-fast-dynamics}
\end{equation}
where $X_{\epsilon}$ is the slow variable, $Y_{\epsilon}$ the fast
variable, $W$ a Wiener process, and $\epsilon$ quantifies the time
scale separation. We assume that the dynamics for $Y_{\epsilon}$
is mixing over timescales of order $\epsilon$. The following discussion
would apply for other classes of dynamics for $Y_{\epsilon}$, beyond
diffusions, with little modifications, for instance for chaotic deterministic
systems with mixing hypothesis.

We are interested in the slow dynamics for $X_{\epsilon}$. Then for
generic hypotheses, with the prescription that $X_{\epsilon}(\tau=0)$ is in the neighborhood of $x(\tau=0)$, we have the large deviation principle
\begin{equation}
\mathbb{P}\left(X_{\epsilon}=x\right)\underset{\epsilon\rightarrow0}{\asymp}\text{e}^{-\frac{1}{\epsilon}\int_{0}^{T}\text{Sup}_{p}\left\{ \dot{x}.p-H(x,p)\right\} \text{d}\tau}\label{eq:Large_Deviations_Slow_Fast}
\end{equation}
\begin{equation}
\text{with }\,\,\,H(x,p)=\lim_{T\rightarrow\infty}\frac{1}{T}\log\mathbb{E}_{x}\left\{ \exp\left[p.\int_{0}^{T}\alpha(x,Y_{x}(t))\text{d}t\right]\right\} ,\label{eq:Hamiltonian_Slow_Fast}
\end{equation}
where $p$ is conjugated to $\dot{x}$, the average $\mathbb{E}_{x}$
is an average over the $Y_{x}$ process with frozen $x$ (the solution
of $\frac{\text{d}Y_{x}}{\text{d}t}=\beta\left(x,Y_{x}\right)+\gamma\left(x,Y_{x}\right)\frac{\mbox{d}W}{\mbox{d}t}$). 

This classical result is proven in the case of stochastic processes
in \cite{FW2012,Veretennikov}. When the slow dynamics is deterministic,
similar results have been proven for instance by Kiffer. A simple
heuristic account for any Markov dynamics is given in \cite{Bouchet_Grafke_Tangarife_Vanden-Eijnden_2015_largedeviations}.
The result (\ref{eq:slow-fast-dynamics}-\ref{eq:Hamiltonian_Slow_Fast})
is easily heuristically understood as $L(x,\dot{x})=\sup_{p}\left\{ \dot{x}.p-H(x,p)\right\} $
appears as a large-time large deviations result, of the Freidlin-Wentzell
type, for the Newton increment of the slow variable 
\[
\frac{X_{\epsilon}(\tau+\Delta\tau)-x}{\Delta\tau}=\frac{1}{\Delta\tau}\int_{0}^{\Delta\tau}\alpha\left(X_{\epsilon}(u),Y_{\epsilon}\left(u\right)\right)\,\text{d}u\simeq\frac{\epsilon}{\Delta\tau}\int_{0}^{\frac{\Delta\tau}{\epsilon}}\alpha\left(x,Y_{x}\left(t\right)\right)\,\text{d}t.
\]
Then formula (\ref{eq:Hamiltonian_Slow_Fast}), with $L(x,\dot{x})=\text{Sup}_{p}\left\{ \dot{x}.p-H(x,p)\right\} $,
appears as a Gärtner--Ellis formula for the large time large
deviations 
\[
\mathbb{E}_{x}\left[\delta\left(\frac{X_{\epsilon}(\tau+\Delta\tau)-x}{\Delta\tau}-\dot{x}\right)\right]\underset{\epsilon\rightarrow0}{\asymp}\text{e}^{-\frac{L\left(x,\dot{x}\right)\Delta\tau}{\epsilon}}.
\]
This last formula is the temporal increment of formula (\ref{sec:Large_Deviations_Slow_Fast}).

We will use formula (\ref{eq:slow-fast-dynamics}-\ref{eq:Hamiltonian_Slow_Fast})
for computing the large deviations of the empirical density, from
the microscopic dynamics, in section \ref{sec:LD-from-microscopic-dynamics}.

\section{Dynamics of plasmas\label{sec:Dynamics-of-plasma}}

In this section we set up the definitions, and present known results
about the kinetic theory of the dynamics of $N$ particles with Coulomb
interactions, in limit of a large plasma parameter (or equivalently
weak coupling). In section \ref{subsec:plasma}, we define the Hamiltonian
dynamics of $N$ particles coupled by a Coulomb pairwise interaction.
In section \ref{subsec:Vlasov-equation}, we introduce the Vlasov
equation that describes the evolution of the empirical density on
timescales of order one. In section \ref{subsec:The-Balescu=002013Guernsey=002013Lenard-equa},
we introduce the Balescu--Guernsey--Lenard equation
that describes the long time relaxation of the empirical density,
from Vlasov stationary solutions to the Maxwell-Boltzmann equilibrium
distribution, and some of its important physical properties. In section
\ref{subsec:The-Landau-equation} we present the Landau equation,
which is an approximation of the Balescu--Guernsey--Lenard
equation which is valid for scales which are small compared to the
Debye length. In section \ref{subsec:The-Balescu=002013Guernsey=002013Lenard-and},
we show that these equations can be seen as non-linear Fokker-Planck
equations.

\subsection{The dynamics of the Coulomb plasma\label{subsec:plasma}}

We consider of a Coulomb plasma of $N$ particles with positions $\left\{ \mathbf{r}_{n}\right\} _{1\leq n\leq N}$
and velocities $\left\{ \mathbf{v}_{n}\right\} _{1\leq n\leq N}$,
and with equal charge $e$ and mass $m$. We consider that $\mathbf{r}_{n}$
belongs to a 3-dimensional torus $\mathbb{T}^{3}$ of size $L^{3}$
(doubly periodic boundary conditions), and $\mathbf{v}_{n}\in\mathbb{R}^{3}$.
However most of our discussion easily generalizables to $\mathbf{r}_{n}\in\mathbb{R}^{3}$,
with slight modifications. The dynamics is a Hamiltonian one with
\begin{equation}
\begin{cases}
{\displaystyle \frac{\text{d}\mathbf{r}_{n}}{\text{d}t}} & =\mathbf{v}_{n}\\
\\
{\displaystyle \frac{\text{d}\mathbf{v}_{n}}{\text{d}t}} & {\displaystyle =-\frac{e^{2}}{4\pi\text{\ensuremath{\epsilon}}_{0}m}\sum_{m\neq n}\frac{\text{d}}{\text{d}\mathbf{r}_{n}}W\left(\mathbf{r}_{n}-\mathbf{r}_{m}\right)}
\end{cases}\label{eq:coulomb_dynamics}
\end{equation}
where $\epsilon_{0}$ is the vacuum permittivity and $W$ is the Coulomb
potential. In both a finite box and an infinite space, $W$ can be
defined through its Fourier transform 
\[
\hat{W}\left(\mathbf{k}\right)=\int\text{d}\mathbf{r}\,\text{e}^{-i\mathbf{k}.\mathbf{r}}W\left(\mathbf{r}\right),
\]
with
\[
\hat{W}\left(\mathbf{k}\right)=\frac{1}{k^{2}},
\]
and where $k=\mathbf{\left|k\right|}$ (this definition is equivalent to $-\Delta W = \delta(\mathbf{r})$). We define the Debye length
$\lambda_{D}=\left(\frac{\epsilon_{0}k_{B}TL^{3}}{e^{2}N}\right)^{1/2}$,
where $k_{B}$ is the Boltzmann constant and $T$ the temperature.
This length is the typical length beyond which Coulomb interaction
are screened \cite{Nicholson_1991}. We also define the plasma electron
frequency $\omega_{pe}=\left(\frac{e^{2}N}{\epsilon_{0}mL^{3}}\right)^{1/2}$,
which is the pulsation of the Langmuir waves in a plasma \cite{Nicholson_1991},
and the thermal velocity $v_{T}=\lambda_{D}\omega_{pe}=\sqrt{k_{B}T/m}$.
Then, if we use the dimensionless variables
\[
\mathbf{\tilde{r}}=\mathbf{r}/\lambda_{D},\,\,\,\tilde{\mathbf{v}}=\mathbf{v}/v_{T}\,\,\,\text{and}\,\,\,\tilde{t}=\omega_{pe}t,
\]
the dimensionless dynamical equations (\ref{eq:coulomb_dynamics})
read
\[
\begin{cases}
{\displaystyle \frac{\text{d}\mathbf{\tilde{r}}_{n}}{\text{d}\tilde{t}}} & =\mathbf{\tilde{v}}_{n}\\
\\
{\displaystyle \frac{\text{d}\mathbf{\tilde{v}}_{n}}{\text{d}\tilde{t}}} & {\displaystyle =-\frac{1}{\Lambda}\sum_{m\neq n}\frac{\text{d}}{\text{d}\mathbf{\tilde{r}}_{n}}\tilde{W}\left(\tilde{\mathbf{r}}_{n}-\mathbf{\tilde{r}}_{m}\right)}
\end{cases}
\]
where $\Lambda\equiv N\left(\lambda_{D}/L\right)^{3}$ is the so-called
plasma parameter. $\Lambda$ is the number of particles in a box of
size of the Debye length. In this new system of units, called plasma
units, $\tilde{\mathbf{r}}_{n}$ belongs to the 3-dimensional torus
$\left(L/\lambda_{D}\right)\mathbb{T}^{3}.$ The dimensionless Coulomb
potential $\tilde{W}$ is defined by

\[
\hat{W}\left(\mathbf{\tilde{k}}\right)=\int\text{d}\mathbf{\tilde{r}}\,\text{e}^{-i\tilde{\mathbf{k}}.\tilde{\mathbf{r}}}\tilde{W}\left(\tilde{\mathbf{r}}\right),
\]
with $\hat{W}\left(\mathbf{\tilde{k}}\right)=\frac{1}{\tilde{k}^{2}}.$
For simplicity, in the following we omit the tildes when referring
to the dimensionless variables. We will work in dimensionless variables,
and give the main results in both dimensionless and physical variables. 

We call $\mu-$space the $\left(\mathbf{r},\mathbf{v}\right)$ space.
The $\mu-$space is of dimension $6$. Let us define $g_{\Lambda}$
the $\mu-$space empirical distribution function for the positions
and velocities of the $N$ particles rescaled by the plasma parameter

\begin{equation}
g_{\Lambda}(\mathbf{r},\mathbf{v},t)=\frac{1}{\Lambda}\sum_{n=1}^{N}\delta(\mathbf{r}-\mathbf{r}_{n}(t))\delta(\mathbf{v}-\mathbf{v}_{n}(t)).\label{eq:g_Lambda}
\end{equation}
In the following we will consider the large plasma parameter limit,
$\Lambda\rightarrow\infty$. Considering that $\Lambda$ is the number
of particles in a box of size of the Debye length, and that in our
non-dimensional units the Debye length is fixed, the scaling $1/\Lambda$
in front of the empirical density (\ref{eq:g_Lambda}) is natural. 

If the box size $L$ is larger than the Debye length $\lambda_{D}$,
the interactions are screened beyond the Debye length and the effective
interaction length scale is $\lambda_{D}$. Otherwise, if the size
of the box is smaller than the Debye length, then the interactions
are not screened in the box and they take place on a length scale
$L$. We call $\ell=\min\left\{ \lambda_{D},L\right\} $ the effective
interaction length scale.

In the following, we study the asymptotic dynamics of $g_{\Lambda}$
as the number of particles in a box of the size of the effective interaction
length scale, e.g. $N\ell^{3}/L^{3}$ goes to infinity. If $L>\lambda_{D}$,
this asymptotic regime is the limit of a large plasma parameter $\Lambda$;
if $L<\lambda_{D}$, it is the limit of a large number of particles
$N$. In this paper, we present detailed results for the case $L>\lambda_{D}$,
and we briefly discuss the slight modifications relevant for the case
$L<\lambda_{D}$ at the end of section \ref{sec:LD-from-microscopic-dynamics}.

\subsection{The Vlasov equation\label{subsec:Vlasov-equation}}

From equation (\ref{eq:coulomb_dynamics}), one immediately obtains
the Klimontovich equation
\begin{equation}
\frac{\partial g_{\Lambda}}{\partial t}+\mathbf{v}\cdot\frac{\partial g_{\Lambda}}{\partial\mathbf{r}}-\frac{\partial V\left[g_{\Lambda}\right]}{\partial\mathbf{r}}\cdot\frac{\partial g_{\Lambda}}{\partial\mathbf{v}}=0,\label{eq:Klimontovich_eq}
\end{equation}
where $V[g_{\Lambda}](\mathbf{r},t)=\int\text{d}\mathbf{v}'\text{d}\mathbf{r}'W(\mathbf{r}-\mathbf{r}')g_{\Lambda}(\mathbf{r}',\mathbf{v}',t)$.
This is an exact equation for the evolution of $g_{\Lambda}$, if
$W$ is regular enough. For the Coulomb interaction, the formal equation
\eqref{eq:Klimontovich_eq} has to be interpreted carefully. In the
following, we do not discuss the divergences that might occur related
to small scale interactions. At a mathematic level, this would be
equivalent to considering a potential which is regularized at small
scales, and smooth. The Klimontovich equation \eqref{eq:Klimontovich_eq}
contains all the information about the trajectories of the $N$ particles.
We would like to build a kinetic theory, that describes the stochastic
process for $g_{\Lambda}$ at a mesoscopic level.

An important first result is that the sequence $\{g_{\Lambda}\text{\}}$
obeys a law of large numbers when $\Lambda\rightarrow+\infty$. More
precisely, if we assume there is a set of initials conditions $\{g_{\Lambda}^{0}\text{\}}$
such that $\lim_{\Lambda\rightarrow+\infty} g_{\Lambda}^{0}\left(\mathbf{r},\mathbf{v}\right)=g^{0}\left(\mathbf{r},\mathbf{v}\right)$,
then over finite time interval $t\in\left[0,T\right]$, the empirical
distribution function $g_{\Lambda}(t)$ converges to $g(t)$ as $\Lambda$
goes to infinity, where $g$ solves the Vlasov equation
\begin{equation}
\frac{\partial g}{\partial t}+\mathbf{v}\cdot\frac{\partial g}{\partial\mathbf{r}}-\frac{\partial V\left[g\right]}{\partial\mathbf{r}}\cdot\frac{\partial g}{\partial\mathbf{v}}=0\,\,\,\text{with}\,\,\,g\left(\mathbf{r},\mathbf{v},t=0\right)=g^{0}\left(\mathbf{r},\mathbf{v}\right).\label{eq:vlasov}
\end{equation}
As the Klimontovich and the Vlasov equations are formally the same,
this is actually a stability result for the Vlasov equation. It has
first been proven for smooth interactions by Braun and Hepp \cite{Braun_Hepp_CommMathPhys_1977}
for smooth enough potential $W$, and \cite{golse2016dynamics} provides a review about the mathematics of this Vlasov limit in various contexts. This Vlasov equation has infinitely
many Casimir conserved quantities. As a consequence, it has an infinite
number of stable stationary states \cite{Yamaguchi_Barre_Bouchet_DR:2004_PhysicaA}.
Any homogeneous distribution $g\left(\text{\ensuremath{\mathbf{r}}},\mathbf{v}\right)=f(\mathbf{v})$
is a stationary solution of the Vlasov equation. In the following,
we will consider homogeneous linearly stable stationary solutions
of the Vlasov equation $f(\mathbf{v})$. The linear stability of such
distributions can be assessed by studying the dielectric susceptibility
$\varepsilon[f](\mathbf{k},\omega)$ \cite{Nicholson_1991,Lifshitz_Pitaevskii_1981_Physical_Kinetics},
defined by
\begin{equation}
\varepsilon[f](\mathbf{k},\omega)=1-\hat{W}\left(\mathbf{k}\right)\int\text{d}\mathbf{v}\frac{\mathbf{k}.\frac{\partial f}{\partial\mathbf{v}}}{\mathbf{k}.\mathbf{v}-\omega-i\tilde{\epsilon}}.\label{eq:dielectric_function}
\end{equation}
Equation (\ref{eq:dielectric_function}) and every other equations
involving $\pm i\tilde{\epsilon}$ have to be understood as the limit
as $\tilde{\epsilon}$ goes to zero with $\tilde{\epsilon}$ positive.
The dielectric susceptibility function $\varepsilon$ plays the role
of a dispersion relation in the linearized dynamics, and a solution
$f$ is stable if $\varepsilon[f]$ has no zeros except for $\omega$
on the real line.

From the point of view of dynamical systems, those homogeneous solutions
might be attractors of the Vlasov equation, with some sort of asymptotic
stability. At a linear level, this convergence for some of the observables,
for instance the potential, is called Landau damping \cite{Nicholson_1991,Lifshitz_Pitaevskii_1981_Physical_Kinetics}.
Such a stability might also be true for the full dynamics. Indeed
some non-linear Landau damping results have recently been proven \cite{Mouhot_Villani:2009}.

In the following we will study the dynamics of $g_{\Lambda}$, when
its initial condition is close to a homogeneous stable state $f(\mathbf{v})$.
On time scales of order one, the distribution is stable and remains
close to $f$ according to the Vlasov equation. However a slow evolution
occurs on a timescale $\tau$ of order $\Lambda$. For this reason,
such $f$ are called quasi-stationary states \cite{Yamaguchi_Barre_Bouchet_DR:2004_PhysicaA}.
In the following section, we explain that this slow evolution is described
by the Balescu--Guernsey--Lenard equation for
most initial conditions. More precisely, after time rescaling $\tau = t/\Lambda$, $g_{\Lambda}(\tau)$ converges
to the solution of the Balescu--Guernsey--Lenard
equation as a law of large numbers.

\subsection{The Balescu--Guernsey--Lenard equation\label{subsec:The-Balescu=002013Guernsey=002013Lenard-equa}}

With the rescaling of time $\tau=t/\Lambda$, we expect a law of large
numbers in the sense that ``for almost all initial conditions''
the empirical distribution function $g_{\Lambda}$ converges to $f$,
with $f$ that evolves according to the Balescu--Guernsey--Lenard
equation

\begin{equation}
\frac{\partial f}{\partial\tau}=\frac{\partial}{\partial\mathbf{v}}.\int\text{d}\mathbf{v}_{2}\,\mathbf{B}\left[f\right](\mathbf{v},\mathbf{v}_{2})\left(-\frac{\partial f}{\partial\mathbf{v}_{2}}f(\mathbf{v})+f(\mathbf{v}_{2})\frac{\partial f}{\partial\mathbf{v}}\right),\label{eq:BLG_eq}
\end{equation}
with 
\begin{equation}
\mathbf{B}\left[f\right](\mathbf{v}_{1},\mathbf{v}_{2})=\pi\left(\frac{\text{\ensuremath{\lambda_{D}}}}{L}\right)^{3}\int_{-\infty}^{+\infty}\text{d}\omega\,\sum_{\mathbf{k}\in2\pi\left(\lambda_{D}/L\right)\mathbb{Z^{*}}^{3}}\frac{\hat{W}\left(\mathbf{k}\right)^{2}\mathbf{k}\mathbf{k}}{\left|\varepsilon[f]\left(\omega,\mathbf{k}\right)\right|^{2}}\delta\left(\omega-\mathbf{k}.\mathbf{v}_{1}\right)\delta\left(\omega-\mathbf{k}.\mathbf{v}_{2}\right).\label{eq:B_GLB}
\end{equation}
The tensor $\mathbf{B}$ is called the collision kernel of the Balescu--Guernsey--Lenard
equation. In equation (\ref{eq:BLG_eq}) and in the sequel, we use the "improper" notation  $\frac{\partial f}{\partial\mathbf{v}_{2}}=\frac{\partial f}{\partial\mathbf{v}}\left(\mathbf{v}_{2}\right)$ to designate the gradient of a function $f$ evaluated in $\mathbf{v}_2$, for the economy of writing.

We know no mathematical proof of such a result. In the theoretical
physics literature, this equation is derived as an exact consequence
of the dynamics once natural hypothesis are made. Two classes of derivations
are known, either the BBGKY hierarchy detailed in \cite{Nicholson_1991}
or the Klimontovich approach presented for instance in \cite{Lifshitz_Pitaevskii_1981_Physical_Kinetics}.
The Klimontovich derivation is the more straightforward from a technical
point of view. We now recall the main steps of the Klimontovich derivation,
that will be useful later.\\

In the following we will consider statistical averages over measures
of initial conditions for the $N$ particle initial conditions $\left\{ \mathbf{r}_{n}^{0},\mathbf{v}_{n}^{0}\right\} $.
We denote $\mathbb{E}_{S}$ the average with respect to this measure
of initial conditions. As an example the measure of initial conditions
could be the product measure $\prod_{n=1}^{N}g^{0}\left(\mathbf{r}_{n}^{0},\mathbf{v}_{n}^{0}\right)\mbox{d}\mathbf{r}_{n}\mbox{d}\mathbf{v}_{n}$.
But we might consider other measures of initial conditions. We recall
that $\Lambda$ is the number of particles in a box of size $\lambda_{D}$.
We will consider the limit $\Lambda\rightarrow\infty$, which is a
large particle number limit. For this reason the limit $\lim_{\Lambda\rightarrow\infty}$
of the empirical density will be called a law of large numbers\footnote{In order to have a discussion of the asymptotic behavior that will
be independent on the box size $L$, for instance in order to consider
infinite box size, it is more natural to discuss the limit $\Lambda\rightarrow\infty$
than $N\rightarrow\infty$.}. We assume that for the statistical ensemble of initial conditions,
the law of large numbers $\lim_{\Lambda\rightarrow\infty} g_{\Lambda}^{0}\left(\mathbf{r},\mathbf{v}\right)=g^{0}\left(\mathbf{r},\mathbf{v}\right)$
is valid at the initial time. This is true for instance for the product
measure. In the following, for simplicity, we restrict the discussions
to cases when the initial conditions are statistically homogenous:
$g^{0}\left(\mathbf{r},\mathbf{v}\right)=f^{0}(\mathbf{v})$. We are
then looking for $\lim_{\Lambda\rightarrow\infty} g_{\Lambda}^{0}\left(\mathbf{r},\mathbf{v},t\right)=f(\mathbf{v},t)$,
valid for any  time $t$ with $\tau = t/\Lambda$ finite. Alternatively, we define $f$ as the
statistical average of $g_{\Lambda}$ over the initial conditions
$f(\mathbf{v},t)=\mathbb{E}_{S}\left(g_{\Lambda}(\mathbf{r},\mathbf{v},t)\right)$.

We define the fluctuations $\delta g_{\Lambda}$ by $g_{\Lambda}(\mathbf{r},\mathbf{v},t)=f(\mathbf{v})+\delta g_{\Lambda}/\sqrt{\Lambda}$.
The scaling $1/\sqrt{\Lambda}$ is natural when we see the Vlasov
equation (\ref{eq:vlasov}) as a law of large numbers for the empirical
distribution. For the potential we obtain $V\left[g_{\Lambda}\right]=V\left[\delta g_{\Lambda}\right]/\sqrt{\Lambda}$,
as $f$ is homogeneous. If we introduce this decomposition in the
Klimontovich equation (\ref{eq:Klimontovich_eq}), we obtain 
\small
\begin{eqnarray}
\frac{\partial f}{\partial t} & = & \frac{1}{\Lambda}\mathbb{E}_{S}\left(\frac{\partial V\left[\delta g_{\Lambda}\right]}{\partial\mathbf{r}}.\frac{\partial\delta g_{\Lambda}}{\partial\mathbf{v}}\right)\label{eq:decomp}\\
\frac{\partial\delta g_{\Lambda}}{\partial t}+\mathbf{v}.\frac{\partial\delta g_{\Lambda}}{\partial\mathbf{r}}-\frac{\partial V\left[\delta g_{\Lambda}\right]}{\partial\mathbf{r}}.\frac{\partial f}{\partial\mathbf{v}} & = & \frac{1}{\sqrt{\Lambda}}\left[\frac{\partial V\left[\delta g_{\Lambda}\right]}{\partial\mathbf{r}}.\frac{\partial\delta g_{\Lambda}}{\partial\mathbf{v}}-\mathbb{E}_{S}\left(\frac{\partial V\left[\delta g_{\Lambda}\right]}{\partial\mathbf{r}}.\frac{\partial\delta g_{\Lambda}}{\partial\mathbf{v}}\right)\right].\label{eq:decomp2}
\end{eqnarray}
\normalsize

In the first equation, the right hand side of the equation $\frac{1}{\Lambda}\mathbb{E}_{S}\left(\frac{\partial V\left[\delta g_{N}\right]}{\partial\mathbf{r}}.\frac{\partial\delta g_{N}}{\partial\mathbf{v}}\right)$
is called the averaged non linear term and is responsible for the
long term evolution of the distribution $f$. The right hand side
of the second equation $\frac{1}{\sqrt{\Lambda}}\left[\frac{\partial V\left[\delta g_{\Lambda}\right]}{\partial\mathbf{r}}.\frac{\partial\delta g_{\Lambda}}{\partial\mathbf{v}}\right.$ $\left.-\mathbb{E}_{S}\left(\frac{\partial V\left[\delta g_{\Lambda}\right]}{\partial\mathbf{r}}.\frac{\partial\delta g_{\Lambda}}{\partial\mathbf{v}}\right)\right]$
describes the fluctuations of the non-linear term. For stable distributions
$f,$ and on timescales much smaller than $\sqrt{\Lambda}$, we can
neglect this term, following Klimontovich and classical textbooks
\cite{Lifshitz_Pitaevskii_1981_Physical_Kinetics}. This closes the
hierarchy of the correlation functions. The Bogoliubov approximation
then amounts at using the time scale separation between the evolution
of $f$ and $\delta g_{\Lambda}$. Then for fixed $f$, the equation
for $\delta g_{\Lambda}$ (\ref{eq:decomp2}) is linear when $f$
is fixed. One computes the correlation function $\mathbb{E}_{S}\left(\frac{\partial V\left[\delta g_{\Lambda}\right]}{\partial\mathbf{r}}.\frac{\partial\delta g_{\Lambda}}{\partial\mathbf{v}}\right)$
resulting from (\ref{eq:decomp2}) with fixed $f$, and argues that
this two point correlation function converges to a stationary quantity
on time scales much smaller than $\sqrt{\Lambda}$. Using this quasi-stationary
correlation function $\mathbb{E}_{S}\left(\frac{\partial V\left[\delta g_{\Lambda}\right]}{\partial\mathbf{r}}.\frac{\partial\delta g_{\Lambda}}{\partial\mathbf{v}}\right)$,
one can compute the right hand side of (\ref{eq:decomp}) as a function
of $f$. After time rescaling $\tau=t/\Lambda$, the closed equation
which is obtained from (\ref{eq:decomp}) is the Balescu--Guernsey--Lenard
equation (\ref{eq:BLG_eq}). We do not reproduce these lengthy and
classical computations that can be found in a plasma physics textbook,
for instance in the chapter 51 of \cite{Lifshitz_Pitaevskii_1981_Physical_Kinetics}.

Based on these computations, a law of large numbers for $\{g_{\Lambda}(\tau)\text{\}}$ is a natural conjecture. More precisely, if we assume there is a set of initials conditions $\{g_{\Lambda}^{0}\text{\}}$
such that $\lim_{\Lambda \rightarrow +\infty} g_{\Lambda}^{0} =f^{0}$, then over finite time interval $\tau \in \left[0,T\right]$, where $\tau = t/\Lambda$, $\lim_{\Lambda \rightarrow +\infty} g_{\Lambda}(\tau)=f(\tau)$, where $g$ solves the Balescu--Guernsey--Lenard equation with $f(\tau =0) = f^{0}$.

\paragraph{Symmetries and conservation properties.}

The Balescu--Guernsey--Lenard equation (\ref{eq:BLG_eq})
has several important physical properties:
\begin{enumerate}
\item It conserves the mass $M[f]$, momentum $\mathbf{P}[f]$ and total
kinetic energy $E[f]$ defined by
\begin{equation}
M[f]=\int\text{d}\mathbf{v}\,f\left(\mathbf{v}\right),\,\,\,\mathbf{P}[f]=\int\text{d}\mathbf{v}\,\textbf{v}f\left(\mathbf{v}\right)\,\,\,\text{and}\,\,\,E[f]=\int\text{d}\mathbf{v\,}\frac{\textbf{v}^{2}}{2}f\left(\mathbf{v}\right).\label{eq:Conservation_Laws}
\end{equation}
\item It increases monotonically the entropy $S[f]$ defined by
\[
S[f]=-\int\text{d}\mathbf{v}\,f\left(\mathbf{v}\right)\log f\left(\mathbf{v}\right).
\]
\item It converges towards the Boltzmann distribution for the corresponding
energy
\[
f_{B}\left(\mathbf{v}\right)=\frac{\beta{}^{3/2}}{\left(2\pi\right)^{3/2}}\exp\left(-\beta\frac{\mathbf{v}^{2}}{2}\right).
\]
\end{enumerate}
The Balescu--Guernsey--Lenard is a good approximation
to describe the long time evolution of system of particles with mean
field interactions but it is quite complicated to handle, especially
because the tensor $\mathbf{B}$ depends on the actual distribution
$f$ in a non-trivial way. The Balescu--Guernsey--Lenard
operator (the right hand side of (\ref{eq:BLG_eq})), is a very complex
non-linear functional of $f$. 

\subsection{The Landau equation\label{subsec:The-Landau-equation}}

Neglecting the collective effects in the Balescu--Guernsey--Lenard
equation, we obtain the Landau equation
\begin{equation}
\frac{\partial f}{\partial\tau}=\frac{\partial}{\partial\mathbf{v}}.\int\text{d}\mathbf{v}_{2}\,\mathbf{B}(\mathbf{v},\mathbf{v}_{2})\left(-\frac{\partial f}{\partial\mathbf{v}_{2}}f(\mathbf{v})+f(\mathbf{v}_{2})\frac{\partial f}{\partial\mathbf{v}}\right),\label{eq:Landau-1}
\end{equation}
where $\mathbf{B}$ for the Landau equation is given by the same expression
as the one for $\mathbf{B}$ in equation (\ref{eq:B_GLB}), but with
$\varepsilon\left(\mathbf{k},\omega\right)=1$: 
\begin{equation}
\mathbf{B}(\mathbf{v}_{1},\mathbf{v}_{2})=\pi\left(\frac{\text{\ensuremath{\lambda_{D}}}}{L}\right)^{3}\int_{-\infty}^{+\infty}\text{d}\omega\,\sum_{\mathbf{k}\in2\pi\left(\lambda_{D}/L\right)\mathbb{Z^{*}}^{3}}\hat{W}\left(\mathbf{k}\right)^{2}\mathbf{k}\mathbf{k}\delta\left(\omega-\mathbf{k}.\mathbf{v}_{1}\right)\delta\left(\omega-\mathbf{k}.\mathbf{v}_{2}\right).\label{eq:B_Landau}
\end{equation}

The Landau approximation of the Balescu--Guernsey--Lenard
equation is valid to describe plasma at scales which are much smaller
than the Debye length (associated with large wavenumbers compared
to $1/\lambda_{D}$), or globally when the effect of those scales
dominate the collision kernel $\mathbf{B}$. Within this approximation,
we can assume that $\varepsilon\left(\mathbf{k},\omega\right)=1$
which means that the dielectric susceptibility does not depend on
the distribution $f$ anymore. This approximation is relevant for
many applications in plasma physics.

\subsection{The Balescu--Guernsey--Lenard and Landau equations
as non-linear Fokker-Planck equations \label{subsec:The-Balescu=002013Guernsey=002013Lenard-and}}

It is possible to consider the Balescu--Guernsey--Lenard
and the Landau equations as non-linear Fokker-Planck equations. Indeed,
introducing the drift and the diffusion terms

\begin{equation}
\begin{cases}
\mathbf{b}\left[f\right](\mathbf{v}) & =\int\text{d}\mathbf{v}_{2}\mathbf{B}\left[f\right](\mathbf{v},\mathbf{v}_{2})\frac{\partial f}{\partial\mathbf{v}_{2}}\\
\mathbf{D}\left[f\right](\mathbf{v}) & =\int\text{d}\mathbf{v}_{2}\mathbf{B}\left[f\right](\mathbf{v},\mathbf{v}_{2})f(\mathbf{v}_{2}),
\end{cases}\label{eq:coeffLandau}
\end{equation}
the Balescu--Guernsey--Lenard and the Landau equations
write

\begin{equation}
\frac{\partial f}{\partial t}=\frac{\partial}{\partial\mathbf{v}}\left\{ -f\mathbf{b}\left[f\right]+\mathbf{D}\left[f\right]\frac{\partial f}{\partial\mathbf{v}}\right\} .\label{eq:Lenard_Balescu_Non_Linear_Fokker_Planck}
\end{equation}
This is the functional form of a Fokker-Planck equation, but by contrast
with the linear Fokker-Planck equation with constant drift and diffusion
coefficient, the drift and diffusion
coefficients depend on $f$.\\

We remark that this equation could be obtained from the dynamics of
$N$ particles governed by the Ito diffusion

\begin{equation}
\text{d}\mathbf{v}_{n}=\mathbf{b}[h_{N}]\left(\mathbf{v}_{n}\right)+\frac{\partial}{\partial\mathbf{v}}.\mathbf{D}[h_{N}]\left(\mathbf{v}_{n}\right)\text{d}t+\sqrt{2}\sigma[h_{N}]\left(\mathbf{v}_{n}\right)\text{d}W_{n,t},\label{eq:N_Diffusions}
\end{equation}
with
\begin{equation}
h_{N}\left(\mathbf{v},t\right)=\frac{1}{N}\sum_{n=1}^{N}\delta\left(\mathbf{v}_{n}(t)-\mathbf{v}\right),
\label{eq:N_Diffusions_h}
\end{equation}
where $\sigma$ is such that $\mathbf{D}[h_{N}]\left(\mathbf{v}_{n}\right)=\sigma[h_{N}]\left(\mathbf{v}_{n}\right)\sigma[h_{N}]\left(\mathbf{v}_{n}\right)^{\top}$,
and $W_{n,t}$ are Wiener processes that satisfy $\mathbb{E}\left(\text{d}W_{m,t}\text{d}W_{n,t'}\right)=\delta_{m,n}\delta\left(t'-t\right)\text{d}t$.
In this equation, the drift and diffusion coefficients $\mathbf{b}\left[h_{N}\right]$
and $\mathbf{D}[h_{N}]$ and the matrix $\sigma$ depend on a mean
field way on the empirical density $h_{N}$. 

There is a link between $h_{N}$ and the empirical density $g_{\Lambda}$
rescaled by the plasma parameter. We define $f_{\Lambda}$ the projection
of $g_{\Lambda}$ over homogeneous distributions over the $\mu-$
space : 
\[
f_{\Lambda}\left(\mathbf{v},t\right)=\left(\frac{\lambda_{D}}{L}\right)^{3}\int_{\left[0,L/\lambda_{D}\right]^{3}}\text{d}\mathbf{r}\,g_{\Lambda}\left(\mathbf{r},\mathbf{v},t\right).
\]
(both $f_{\Lambda}$ and $g_{\Lambda}$ are distributions over the
$\mu-$ space). We note that $f_{\Lambda}$, which is a homogeneous
distribution over the $\mu-$space can also be interpreted as a distribution
over the velocity space. Then using the relation between $N$ and
$\Lambda$: $\Lambda L^{3}/\lambda_{D}^{3}=N$, one can check that
$f_{\Lambda}=h_{N}$. 

The law of large numbers for the empirical density $h_{N}$ for these
$N$ particles with mean field coupling insures that $\lim_{N\rightarrow\infty}h_{N}=f$
where $f$ satisfies the Balescu--Guernsey--Lenard
equation (\ref{eq:Lenard_Balescu_Non_Linear_Fokker_Planck}). From
this remark, a natural question is whether the dynamical large deviations
for the empirical distribution $h_{N}$ in (\ref{eq:N_Diffusions}-\ref{eq:N_Diffusions_h})
are the same as the dynamical large deviations of $N$ particles with
Coulomb interactions (the large deviation of the Balescu--Guernsey--Lenard
equation). We address this very natural question in the following
section.

\section{Large deviations for $N$ independent diffusions and $N$ diffusions
with mean field coupling\label{sec:N independent diffusions and N diffusions with mean field coupling}}

The aim of this section is to address the following question: are
the dynamical large deviations (\ref{eq:N_Diffusions}) for the empirical
distribution $h_{N}$ in (\ref{eq:N_Diffusions}) the same as the
dynamical large deviations of $N$ particles with mean field interactions
(the large deviations for the Balescu--Guernsey--Lenard
or the Landau equations)? In section \ref{subsec:Large-deviations-N-markov-processes}
we derive the large deviation rate function for the empirical density
defined as $h_{N}(\mathbf{v},t)=\frac{1}{N}\sum_{n=1}^{N}\delta\left(\mathbf{v}_{n}(t)-\mathbf{v}\right)$
of $N$ independent particles, where each $\mathbf{v}_{n}(t)$ is
governed by a Markov dynamics with infinitesimal generator $G$.

In section \ref{subsec:LD-for-N-diff} we apply this to the case when
the $N$ independent Markov dynamics are diffusions, and in section
\ref{subsec:LD-N-coupled-diff} when the particles are not independent
anymore but are coupled in a mean field way, as in (\ref{eq:N_Diffusions}).
For each of these cases we prove that with the prescription that $h_N(t=0)$ is in the neighborhood of $h(t=0)$
\begin{equation}
\mathbf{P}\left(\left\{ h_{N}(t)\right\} _{0\leq t\leq T}=\left\{ h(t)\right\} _{0\leq t\leq T}\right)\underset{N\rightarrow\infty}{\asymp}\text{e}^{-N\text{Sup}_{p}\int_{0}^{T}\left\{ \int\text{d}\mathbf{v}\,\dot{h}p-H[h,p]\right\} },\label{eq:LDP-n-diff}
\end{equation}
where the corresponding $H$ are given by formula (\ref{eq:Hamiltonien_N_Markov_Processes}),
(\ref{eq:N_Diff_Hamiltonian}) and (\ref{eq:Mean_field_Hamiltonian-1}),
respectively.

In section \ref{subsec:LD-N-coupled-diff}, we prove that the large
deviations of the Balescu--Guernsey--Lenard or
the Landau equations are \textbf{not} the large deviations of $N$
diffusing particles with mean field coupling (\ref{eq:N_Diffusions}),
as might have been naturally hypothesized.

\subsection{Large deviations for the empirical density of $N$ independent Markov
processes\label{subsec:Large-deviations-N-markov-processes}}

We consider $N$ continuous time independent Markov processes $\left\{ \mathbf{v}_{n}(t)\right\} _{t\in[0,T],1\leq n\leq N}$,
where each $\mathbf{v}_{n}(t)$ is governed by a Markov dynamics with
infinitesimal generator $G$. $G$ acts on functions $\phi:\mathbb{R}^{3}\rightarrow\mathbb{R}$
and is defined by
\begin{equation}
G\left[\phi\right]\left(\mathbf{v}_{0}\right)=\lim_{\Delta T\rightarrow0}\frac{\mathbb{E}_{\mathbf{v}_{0}}\left[\phi\left(\mathbf{\mathbf{v}}(\Delta T)\right)\right]-\phi\left(\mathbf{v}_{0}\right)}{\Delta T}.\label{eq:GenInfinites}
\end{equation}
Then, with the prescription that $h_N(t=0)$ is in the neighborhood of $h(t=0)$, the empirical density $h_{N}$ satisfies a large deviation principle
\begin{equation}
\mathbf{P}(h_{N}=h)\underset{N\rightarrow\infty}{\asymp}\text{e}^{-N\text{Sup}_{p}\int_{0}^{T}\left\{ \int\text{d}\mathbf{v}\,\dot{h}p-H[h,p]\right\} }\label{eq:LDP_Hamiltonian}
\end{equation}
where
\begin{equation}
H[h,p]=\int\text{d}\mathbf{v}h(\mathbf{v})G_{\text{}}\left[\text{e}^{p(\cdot)}\right](\mathbf{v})\text{e}^{-p(\mathbf{v})},\label{eq:Hamiltonien_N_Markov_Processes}
\end{equation}
in this expression, the variable $p$ is the conjugate momentum to
$h$, and it is a scalar function of the velocity $\mathbf{v}$.

\paragraph{Formal proof}

The empirical density $h_{N}$ is also itself a continuous time Markov
process. We denote $G_{h}$ its infinitesimal generator, defined by
\[
G_{h}\left[\psi\right]\left(h_{0}\right)=\lim_{\Delta T\rightarrow0}\frac{\mathbb{E}_{h_{0}}\left[\psi\left(h(\Delta T)\right)\right]-\psi\left(h_{0}\right)}{\Delta T},.
\]
where $\psi$ is a functional. Then, from the result explained 
in section \ref{sec:Large_Deviations_Generator-1}, we know that if
the limit 

\[
H[h,p]=\lim_{N\rightarrow\infty}\frac{1}{N}\text{e}^{-N\int\text{d}\mathbf{v}\,ph}G_{h}\left[\text{e}^{N\int\text{d}\mathbf{v}ph}\right],
\]
exists (see  (\ref{eq:H-Generator})), then we have the large deviation principle (\ref{eq:LDP_Hamiltonian}).
Using the definition of the empirical density, we find

\begin{eqnarray*}
G_{h}\left[\text{e}^{N\int\text{d}\mathbf{v}ph_{N}}\right] & = & G_{h}\left[\text{e}_{n=1}^{\sum^{N}p(\mathbf{v}_{n})}\right]\\
 & = & \lim_{\Delta T\rightarrow0}\frac{1}{\Delta T}\left(\mathbb{E}\left(\text{e}^{\sum_{n=1}^{N}p(\mathbf{v}_{n}\left(\Delta T\right))}\right)-\text{e}^{\sum_{n=1}^{N}p(\mathbf{v}_{n}\left(0\right))}\right).
\end{eqnarray*}
Then, using that the particles are independent

\[
H[h,p]=\lim_{N\rightarrow\infty}\lim_{\Delta T\rightarrow0}\frac{1}{N\Delta T}\left(\prod_{n=1}^{N}\mathbb{E}\left(\text{e}^{\Delta p(\mathbf{v}_{n})}\right)-1\right),
\]
where $\mathbb{E}\left(\text{e}^{\Delta p(\mathbf{v}_{n})}\right)=\mathbb{E}\left(\text{e}^{p(\mathbf{v}_{n}\left(\Delta T\right))}\right)\text{e}^{-p(\mathbf{v}_{n}(0))}$.
Furthermore, using the definition of the infinitesimal generator for
the diffusion process (\ref{eq:GenInfinites}), we have

\[
\mathbb{E}\left(\text{e}^{\Delta p(\mathbf{v}_{n})}\right)=1+\Delta TG\left[\text{e}^{p(\mathbf{v}_{n}(0))}\right]\text{e}^{-p(\mathbf{v}_{n}(0))}+o(\Delta T)\,\,\,\left(\Delta T\rightarrow0\right).
\]
To the same precision we can compute the product for $1\leq n\leq N$
\[
\prod_{n=1}^{N}\mathbb{E}\left(\text{e}^{\Delta p(\mathbf{v}_{n})}\right)-1=\Delta T\sum_{n=1}^{N}G\left[\text{e}^{p(\mathbf{v}_{n}(0))}\right]\text{e}^{-p(\mathbf{v}_{n}(0))}+o(\Delta T)\,\,\,\left(\Delta T\rightarrow0\right).
\]
From this expansion, it is possible to compute the limit as $\Delta T$
goes to $0$
\[
\lim_{\Delta T\rightarrow0}\frac{1}{N\Delta T}\left(\prod_{n=1}^{N}\mathbb{E}\left(\text{e}^{\Delta p(\mathbf{v}_{n})}\right)-1\right)=\sum_{n=1}^{N}G\left[\text{e}^{p(\mathbf{v}_{n}(0))}\right]\text{e}^{-p(\mathbf{v}_{n}(0))}.
\]
It is important to note that the order of the limits $N\rightarrow\infty$
and $\Delta T\rightarrow0$ is crucial. From there, it comes easily
that 
\[
H[h,p]=\lim_{N\rightarrow\infty}\frac{1}{N}\sum_{n=1}^{N}G\left[\text{e}^{p(\mathbf{v}_{n}(0))}\right]\text{e}^{-p(\mathbf{v}_{n}(0))}=\int\text{d}\mathbf{v}h(\mathbf{v})G\left[\text{e}^{p(\cdot)}\right](\mathbf{v})\text{e}^{-p(\mathbf{v})}.
\]
 \\

We remark that the Hamiltonian (\ref{eq:Hamiltonien_N_Markov_Processes})
is in general not quadratic in $p$, reflecting the fact that the
large deviations are not Gaussian, although they arise from the sum
of $N$ independent contributions.

\subsection{Large deviations for the empirical density of $N$ independent diffusions\label{subsec:LD-for-N-diff}}

From equation (\ref{eq:Hamiltonien_N_Markov_Processes}), it is straightforward
to compute the Hamiltonian that describes the large deviations for
the empirical density of $N$ particles with independent diffusions.

Let us consider $N$ particles with velocities $\left\{ \mathbf{v}_{n}\right\} _{1\leq n\leq N}$
with the following Ito diffusion dynamics

\begin{equation}
\text{d}\mathbf{v}_{n}=\left[\mathbf{b}\left(\mathbf{v}_{n}\right)+\frac{\partial}{\partial\mathbf{v}}.\mathbf{D}\left(\mathbf{v}_{n}\right)\right]\text{d}t+\sqrt{2}\sigma\left(\mathbf{v}_{n}\right)\text{d}W_{n,t}.\label{eq:Diffusion_Particules_Independantes}
\end{equation}
We define $\mathbf{D}$ the diffusion tensor as $\mathbf{D}=\sigma\sigma^{\top}$.
We call $h$ the probability density function of $\mathbf{v}_{n}$
for some $n$. It does not depend on $n$ as we consider $N$ non-interacting
particles, we can write the Fokker-Planck equation associated with
the diffusion of a particle

\[
\frac{\partial h}{\partial t}=\frac{\partial}{\partial\mathbf{v}}\cdot\left\{ -h\mathbf{b}+\mathbf{D}\frac{\partial h}{\partial\mathbf{v}}\right\} .
\]
Now, we define $h_{N}$ the empirical density of the velocity distribution
\[
h_{N}(\mathbf{v},t)=\frac{1}{N}\sum_{n=1}^{N}\delta\left(\mathbf{v}_{n}(t)-\mathbf{v}\right).
\]
We want to compute $H[h,p]$ the Hamiltonian associated with the large
deviation principle for the empirical density

\[
\mathbf{P}(h_{N}=h)\underset{N\rightarrow\infty}{\asymp}\text{e}^{-N\text{Sup}_{p}\int_{0}^{T}\left\{ \int\text{d}\mathbf{v}\,\dot{h}p-H[h,p]\right\} }.
\]
We showed in section \ref{subsec:Large-deviations-N-markov-processes}
that $H[h,p]$ is given by

\[
H[h,p]=\int\text{d}\mathbf{v}h(\mathbf{v})G\left[\text{e}^{p(\cdot)}\right](\mathbf{v})\text{e}^{-p(\mathbf{v})}.
\]
It is a classical result in stochastic analysis that the infinitesimal
generator $G$ of the diffusion stochastic process is 
\[
G=\mathbf{b}.\frac{\partial}{\partial\mathbf{v}}+\frac{\partial}{\partial\mathbf{v}}\cdot\left(\mathbf{D}\frac{\partial}{\partial\mathbf{v}}\right),
\]
the adjoint of the Fokker-Planck operator. This leads to the Hamiltonian
associated with the empirical density of $N$ particles diffusing
independently

\begin{equation}
H\left[h,p\right]=\int\text{d}\mathbf{v\,}h\left\{ \mathbf{b}.\frac{\partial p}{\partial\mathbf{v}}+\frac{\partial}{\partial\mathbf{v}}\left(\mathbf{D}\frac{\partial p}{\partial\mathbf{v}}\right)+\mathbf{D}:\frac{\partial p}{\partial\mathbf{v}}\frac{\partial p}{\partial\mathbf{v}}\right\} ,\label{eq:N_Diff_Hamiltonian}
\end{equation}
where the symbol ``$:$'' means the contraction of two second order
symmetric tensors : $\mathbf{M}:\mathbf{N}=\text{Tr}\left(\mathbf{M}\mathbf{N}\right)=\sum_{ij}M_{ij}N_{ij}$.\\

We remark that the Hamiltonian (\ref{eq:Hamiltonien_N_Markov_Processes})
is quadratic in $p$. This means that the large deviations are Gaussian.
This reflects the fact that the large deviations arise from the sum
of $N$ independent Gaussian increments. Because of this property,
we can also recover from the Hamiltonian an equivalent stochastic
differential equation for the empirical $h_{N}$ that involves a Gaussian
noise. More precisely, a quadratic Hamiltonian
\[
H\left[h,p\right]=\int\text{d}\mathbf{v}\,A[h]\left(\mathbf{v}\right)p\left(\mathbf{v}\right)+\iint\text{d}\mathbf{v}\text{d}\mathbf{v}'p\left(\mathbf{v}\right)C\left[h\right]\left(\mathbf{v},\mathbf{v}'\right)p\left(\mathbf{v}'\right)
\]
is the Hamiltonian that describes the dynamical large deviations of
the stochastic differential equation
\[
\frac{\partial h_{N}}{\partial t}=A\left[h_{N}\right]\left(\mathbf{v}\right)+\sqrt{\frac{2}{N}}\eta\left(\mathbf{v},t\right)
\]
with
\[
\mathbb{E}\left(\eta\left(\mathbf{v},t\right)\eta\left(\mathbf{v}',t'\right)\right)=C\left[h_{N}\right]\left(\mathbf{v},\mathbf{v}'\right).
\]
Using partial integration, we can identify $A[h]$ and $C\left[h\right]\left(\mathbf{v},\mathbf{v}'\right)$
for the Hamiltonian (\ref{eq:N_Diff_Hamiltonian}). The associated
stochastic differential equation for the empirical density is
\begin{equation}
\frac{\partial h_{N}}{\partial t}=\frac{\partial}{\partial\mathbf{v}}\cdot\left\{ -h_{N}\mathbf{b}+\mathbf{D}\frac{\partial h_{N}}{\partial\mathbf{v}}\right\} +\sqrt{\frac{2}{N}}\eta\left(\mathbf{v},t\right)\label{eq:SDE_ass_w/H}
\end{equation}
with,
\[
\mathbb{E}\left(\eta\left(\mathbf{v},t\right)\eta\left(\mathbf{v}',t'\right)\right)=\frac{\partial^{2}}{\partial\mathbf{v}\partial\mathbf{v}'}:\left(h_{N}(\mathbf{v})\delta\left(\mathbf{v}-\mathbf{v}'\right)\mathbf{D}\right)\delta\left(t-t'\right).
\]
Recalling that $\mathbf{D}=\sigma\sigma^{\top}$, we can rewrite equation
(\ref{eq:SDE_ass_w/H}) as a conservative equation
\[
\frac{\partial h_{N}}{\partial t}=\frac{\partial}{\partial\mathbf{v}}\cdot\left\{ -h_{N}\mathbf{b}+\mathbf{D}\frac{\partial h_{N}}{\partial\mathbf{v}}+\sqrt{\frac{2}{N}h_{N}}\sigma\mathbf{\xi}\left(\mathbf{v},t\right)\right\} ,
\]
with $\xi$ a tridimensional Gaussian noise that satisfies 
\[\mathbb{E}\left(\xi^{i}\left(\mathbf{v},t\right)\xi^{j}\left(\mathbf{v}',t'\right)\right)=\delta^{ij}\delta\left(\mathbf{v}-\mathbf{v}'\right)\delta\left(t-t'\right).
\]

\subsection{Large deviations for $N$ diffusions with mean field coupling\label{subsec:LD-N-coupled-diff}}

In the previous section, we have derived the large deviation Hamiltonian
for the empirical density of $N$ independent particles driven by
the diffusion (\ref{eq:Diffusion_Particules_Independantes}). We now
consider the case when the drift and diffusion coefficients depend
on the empirical density itself: 

\begin{equation}
\text{d}\mathbf{v}_{n}=\mathbf{b}[h_{N}]\left(\mathbf{v}_{n}\right)+\frac{\partial}{\partial\mathbf{v}}.\mathbf{D}[h_{N}]\left(\mathbf{v}_{n}\right)\text{d}t+\sqrt{2}\sigma[h_{N}]\left(\mathbf{v}_{n}\right)\text{d}W_{n,t},\label{eq:N_Diffusions-1}
\end{equation}
with $h_{N}(\ensuremath{\mathbf{v}},t)=\ensuremath{\frac{1}{N}\sum_{n=1}^{N}\delta\left(\mathbf{v}_{n}(t)-\mathbf{v}\right)}.$
We denote $\mathbf{D}\left[h\right]=\sigma\left[h\right]\sigma\left[h\right]^{\top}$.
For this case, the particles are no more statistically independent.
However, for such a mean field coupling, it is an easy exercise\textbf{
}to adapt the derivation that leads to the Hamiltonian (\ref{eq:Hamiltonien_N_Markov_Processes})
in section \ref{subsec:Large-deviations-N-markov-processes} to this
specific case. We find that the Hamiltonian that describes the large
deviation of the empirical density is 
\begin{equation}
H_{MF,h}\left[h,p\right]=\int\text{d}\mathbf{v}h\left\{ \mathbf{b}\left[h\right].\frac{\partial p}{\partial\mathbf{v}}+\frac{\partial}{\partial\mathbf{v}}\left(\mathbf{D}\left[h\right]\frac{\partial p}{\partial\mathbf{v}}\right)+\mathbf{D}\left[h\right]:\frac{\partial p}{\partial\mathbf{v}}\frac{\partial p}{\partial\mathbf{v}}\right\} .\label{eq:Mean_field_Hamiltonian-1}
\end{equation}
The subscript $MF,h$ denotes that this is the Hamiltonian for a mean
field dynamics without spatial structure. We note that this Hamiltonian
is the same as (\ref{eq:N_Diff_Hamiltonian}), but with drift and
diffusion constant that depend of $h$. The corresponding stochastic
dynamics is 
\begin{equation}
\frac{\partial h}{\partial t}=\frac{\partial}{\partial\mathbf{v}}\cdot\left\{ -h\mathbf{b}\left[h\right]+\mathbf{D}\left[h\right]\frac{\partial h}{\partial\mathbf{v}}+\sqrt{\frac{2}{N}h}\sigma\left[h\right]\mathbf{\xi}\left(\mathbf{v},t\right)\right\} ,\label{eq:Stochastic_Dynamics_Mean_Field_Coupling}
\end{equation}
with $\xi$ a tridimensional Gaussian noise that satisfies 
\[
\mathbb{E}\left(\xi^{i}\left(\mathbf{v},t\right)\xi^{j}\left(\mathbf{v}',t'\right)\right)=\delta^{ij}\delta\left(\mathbf{v}-\mathbf{v}'\right)\delta\left(t-t'\right).\]

Now let us get back to the remark of section \ref{subsec:The-Balescu=002013Guernsey=002013Lenard-and}.
In section \ref{subsec:The-Balescu=002013Guernsey=002013Lenard-and},
we have noticed that one could see the Balescu--Guernsey--Lenard
and Landau equations as non-linear Fokker--Planck equation
for $N$ diffusions with mean field coupling defined by \eqref{eq:N_Diffusions}.
As we already stated, the law of large numbers for the empirical density
indicates $\lim_{N\rightarrow\infty}h_{N}=h$, where $h$ solves the
Balescu--Guernsey--Lenard equation. For this dynamics
and for the empirical measure $h_{N}$, we can derive a large deviation
principle with computations that are analogous to the one we did to
obtain the large deviation principle \eqref{eq:LDP-n-diff}-\eqref{eq:Mean_field_Hamiltonian-1}.
The result reads
\begin{equation}
\mathbf{P}\left(\left\{ h_{N}(t)\right\} _{0\leq t\leq T}=\left\{ h(t)\right\} _{0\leq t\leq T}\right)\underset{N\rightarrow\infty}{\asymp}\text{e}^{-N\int_{0}^{T}\text{d}t\,\text{Sup}_{p}\left\{ \int\text{d}\mathbf{v}\,\dot{h}p-H_{MF,h}[h,p]\right\} }.\label{eq:PGD_mean_field_indep_diff}
\end{equation}
In the following, we examine the properties of this large deviation
Hamiltonian and we conclude that it cannot describe the large deviations
associated to the Balescu--Guernsey--Lenard kinetic
theory.

\paragraph{Relaxation paths and most probable evolutions}

For the dynamics (\ref{eq:N_Diffusions-1}), we also noted at the
end of section \ref{subsec:The-Balescu=002013Guernsey=002013Lenard-and}
that the evolution of the average of the empirical density is given
asymptotically by the Balescu--Guernsey--Lenard. As a consequence we expect the most probable evolution for
the Hamiltonian (\ref{eq:Mean_field_Hamiltonian-1}), also called
a relaxation path to be the Balescu--Guernsey--Lenard
equation (\ref{eq:BLG_eq}). Using the equation for relaxation paths
(equation \eqref{eq:Most-Probab_Evo} in section \ref{subsec:LD_Kinetic_theories})
we check that indeed
\begin{equation}
\frac{\partial h}{\partial t}=\frac{\delta H_{MF,h}}{\delta p}[h,p=0]=\frac{\partial}{\partial\mathbf{v}}\left\{ -h\mathbf{b}\left[h\right]+\mathbf{D}\left[h\right]\frac{\partial h}{\partial\mathbf{v}}\right\} .\label{eq:mostprobab_markov}
\end{equation}

\paragraph{Relative entropy and quasipotential}

In section \ref{sec:Large_Deviations_Generator-1} we define the quasipotential for the empirical distribution $h_{N}$. It is defined as $\mathbb{P}\left(h_{N}=h\right)\underset{N\rightarrow\infty}{\asymp}\text{e}^{-NU\left[h\right]}$.
As the $N$ particles are coupled only in a mean field way, in view
of Sanov's theorem adapted for this case, it is natural to conjecture
that the quasipotential for the dynamics of the empirical density
is $U\left[h\right]=-S\left[h\right]$ where $S$ is the relative
entropy 
\[
S_{\text{rel}}\left[h\right]=-\int\text{d}\mathbf{v}h\log\left(h/h_{\text{eq}}\right),
\]
where $h_{\text{eq}}$ is the stationary solution of the Balescu--Guernsey--Lenard
equation. A necessary condition for the $-S$ to be the quasipotential
is the stationary Hamilton-{}-Jacobi equation $H_{MF,h}\left[h,-\delta S/\delta h\right]=0$.
We check in the appendix \ref{sec:Quasipotential-and-relative} that
this stationary Hamilton--Jacobi equation is indeed verified
when $\mathbf{b}\left[h\right]=\mathbf{b}$ and $\mathbf{D}\left[h\right]=\mathbf{D}$
do not depend on $h$, i.e. when the $N$ diffusions are independent
from each other. However, we also check that this is no more the case
in general if $\mathbf{b}\left[h\right]$ and $\mathbf{D}\left[h\right]$
actually depend on $h$. This remark is enough to conclude that the
Hamiltonian (\ref{eq:Mean_field_Hamiltonian-1}) cannot be the correct
Hamiltonian system for empirical measure of $N$ interacting particles
with mean field interactions, or for particles with Coulomb interactions. 

Moreover an easy direct computation shows that 
\[
\int\text{d}\mathbf{v}\,\frac{\delta H_{MF,h}}{\delta p}\left[f,p\right]\frac{\delta E}{\delta f}\neq0,
\]
where $E$ is the kinetic energy (\ref{eq:Conservation_Laws}). As
explained in section \ref{sec:Large_Deviations_Generator-1}, $\int\text{d}\mathbf{v}\,\frac{\delta H_{MF,h}}{\delta p}\left[f,p\right]\frac{\delta E}{\delta f}=0$
is the energy conservation formula. Equivalently the noise in equation
(\ref{eq:Stochastic_Dynamics_Mean_Field_Coupling}) is not an energy
conserving noise, and thus cannot describe the empirical density of
particles with Coulomb interactions (\ref{eq:coulomb_dynamics}). 

We thus conclude the Hamiltonian (\ref{eq:Mean_field_Hamiltonian-1})
is not the Hamiltonian for the large deviations of systems of particles
for $N$ interacting particles with mean field interactions, or for
particles with Coulomb interactions. In the next sections we derive
in two different ways the large deviation Hamiltonian for the Landau
equation.

\section{Large deviations associated with the Landau kinetic theory from the
Boltzmann kinetic theory\label{sec:Large-dev-from-boltzmann}}

The Landau equation has been presented in section \ref{subsec:The-Landau-equation}
as an approximation of the Balescu--Guernsey--Lenard
equation. However it also has a strong link with the Boltzmann equation
that describes a dilute gas of particles in the Boltzmann--Grad
limit. One can look for instance in \cite{Lifshitz_Pitaevskii_1981_Physical_Kinetics}
for this connection. Moreover, the large deviation Hamiltonian for
the Boltzmann equation has already been obtained, for toy models which
are analogue to the dilute gas dynamics \cite{rezakhanlou1998} or
for the dilute gas dynamics \cite{bodineau2020fluctuation,Bouchet_Boltzmann_JSP}.
The aim of this section is to derive the large deviation Hamiltonian
associated with the Landau equation from the large deviation Hamiltonian
associated with the Boltzmann equation. 

In section \ref{subsec:The-Boltzmann-equation}, we introduce the
notations for the Boltzmann equation and the large deviation Hamiltonian
for a dilute gas in the Boltzmann--Grad limit. In section
\ref{subsec:Formal-derivation-of}, following \cite{Lifshitz_Pitaevskii_1981_Physical_Kinetics},
we derive the Landau equation from the Boltzmann equation using the
grazing collision limit. Using the same limit but for the large deviation
Hamiltonian, rather than for the kinetic equation, we derive the large
deviation Hamiltonian for the Landau equation (\ref{eq:ham_from_boltz})
in section \ref{subsec:Deriving-Landau's-large}. In section \ref{subsec:Verifications-of-the},
we show that this Hamiltonian satisfies all the expected symmetries
and conservation properties. In section \ref{subsec:Gradient-flow-structure-associat},
we derive the gradient flow structure of the Landau equation associated
with this Hamiltonian. In section \ref{subsec:from_Landau_to_BGL},
we conjecture the Hamiltonian associated with Balescu--Lenard--Guernsey
equation from the Landau equation Hamiltonian.

\subsection{The Boltzmann equation for a dilute gas\label{subsec:The-Boltzmann-equation}}

We consider the dynamics of a dilute gas composed of atoms or molecules.
We neglect any internal degrees of freedom. We assume that the $N$
particles evolve through a Hamiltonian dynamics with short range two
body interactions, for instance hard sphere collisions. 

Let us first define the collision kernel and the collision cross-section.
We consider a thread of particles with velocities $\mathbf{v}_{1}$
that meets a thread of particles with velocities $\mathbf{v}_{2}$.
We assume that particles of each velocity type are distributed according
to a homogeneous Poisson point process with densities $\ensuremath{\varrho}(\mathbf{v}_{1})\text{d}\mathbf{v}_{1}$
and $\varrho(\mathbf{v}_{2})\text{d}\mathbf{v}_{2}$, respectively.
These particle distributions will give rise to collisions where $(\mathbf{v}_{1},\mathbf{v}_{2})$
particle pairs undergo a random change towards pairs of the type $(\mathbf{v}'_{1},\mathbf{v}'_{2})$,
up to $(\mbox{d}\mathbf{v}'_{1},\mbox{d}\mathbf{v}'_{2})$. This occurs
at a rate per unit of time and unit of volume which is proportional
to the $\mathbf{v}_{1}$ incident particle number $\varrho(\mathbf{v}_{1})\text{d}\mathbf{v}_{1}$,
the $\mathbf{v}_{2}$ incident particle number $\varrho(\mathbf{v}_{2})\text{d}\mathbf{v}_{2}$,
$\mbox{d}\mathbf{v}'_{1},$ and $\mbox{d}\mathbf{v}'_{2}$. The proportionality
coefficient is called the collision kernel and is denoted 
\begin{equation}
w_{0}\left(\mathbf{v}'_{1},\mathbf{v}'_{2};\mathbf{v}_{1},\mathbf{v}_{2}\right)/2.\label{eq:Microscopic_rate}
\end{equation}
The local conservation of momentum and energy implies that
\begin{equation}
w_{0}(\mathbf{v}'_{1},\mathbf{v}'_{2};\mathbf{v}_{1},\mathbf{v}_{2})=\sigma_{0}(\mathbf{v}'_{1},\mathbf{v}'_{2};\mathbf{v}_{1},\mathbf{v}_{2})\delta\left(\mathbf{v}_{1}+\mathbf{v}_{2}-\mathbf{v}'_{1}-\mathbf{v}'_{2}\right)\delta\left(\mathbf{v}_{1}^{2}+\mathbf{v}_{2}^{2}-\mathbf{v'}_{1}^{2}-\mathbf{v'}{}_{2}^{2}\right),\label{eq:cross_section-1}
\end{equation}
where $\sigma_{0}$ is the diffusion cross-section. $\sigma_{0}$
is of the order of $a^{2}$ where $a$ is a typical atom size. We
detail the different symmetry properties of the collision kernel in
annex \ref{sec:Symmetries-and-conservation}.

Several length scales are important to describe a dilute gas: a typical
atom size $a$, that we will defined more precisely below in relation
with the diffusion cross-section, a typical interparticle distance
$1/\rho^{1/3}$ where $\rho$ is the averaged gas density, the mean
free path which is the averaged length a particle travels between
two collisions, and a typical box size $L$. The mean free path is
given by $l=c/a^{2}\rho$, where $c$ is a non-dimensional number
that depends on the collision kernel. The gas is said dilute if we
have the following relation between those scales 
\[
a\ll\frac{1}{\rho^{1/3}}\ll l.
\]
A limit in which those inequalities are satisfied is called a Boltzmann--Grad
limit. We consider the 4 physically independent parameters $a$, $L$,
$N$ and the inverse temperature $\beta$ ($\rho=N/L^{3}$). From
those four, we can choose two independent non-dimensional parameters.
In the following we choose $N$ and the Knudsen number $\alpha=l/L$
as those two independent parameters. The inverse of the number of
particles in a volume of the size $l$ is then $\epsilon=1/l^{3}\rho=a^{2}/l^{2}=a^{6}\rho^{2}$
and is another non-dimensional parameter. 

We will use the large deviation result in the limit $N\rightarrow\infty$
with fixed Knudsen number $\alpha$ . In this limit, from $l=c/a^{2}\rho$
we see that $a^{2}=c/\alpha N$. As the diffusion cross-section $\sigma_{0}$
is of the order of $a^{2}$, in the limit $N\rightarrow\infty$, it
is thus natural to consider the rescaled cross-section $\sigma=N\sigma_{0}.$
Moreover, in the following it will be convenient to consider momentum
exchange. We thus use the following definition of $w$ 
\begin{equation}
w\left(\mathbf{v}_{1}+\frac{1}{2}\mathbf{q},\mathbf{v}_{2}-\frac{1}{2}\mathbf{q};\mathbf{q}\right)=\gamma Nw_{0}(\mathbf{v}_{1}+\mathbf{q},\mathbf{v}{}_{2}-\mathbf{q};\mathbf{v}_{1},\mathbf{v}_{2}),\label{eq:Nouvelles_Notations}
\end{equation}
where $\mathbf{q}$ is the momentum transfer between the incident
particles with momenta $(\mathbf{v}_{1},\mathbf{v}_{2})$ and the
scattered particles with momenta $\mathbf{(v}_{1}+\mathbf{q},\mathbf{v}_{2}-\mathbf{q})$.
Writing the collision kernel this way automatically takes into account
momentum conservation during the collision process. In this reasoning,
the coefficient $\gamma$ is any non-dimensional coefficient which
is held fixed in the limit $N\rightarrow\infty$. In the following
sections, for the specific case of the Coulomb interaction, we will
consider 
\[
\gamma=\left(\frac{\lambda_{D}}{L}\right)^{3},
\]
where $\lambda_{D}$ is the Debye length and $L$ the size of the
box.

We define a rescaled empirical density 
\begin{equation}
g_{\gamma}\left(\mathbf{r},\mathbf{v},t\right)=\left(\gamma N\right)^{-1}\sum_{n=1}^{N}\delta(\mathbf{v}-\mathbf{v}_{n}(t))\delta\left(\mathbf{r}-\mathbf{r}_{n}(t)\right).\label{eq:g_gamma}
\end{equation}
We note that with $\gamma=\left(\lambda_{D}/L\right)^{3}$, $g_{\gamma}$
coincides with $g_{\Lambda}\left(\mathbf{r},\mathbf{v},t\right)=\Lambda^{-1}\sum_{n=1}^{N}\delta(\mathbf{v}-\mathbf{v}_{n}(t))\delta\left(\mathbf{r}-\mathbf{r}_{n}(t)\right)$
(see \eqref{eq:g_Lambda}, page \pageref{eq:g_Lambda}). When these
$N$ particles undergo a dilute gas dynamics, the empirical density
$g_{\gamma}$ has a law of a large numbers. More precisely, if we
assume that for a set of initial conditions, an initial law of large
numbers holds: $\lim_{N\rightarrow\infty} g_{\gamma} \left(\mathbf{r},\mathbf{v}, 0 \right)=g^{0}\left(\mathbf{r},\mathbf{v}\right)$,
then we have at a time $t$ the law of large numbers $\lim_{N\rightarrow\infty} g_{\gamma}\left(\mathbf{r},\mathbf{v},t\right)=g\left(\mathbf{r},\mathbf{v},t\right)$,
where $g$ is a solution of the Boltzmann equation.
\small
\begin{equation}
\frac{\partial g}{\partial t}+\mathbf{v}.\frac{\partial g}{\partial\mathbf{r}}=\int\mbox{d}\mathbf{v}{}_{2}\text{d}\mathbf{q}\,w\left(\mathbf{v}+\frac{1}{2}\mathbf{q},\mathbf{v}_{2}-\frac{1}{2}\mathbf{q};\mathbf{q}\right)\left[g\left(\mathbf{\mathbf{v}}+\mathbf{q},\mathbf{r}\right)g\left(\mathbf{v}{}_{2}-\mathbf{q},\mathbf{r}\right)-g\left(\mathbf{\mathbf{v}},\mathbf{r}\right)g\left(\mathbf{v}_{2},\mathbf{r}\right)\right],\label{eq:Boltzmann_eq-1}
\end{equation}
\normalsize
with initial condition $g\left(\mathbf{r},\mathbf{v},0\right)=g^{0}\left(\mathbf{r},\mathbf{v}\right)$.
We refer to classical textbooks in kinetic theories, for instance
\cite{Lifshitz_Pitaevskii_1981_Physical_Kinetics}, or \cite{Bouchet_Boltzmann_JSP}
for a detailed presentation of an heuristic derivation of the Boltzmann
equation. 

In \cite{Bouchet_Boltzmann_JSP}, a large deviation principle for
the empirical density is derived (equations (1) to (3) in \cite{Bouchet_Boltzmann_JSP}).
This large deviation is derived in the limit $\epsilon=1/N\alpha^{3}\rightarrow0$.
In this paper, we will consider the limit $\gamma N\rightarrow\infty$,
with fixed Knudsen number and fixed $\gamma$. In this limit, we have
$\epsilon=1/N\alpha^{3}\rightarrow0$. Then the large deviation result
justified in \cite{Bouchet_Boltzmann_JSP} can be directly used in
this paper. After adapting equations (1) to (3) in \cite{Bouchet_Boltzmann_JSP}
to the notations (\ref{eq:Nouvelles_Notations}) and \eqref{eq:g_gamma}, with the prescription that $g_{\gamma}(t=0)$ is in the neighborhood of $g(t=0)$,  
we have 
\begin{equation}
\mathbf{P}\left(\left\{ g_{\gamma}(\mathbf{r},\mathbf{v},t)\right\} _{0\leq t\leq T}=\left\{ g(\mathbf{r},\mathbf{v},t)\right\} _{0\leq t\leq T}\right)\underset{N\rightarrow\infty}{\asymp}\text{e}^{-\gamma N\int_{0}^{T}\text{Sup}_{p}\left\{ \int\text{d}\mathbf{r}\text{d}\mathbf{v}\,\dot{g}p-H_{B}[g,p]\right\} },\label{eq:LDP_Boltzmann-1}
\end{equation}
where 
\begin{equation}
H_{B}\left[g,p\right]=H_{C}\left[g,p\right]+H_{T}\left[g,p\right],\label{eq:H_Boltzmann}
\end{equation}
and with the collision Hamiltonian 
\begin{multline}
H_{C}\left[g,p\right]=\frac{1}{2}\int\mbox{d}\mathbf{v}{}_{1}\mbox{d}\mathbf{v}{}_{2}\text{d}\mathbf{q}\mbox{d}\mathbf{r}\,w\left(\mathbf{v}_{1}+\frac{1}{2}\mathbf{q},\mathbf{v}_{2}-\frac{1}{2}\mathbf{q};\mathbf{q}\right)\\\times g(\mathbf{r},\mathbf{\mathbf{v}_{1})}g\left(\mathbf{r},\mathbf{v}_{2}\right)\left\{ \mbox{e}^{\left[-p\left(\mathbf{r},\mathbf{v}_{1}\right)-p\left(\mathbf{r},\mathbf{v}_{2}\right)+p\left(\mathbf{r},\mathbf{v}{}_{1}+\mathbf{q}\right)+p\left(\mathbf{r},\mathbf{v}{}_{2}-\mathbf{q}\right)\right]}-1\right\} ,\label{eq:LDP_Boltzmann_HC-1}
\end{multline}
and the free transport Hamiltonian 
\begin{equation}
H_{T}\left[g,p\right]=-\int\mbox{d}\mathbf{r}\mbox{d}\mathbf{v}\,p(\mathbf{r},\mathbf{v})\mathbf{v}.\frac{\partial g}{\partial\mathbf{r}}(\mathbf{r},\mathbf{v}).\label{eq:LDP_Boltzmann_HT-1}
\end{equation}

\subsection{From the Boltzmann to the Landau equations \label{subsec:Formal-derivation-of}}

In the case of long-range interactions between particles, e.g. Coulomb
type interactions, the two-particle collisions are dominated by small-angle
scattering events. This allows some simplification. The related limit
is called the \textbf{grazing collision limit}. In this section we
justify that in the grazing collision limit and for a homogeneous
gas, from the Boltzmann equation one obtains the Landau equation

\begin{equation}
\frac{\partial f}{\partial t}=\frac{1}{\Lambda}\frac{\partial}{\partial\mathbf{v}}\int\text{d}\mathbf{v}_{2}\,\mathbf{B}(\mathbf{v},\mathbf{v}_{2})\left(-\frac{\partial f}{\partial\mathbf{v}_{2}}f(\mathbf{v})+\frac{\partial f}{\partial\mathbf{v}}f(\mathbf{v}_{2})\right),\label{eq:Landau}
\end{equation}
where the tensor $\mathbf{B}$ is defined by \eqref{eq:B_Landau},
page \pageref{eq:B_Landau}. In equation \eqref{eq:Landau-1} of section
\eqref{subsec:The-Landau-equation}, we expressed this equation with
the time variable $\tau=t/\Lambda$ rescaled by the plasma parameter.
This is why there is no factor $\Lambda^{-1}$ in the right hand side
of equation \eqref{eq:Landau-1}.

The following derivation of the Landau equation from the Boltzmann
equation is strongly inspired by the paragraph \S 42  of \cite{Lifshitz_Pitaevskii_1981_Physical_Kinetics}.
However, here we present a slightly different derivation. First, we
consider homogenous solutions of the Boltzmann equation $g\left(\mathbf{r},\mathbf{v},t\right)=f\left(\mathbf{v},t\right)$
that do not depend on the position variable. The homogeneous Boltzmann
equation reads

\begin{equation}
\frac{\partial f}{\partial t}=\underset{I(\mathbf{v})}{\underbrace{\int\text{d}\mathbf{v}_{2}\text{d}\mathbf{q\,}w\left(\mathbf{v}+\frac{1}{2}\mathbf{q},\mathbf{v}_{2}-\frac{1}{2}\mathbf{q};\mathbf{q}\right)\left[f(\mathbf{v}+\mathbf{q})f(\mathbf{v}_{2}-\mathbf{q})-f(\mathbf{v})f(\mathbf{v}_{2})\right]}}.\label{eq:definition_I}
\end{equation}

From there, we will work in the grazing collision limit, meaning that
we will only take into account collisions that imply small transfer
of momentum. More precisely, we consider only collisions with $\left|\mathbf{q}\right|\ll\mathbf{\left|v\right|},\left|\mathbf{v}_{2}\right|$.
This approximation is relevant and often used in plasma physics, where
Coulomb interactions tend to make collisions with small scattering
angles more numerous and more influential than the other ones, see
the first chapter of \cite{Nicholson_1991} for quantitative arguments.
In order to understand at which precision we shall use this approximation,
let us first give the relation between $\mathbf{B}$ and the collision
kernel: 

\begin{equation}
\mathbf{B}(\mathbf{v}_{1},\mathbf{v}_{2})=\frac{1}{2}\Lambda\int\text{d}\mathbf{q}\,w(\mathbf{v}_{1},\mathbf{v}_{2};\mathbf{q})\mathbf{q}\otimes\mathbf{\mathbf{q}},\label{eq:tensor_B}
\end{equation}
where $\mathbf{q}_{1}\otimes\mathbf{q}_{2}$ is the tensor product
of the two vectors $\mathbf{q}_{1}$ and $\mathbf{q}_{2}$ (a tensor
of rank 2). In appendix \ref{sec:Lien_tenseur_B_boltz_LB}, we prove
that for Coulomb interaction the two expression for $\mathbf{B}$,
\eqref{eq:tensor_B} and \eqref{eq:B_Landau} are equal. In the following,
we will omit the tensor product symbol, and a product of vector without
a dot should be understood as a tensor product: $\mathbf{q}_{1}\mathbf{q}_{2}\equiv\mathbf{q}_{1}\otimes\mathbf{\mathbf{q}}_{2}$.
In the case of the Landau equation, the tensor $\mathbf{B}$ is well
known and has a list of properties related to the geometry and the
physics of the collisions (conservation laws and symmetry properties).
For our study, we will retain that $\mathbf{B}$ is a symmetric tensor,
that $\mathbf{B}$ is symmetric with respect to the exchange of its
two arguments: $\mathbf{B}(\mathbf{v}_{1},\mathbf{v}_{2})=\mathbf{B}(\mathbf{v}_{2},\mathbf{v}_{1})$,
and that $\mathbf{B}(\mathbf{v}_{1},\mathbf{v}_{2}).(\mathbf{v}_{1}-\mathbf{v}_{2})=\overrightarrow{0}$,
we prove these properties in appendix \ref{subsec:The-Landau-collision}.
We will make a link between those properties and the symmetries of
the Landau equation (\ref{eq:Landau}) in section \ref{subsec:Conservation-laws}.

In appendix \ref{subsec:Asymptotic-expansions-Landau}, we develop
$I$ in the Boltzmann equation (\ref{eq:definition_I}) at order 2
in $\text{\ensuremath{\mathbf{q}}}$ and we obtain the Landau equation
(\ref{eq:Landau}). We have thus justified the Landau equation as
an approximation of the Boltzmann equation in the grazing collision
limit.

\subsection{Deriving Landau's large deviation principle from Boltzmann's large
deviation principle\label{subsec:Deriving-Landau's-large}}

In this section we derive the Hamiltonian for the path large deviations
of the Landau equation from the Hamiltonian for the path large deviations
of the Boltzmann equation, using the grazing collision limit. 

We start from the large deviation principle discussed in section \eqref{subsec:The-Boltzmann-equation}.
Adapting the discussion of section \eqref{subsec:The-Boltzmann-equation},
with 
\[
g_{\Lambda}\left(\mathbf{r},\mathbf{v},t\right)=\Lambda^{-1}\sum_{n=1}^{N}\delta(\mathbf{v}-\mathbf{v}_{n}(t))\delta\left(\mathbf{r}-\mathbf{r}_{n}(t)\right),
\]
and with $\gamma=\left(\lambda_{D}/L\right)^{3}$, with the prescription that $g_{\Lambda}(\tau=0)$ is in the neighborhood of $g(\tau=0)$, we have 

\[
\mathbf{P}\left(\left\{ g_{\Lambda}(\mathbf{r},\mathbf{v},\tau)\right\} _{0\leq\tau\leq T}=\left\{ g(\mathbf{r},\mathbf{v},\tau)\right\} _{0\leq\tau\leq T}\right)\underset{\Lambda\rightarrow\infty}{\asymp}\text{e}^{-\Lambda\int_{0}^{T}\text{Sup}_{p}\left\{ \int\text{d}\mathbf{r}\text{d}\mathbf{v}\,\dot{g}p-\Lambda H_{B}[g,p]\right\} \text{d}\tau},
\]
where $H_{B}$ is given by \eqref{eq:H_Boltzmann} and where we used
the rescaled time variable $\tau=t/\Lambda$ by the plasma parameter
$\Lambda$ in the large deviation action.

In the following we will be interested in the case of homogeneous
distributions, i.e. distributions that only depend on the velocity
variable, denoted by the letter $f$: $g\left(\mathbf{r},\mathbf{v},\tau \right)=f\left(\mathbf{v},\tau \right)$.
Then the large deviation principle reads
\begin{equation}
\mathbf{P}(g_{\Lambda}=f)\underset{N\rightarrow\infty}{\asymp}\text{e}^{-\Lambda\int_{0}^{T}\text{Sup}_{p}\left\{ \int\text{d}\mathbf{r}\text{d}\mathbf{v}\,\dot{f}p-H[f,p]\right\} \text{d}\tau},\label{eq:PGD-boltzmann_homogene}
\end{equation}
with the prescription that $g_{\Lambda}(\tau=0)$ is in the neighborhood of $f(\tau=0)$, and with 
\begin{multline}
H\left[f,p\right]=\frac{\Lambda}{2}\int\text{d}\mathbf{r}\mbox{d}\mathbf{v}{}_{1}\mbox{d}\mathbf{v}{}_{2}\text{d}\mathbf{q}\,w\left(\mathbf{v}_{1}+\frac{1}{2}\mathbf{q},\mathbf{v}_{2}-\frac{1}{2}\mathbf{q};\mathbf{q}\right)\\\times f(\mathbf{v}_{1})f\left(\mathbf{v}_{2}\right)\left\{ \mbox{e}^{\left[-p\left(\mathbf{v}_{1}\right)-p\left(\mathbf{v}_{2}\right)+p\left(\mathbf{v}{}_{1}+\mathbf{q}\right)+p\left(\mathbf{v}{}_{2}-\mathbf{q}\right)\right]}-1\right\} .
\label{eq:H_Boltzmann_Homogene}
\end{multline}
\\

The idea to obtain the large deviation Hamiltonian for the Landau
equation, is to use the same hypothesis of grazing collisions used
in section \eqref{subsec:Formal-derivation-of}. As in section \eqref{subsec:Formal-derivation-of},
we will make a Taylor expansion in $\mathbf{q}$ at order 2. Rather
than doing this expansion for the Boltzmann equation, we do it in
the large deviation Hamiltonian \eqref{eq:H_Boltzmann_Homogene}.
The full computation is detailed in appendix \ref{subsec:Asymptotic-expansions-Landau-Hamiltonian},
and we find that the large deviation Hamiltonian $H_{\text{Landau}}[f,p]$
for the Landau equation is
\begin{equation}
H_{\text{Landau}}[f,p]=H_{MF}\left[f,p\right]+H_{I}\left[f,p\right],\label{eq:ham_from_boltz}
\end{equation}
with 
\[
H_{MF}\left[f,p\right]=\int\text{d}\mathbf{r}\text{d}\mathbf{v}_{1}f\left\{ \mathbf{b}\left[f\right].\frac{\partial p}{\partial\mathbf{v}_{1}}+\frac{\partial}{\partial\mathbf{v}_{1}}\left(\mathbf{D}\left[f\right]\frac{\partial p}{\partial\mathbf{v}_{1}}\right)+\mathbf{D}\left[f\right]:\frac{\partial p}{\partial\mathbf{v}_{1}}\frac{\partial p}{\partial\mathbf{v}_{1}}\right\} ,
\]
and 
\[
H_{I}\left[f,p\right]=-\int\text{d}\mathbf{r}\text{d\ensuremath{\mathbf{v}_{1}}}\text{d}\mathbf{v}_{2}f(\mathbf{v}_{1})f(\mathbf{v}_{2})\frac{\partial p}{\partial\mathbf{v}_{1}}\frac{\partial p}{\partial\mathbf{v}_{2}}:\mathbf{B}\left(\mathbf{v}_{1},\mathbf{v}_{2}\right),
\]
 where $\mathbf{b}\left[f\right]$ and $\mathbf{D}\left[f\right]$
are defined in equation (\ref{eq:coeffLandau}), and in which we recognize
$H_{MF}=\int\text{d}\mathbf{r}\,H_{MF,h}$ where $H_{MF,h}$ is the
mean field Hamiltonian (\ref{eq:Mean_field_Hamiltonian-1}) and a
new additional term $H_{I}$.

We have thus justified a large deviation principle for the rescaled
empirical density $g_{\Lambda}$ in the limit of a large plasma parameter
$\Lambda$. It reads
\begin{equation}
\mathbf{P}\left(\left\{ g_{\Lambda}(\mathbf{r},\mathbf{v},\tau)\right\} _{0\leq\tau\leq T}=\left\{ f(\mathbf{v},\tau)\right\} _{0\leq\tau\leq T}\right)\underset{\Lambda\rightarrow\infty}{\asymp}\text{e}^{-\Lambda\text{Sup}_{p}\int_{0}^{T}\text{d}\tau\,\left\{ \int\text{d}\mathbf{r}\text{d}\mathbf{v}\,\dot{f}p-H_{\text{Landau}}[f,p]\right\} },\label{eq:LDP_LAndau_from_Boltz}
\end{equation}
with the prescription that $g_{\Lambda}(\tau=0)$ is in the neighborhood of $f(\tau=0)$, and where $H_{\text{Landau}}$ is defined in \eqref{eq:ham_from_boltz}.

We note that this Hamiltonian is quadratic in its conjugate momentum
$p$. Then, in the grazing collision limit, the large deviations are
Gaussian. This is a consequence of neglecting the collisions that
involve large changes of velocity for the particles. This constrains
the fluctuations of the empirical density $g_{\Lambda}$ in a reduced
range where they can be considered as Gaussian fluctuations. As mentioned
in section \ref{subsec:LD-for-N-diff}, a quadratic Hamiltonian can
be associated with a stochastic differential equation involving a
Gaussian noise. In this case, 
\begin{equation}
\frac{\partial g_{\Lambda}}{\partial\tau}=\frac{\partial}{\partial\mathbf{v}}\cdot\left\{ -g_{\Lambda}\mathbf{b}+\mathbf{D}\frac{\partial g_{\Lambda}}{\partial\mathbf{v}}\right\} +\sqrt{\frac{2}{\Lambda}}\eta\left(\mathbf{v},\tau\right),\label{eq:SDE_Landau}
\end{equation}
with 
\small
\[
\mathbb{E}\left(\eta\left(\mathbf{r},\mathbf{v},\tau\right)\eta\left(\mathbf{r'},\mathbf{v}',\tau'\right)\right)=\frac{\partial^{2}}{\partial\mathbf{v}\partial\mathbf{v}'}:\left(g_{\Lambda}(\mathbf{v})\delta\left(\mathbf{v}-\mathbf{v}'\right)\mathbf{D}-g_{\Lambda}(\mathbf{v})g_{\Lambda}(\mathbf{v}')\mathbf{B}\left(\mathbf{v},\mathbf{v}'\right)\right)\delta\left(\mathbf{r}-\mathbf{r}'\right)\delta\left(\tau-\tau'\right).
\]
\normalsize
The Gaussian fluctuations have a non-trivial correlation structure.

\subsection{Verifications of the properties of the Hamiltonian\label{subsec:Verifications-of-the}}

Let us check all the expected properties for the Hamiltonian (\ref{eq:ham_from_boltz}).

\subsubsection{Most probable evolution}

First, we should verify that the most probable evolution associated
with this Hamiltonian is the Landau equation, i.e. that 
\begin{equation}
\frac{\partial f}{\partial\tau}=\frac{\delta H_{\text{Landau}}}{\delta p}[f,p=0]=\frac{\partial}{\partial\mathbf{v_{1}}}\left\{ -f\mathbf{b}\left[f\right]+\mathbf{D}\left[f\right]\frac{\partial f}{\partial\mathbf{v}_{1}}\right\} .\label{eq:mostprobab}
\end{equation}
We already know from equation (\ref{eq:mostprobab_markov})
\[
\frac{\delta H_{MF}}{\delta p}[f,p=0]=\frac{\partial}{\partial\mathbf{v}_{1}}\left\{ -f\mathbf{b}\left[f\right]+\mathbf{D}\left[f\right]\frac{\partial f}{\partial\mathbf{v}_{1}}\right\} .
\]
In addition to this, 
\[
\frac{\delta H_{I}}{\delta p}[f,p]=-2\frac{\partial}{\partial\mathbf{v}_{1}}\left\{ \int\text{d}\mathbf{v}_{2}\,f(\mathbf{v}_{1})f(\mathbf{v}_{2})\mathbf{B}(\mathbf{v}_{1},\mathbf{v}_{2})\frac{\partial p}{\partial\mathbf{v}_{2}}\right\} ,
\]
in particular, $\frac{\delta H_{I}}{\delta p}[f,p=0]=0$. Thus, property
(\ref{eq:mostprobab}) is verified. It is important to notice that,
since we rescaled the time variable $\tau=t/\Lambda$ by the plasma
parameter, there is no factor $\Lambda^{-1}$ in the right hand side
of \eqref{eq:mostprobab}.

\subsubsection{Conservation laws\label{subsec:Conservation-laws}}

From the result (\ref{eq:Conservation_Law_H-1}) of section \ref{sec:Dynamical-large-deviations},
we know that a functional $C[f]$ is a conserved quantity if and only
if $\int\text{d}\mathbf{r}\text{d}\mathbf{v}\frac{\delta H_{\text{Landau}}}{\delta p}\frac{\delta C}{\delta f}=0$
or equivalently, if for any $f$, $p$ and $\alpha$: $H_{\text{Landau}}[f,p]=H_{\text{Landau}}[f,p+\alpha\frac{\delta C}{\delta f}]$.

\paragraph{Mass conservation}

It is easily checked that the mass $M[f]$ defined as $M[f]=\int\text{d}\mathbf{v}f$
is conserved. Indeed, $\frac{\delta M}{\delta f}=1$ and $H_{\text{Landau}}\left[f,p+\alpha\right]=H_{\text{Landau}}\left[f,p\right]$
as $H$ does not depend explicitly on $p$ but only on its derivatives.

\paragraph{Momentum conservation}

Let us check the conservation of $\mathbf{P}$ the momentum defined
as $\mathbf{P}[f]=\int\text{d}\mathbf{v}\mathbf{v}f$. First, we notice
that $\frac{\delta\mathbf{P}}{\delta f}=\mathbf{v}$. The functional
derivative of $H$ is
\small
\[
\frac{\delta H_{\text{Landau}}}{\delta p}=\int\text{d}\mathbf{v}_{2}\,\frac{\partial}{\partial\mathbf{v}}\left\{ -\mathbf{B}(\mathbf{v},\mathbf{v}_{2})\left[\frac{\partial f}{\partial\mathbf{v}_{2}}f(\mathbf{v})-\frac{\partial f}{\partial\mathbf{v}}f(\mathbf{v}_{2})+2f(\mathbf{v})f(\mathbf{v}_{2})\left(\frac{\partial p}{\partial\mathbf{v}}-\frac{\partial p}{\partial\mathbf{v}_{2}}\right)\right]\right\} .
\]
\normalsize
Hence, integrating by parts we have
\small
\[
\int\text{d}\mathbf{r}\text{d}\mathbf{v}\,\frac{\delta H_{\text{Landau}}}{\delta p}\frac{\delta\mathbf{P}}{\delta f}=\int\text{d}\mathbf{r}\text{d}\mathbf{v}\text{d}\mathbf{v}_{2}\,\mathbf{B}(\mathbf{v},\mathbf{v}_{2})\left[\frac{\partial f}{\partial\mathbf{v}_{2}}f(\mathbf{v})-\frac{\partial f}{\partial\mathbf{v}}f(\mathbf{v}_{2})+2f(\mathbf{v})f(\mathbf{v}_{2})\left(\frac{\partial p}{\partial\mathbf{v}}-\frac{\partial p}{\partial\mathbf{v}_{2}}\right)\right].
\]
\normalsize
Then, using the fact that $\mathbf{B}(\mathbf{v},\mathbf{v}_{2})=\mathbf{B}(\mathbf{v}_{2},\mathbf{v})$,
we find
\[
\int\text{d}\mathbf{r}\text{d}\mathbf{v}\,\frac{\delta H_{\text{Landau}}}{\delta p}\frac{\delta\mathbf{P}}{\delta f}=0.
\]
This means that the total momentum $\mathbf{P}$ is conserved by the
dynamics. During this calculation, it is interesting to notice that
both the first two terms and the last two terms of $H$ preserve the
momentum independently. This means that both the deterministic part
of $H$ and the noise part of $H$ preserve the momentum independently.
More precisely, the last term that came up with our approach, which
did not appear in the naive mean field approach, compensates the contribution
of the last term of $H_{MF}$. Another interesting property, is that
a necessary condition for the deterministic part of the Hamiltonian
to conserve the momentum is the following relation between the deterministic
drift $\mathbf{b}$ and the deterministic diffusion coefficient $\mathbf{D}$:
$\int\text{d}\mathbf{v}\,f\left(\mathbf{v}\right)\left\{ \mathbf{b}\left[f\right]+\frac{\partial}{\partial\mathbf{v}}.\mathbf{D}\left[f\right]\right\} =0.$

\paragraph{Energy conservation}

Now we should check that the total kinetic energy $E$ is conserved,
with $E[f]=\frac{1}{2}\int\text{d}\mathbf{v}\,\mathbf{v}^{2}f$. Here,
$\frac{\delta E}{\delta f}=\frac{1}{2}\mathbf{v}^{2}$. Using an integration
by part we can write

\begin{multline*}
\int\text{d}\mathbf{v}\,\frac{\delta H_{\text{Landau}}}{\delta p}\frac{\delta E}{\delta f}=\int\text{d}\mathbf{v}\text{d}\mathbf{v}_{2}\,\mathbf{B}(\mathbf{v},\mathbf{v}_{2})\left\{ \left(\frac{\partial f}{\partial\mathbf{v}_{2}}f(\mathbf{v})-\frac{\partial f}{\partial\mathbf{v}}f(\mathbf{v}_{2})\right).\mathbf{v}\right.\\\left.+2\left(f(\mathbf{v})f(\mathbf{v}_{2})\left(\frac{\partial p}{\partial\mathbf{v}}-\frac{\partial p}{\partial\mathbf{v}_{2}}\right)\right).\mathbf{v}\right\} ,
\end{multline*}
and because $\mathbf{B}(\mathbf{v},\mathbf{v}_{2})=\mathbf{B}(\mathbf{v}_{2},\mathbf{v})$,
we have

\[
\int\text{d}\mathbf{v}\,\frac{\delta H_{\text{Landau}}}{\delta p}\frac{\delta E}{\delta f}=\int\text{d}\mathbf{v}\text{d}\mathbf{v}_{2}\,\left\{ \frac{\partial f}{\partial\mathbf{v}_{2}}f(\mathbf{v})+2f(\mathbf{v})f(\mathbf{v}_{2})\frac{\partial p}{\partial\mathbf{v}}\right\} \mathbf{B}(\mathbf{v},\mathbf{v}_{2}).(\mathbf{v}-\mathbf{v_{2}}).
\]
We have seen in appendix \eqref{subsec:The-Landau-collision}, that
$\mathbf{B}(\mathbf{v},\mathbf{v}_{2}).(\mathbf{v}-\mathbf{v}_{2})=\overrightarrow{0}$,
as a consequence of energy conservation in each collision. Then the
integrand of the last formula is zero and we find that the total kinetic
energy is conserved. Here too, both the deterministic part and the
noise part of $H$ preserve energy independently. 

\subsubsection{Entropy, quasipotential and time reversal symmetry}

\paragraph{Entropy and quasipotential}

We define $S[f]$ the entropy functional:
\begin{equation}
S[f]=-\int\text{d}\mathbf{v}f\log f\label{eq:Entropy}
\end{equation}
Using results from section \ref{sec:Dynamical-large-deviations},
we are going to check that $-S$ is a quasipotential as long as the
conservation laws of mass, momentum and energy hold. Here, we only
check the necessary condition which is that $-S$ satisfies the Hamilton-Jacobi
equation, more precisely that: $H_{\text{Landau}}\left[f,-\frac{\delta S}{\delta f}\right]=0$. 

Given the definition of $S$, $\frac{\delta S}{\delta f}=-\log f+c$
where $c$ is a constant which, because of the mass conservation,
has no effect and we have
\[
H_{\text{Landau}}\left[f,-\frac{\delta S}{\delta f}\right]=\int\text{d}\mathbf{r}\text{d}\mathbf{v}\text{d}\mathbf{v}_{2}\,\left(f(\mathbf{v})f(\mathbf{v}_{2})\frac{\partial^{2}\mathbf{B}}{\partial\mathbf{v}\partial\mathbf{v}_{2}}-\frac{\partial f}{\partial\mathbf{v}}\frac{\partial f}{\partial\mathbf{v}_{2}}\mathbf{B}\right).
\]
Integrating by parts twice the second term, we find out that the integrand
is zero and that $-S$ satisfies the Hamilton-Jacobi equation: $H\left[f,-\frac{\delta S}{\delta f}\right]=0$.

\paragraph{Time reversal symmetry}

We define the time reversal operator $I$ by $I[f](\mathbf{v})=f(-\mathbf{v})$.
One can easily check that $H_{\text{Landau}}\left[I[f],-I[p]\right]=H_{\text{Landau}}\left[f,p-\frac{\delta S}{\delta f}\right]$.
The computation is very close to the one above, that was performed
to prove that the entropy is the negative of the quasipotential up
to conservation laws. 

We stated in section \ref{eq:Grandes_Deviations_Theorie_Cinetique}
that $H_{\text{Landau}}\left[I[f],-I[p]\right]=H_{\text{Landau}}\left[f,p-\frac{\delta S}{\delta f}\right]$
implies a time reversal symmetry of the path $\{f(t)\}_{0\leq t\leq T}$
at the level of large deviations. The fluctuation paths are thus the
time reversed of the relaxation paths.

Moreover, from results (\ref{increase}) and (\ref{decrease}) of
section \ref{sec:Dynamical-large-deviations}, we deduce that entropy
increases along the relaxation paths. Thanks to the time reversal
symmetry of the large deviation structure, we can also conclude that
the entropy decreases along the fluctuation paths.\\

As a conclusion, we have derived the Hamiltonian for the Landau equation
and we have checked all its expected properties.

\subsection{The gradient flow structure of the Landau equation derived from the
large deviation Hamiltonian\label{subsec:Gradient-flow-structure-associat}}

It is customary and classical to observe that many dynamical models
related to kinetic theories and mesoscopic systems in interaction
with thermal baths have a gradient-transverse structure 
\begin{equation}
\frac{\partial f}{\partial t}=-\text{Grad}_{f}\mathcal{U}\left[f\right]+\mathcal{G}\left[f\right],\label{eq:Gradient-flow}
\end{equation}
where $\mathcal{U}$ might be the free energy or minus the entropy,
where for any $f$ $\left(\text{Grad}_{f}\mathcal{U},\mathcal{G}\right)=0.$
$\text{Grad}_{f}$ is the gradient with respect to a $f$-dependent
norm $(p,C\left[f\right]p)$, where $C$ is a quadratic form: $\text{Grad}_{f}\mathcal{U}\left[f\right]=C\left[f\right]\frac{\delta U}{\delta f}$.
$\mathcal{G}$ is often associated to the microscopic reversible dynamics
or the free transport. 

For example, for the Fourier law $\frac{\partial\rho}{\partial t}=D\frac{\partial^{2}\rho}{\partial\mathbf{r}^{2}}$,
has this structure \cite{otto2001geometry,villani2008optimal}, where
$\mathcal{U}=\int\text{d}\mathbf{r}\,\rho\log\rho$ is the negative
of the relative entropy, the metric used to compute the gradient is
the Wasserstein distance with $C\left[f\right]\left(\mathbf{r},\mathbf{r}'\right)=D\frac{\partial^{2}}{\partial\mathbf{r}^{2}}\left(\rho(\mathbf{r})\delta\left(\mathbf{r}-\mathbf{r}'\right)\right)$,
and $\mathcal{G}=0$. Another classical example is the McKean-Vlasov
equation \cite{otto2001geometry,villani2008optimal}. 

For the Landau equation, such a gradient structure has recently been
described by \cite{carrillo2020landau}. In this section, we explain
the connection of this structure with the large deviation formalism.\\

Even if this gradient-transverse structure is customarily observed,
it is not always easy to determine the quadratic form $C$. Moreover
a general explanation of the source of this structure is of interest.
In section 5 of \cite{Bouchet_Boltzmann_JSP}, we explain simply,
following \cite{mielke2014relation}, that there is a close relation
between the large deviations of the empirical density of particle
system with detailed balance, and the gradient-transverse flow structure
of the partial differential equations that describe kinetic theories.
Whenever the detailed balance condition \eqref{eq:detailed balance}
is satisfied at the large deviations level, and whenever the large
deviation Hamiltonian is quadratic in $p$, $\mathcal{U}$ is the
quasipotential, and the metric used to compute the gradient in (\ref{eq:Gradient-flow})
is given by the quadratic part of the large deviation Hamiltonian.

If we apply this general result to the Landau equation, using the
large deviation principle that we just derived (equations \eqref{eq:ham_from_boltz}-\eqref{eq:LDP_LAndau_from_Boltz}),
we can conclude that the Landau equation has a gradient flow structure
$\frac{\partial f}{\partial t}=-\text{Grad}_{f}\mathcal{U}\left[f\right]$
(in this case $\mathcal{G}=0$ for homogeneous distribution). It reads
\begin{equation}
\frac{\partial f}{\partial t}=\int\text{d}\mathbf{v}'\,C\left[f\right]\left(\mathbf{v},\mathbf{v}'\right)\frac{\delta S}{\delta f}\left(\mathbf{v}'\right)\label{eq:Gradient-flow-landau}
\end{equation}
where $S\left[f\right]=-\int\text{d}\mathbf{v}f\log f$ is the Boltzmann
entropy functional (the negative of the quasipotential), and $C\left[f\right]$
is the quadratic part of the Hamiltonian (\ref{eq:ham_from_boltz})
and reads
\begin{equation}
C\left[f\right]\left(\mathbf{v},\mathbf{v}'\right)=\frac{\partial^{2}}{\partial\mathbf{v}\partial\mathbf{v}'}:\left(f(\mathbf{v})\delta\left(\mathbf{v}-\mathbf{v}'\right)\mathbf{D}\left[f\right]\left(\mathbf{v}\right)-f(\mathbf{v})f(\mathbf{v}')\mathbf{B}\left[f\right]\left(\mathbf{v},\mathbf{v}'\right)\right).\label{eq:C}
\end{equation}
As discussed before, for independent particles, for instance independent
Brownian motion leading to the Fourier law, the gradient is computed
with respect to the Wasserstein distance. For particles with mean
field interactions, for instance leading to the McKean--Vlasov
equation, the relevant metric is still the Wasserstein one. More generally
for particles with mean field interaction with a diffusion coefficient
that might be non-uniform and $f$ dependent, as described in section
\ref{subsec:LD-N-coupled-diff}, from the quadratic part of the Hamiltonian
one finds $C\left[f\right]\left(\mathbf{v},\mathbf{v}'\right)=\frac{\partial^{2}}{\partial\mathbf{v}\partial\mathbf{v}'}:\left(f(\mathbf{v})\delta\left(\mathbf{v}-\mathbf{v}'\right)\mathbf{D}\left[f\right]\left(\mathbf{v}\right)\right)$.
This metric is still a kind of deformed Wasserstein one, that involves
a $f$ dependent diffusion coefficient. However for plasma in the
weak coupling limit, and the Landau equation, one can see from equation
\eqref{eq:C} that the metric is no more simply related to the Wasserstein
distance. One see in equation \eqref{eq:C}, that to the Wasserstein
like term linear in $f$ associated to independent motion of particles,
one has to add a quadratic term in $f$ related to the weak two-body
interactions. This is an interesting remark. 

\subsection{A possible candidate for the large deviations Hamiltonian for the
Balescu--Guernsey--Lenard equation\label{subsec:from_Landau_to_BGL}}

The Landau equation is also an approximation of the Balescu--Guernsey--Lenard
equation

\[
\frac{\partial f}{\partial\tau}=\frac{\partial}{\partial\mathbf{v}}\int\text{d}\mathbf{v}_{2}\,\mathbf{B}[f](\mathbf{v},\mathbf{v}_{2})\left(-\frac{\partial f}{\partial\mathbf{v}_{2}}f(\mathbf{v})+\frac{\partial f}{\partial\mathbf{v}}f(\mathbf{v}_{2})\right),
\]
which only differs from the Landau equation by the definition of the
tensor $\mathbf{B}$:

\begin{equation}
\mathbf{B}(\mathbf{v}_{1},\mathbf{v}_{2})=\pi\left(\frac{\lambda_{D}}{L}\right)^{3}\int_{-\infty}^{+\infty}\text{d}\omega\,\sum_{\mathbf{k}}\frac{\mathbf{k}\mathbf{k}\hat{W}(\mathbf{k})^{2}}{\left|\varepsilon[f]\left(\omega,\mathbf{k}\right)\right|^{2}}\delta\left(\omega-\mathbf{k}.\mathbf{v}_{1}\right)\delta\left(\omega-\mathbf{k}.\mathbf{v}_{2}\right).\label{eq:B_plasma}
\end{equation}
where $\varepsilon[f]\left(\omega,\mathbf{k}\right)$ is the dielectric
susceptibility defined by
\[
\varepsilon[f](\mathbf{k},\omega)=1-\hat{W}(\mathbf{k)}\int\text{d}\mathbf{v}\frac{\mathbf{k}.\frac{\partial f}{\partial\mathbf{v}}}{\mathbf{k}.\mathbf{v}-\omega-i\tilde{\epsilon}},
\]
which depends on the actual distribution of $f$. We recover the Landau
equation by setting $\varepsilon\left(\omega,\mathbf{k}\right)$ to
$1$. It is easy to check that the tensor $\mathbf{B}$ for the Balescu--Guernsey--Lenard
equation satisfies the same property as the tensor $\mathbf{B}$ for
the Landau equation:
\begin{enumerate}
\item $\mathbf{B}(\mathbf{v}_{1},\mathbf{v}_{2})$ is a symmetric tensor
for all pair of momenta $(\mathbf{v}_{1},\mathbf{v}_{2}),$
\item $\mathbf{\mathbf{B}}(\mathbf{v}_{1},\mathbf{v}_{2})=\mathbf{\mathbf{B}}(\mathbf{v}_{2},\mathbf{v}_{1}),$
\item $\mathbf{\mathbf{B}}(\mathbf{v}_{1},\mathbf{v}_{2}).(\mathbf{v}_{1}-\mathbf{v}_{2})=\overrightarrow{0}.$
\end{enumerate}
Furthermore, in our derivation of the Hamiltonian for the Landau equation,
we never used the explicit expression of $\mathbf{B}$ or the fact
it did not depend on $f$. 

All these remarks lead to the natural conjecture that the Hamiltonian
(\ref{eq:ham_from_boltz}) could describe the large deviations for
the Balescu--Guernsey--Lenard equation as long
as we replace the Landau expression of tensor $\mathbf{B}$ (\ref{eq:tensor_B})
by the Lenard-Balescu expression of tensor $\mathbf{B}$ (\ref{eq:B_plasma}).
In other words, we might conjecture that the large deviations of the
Balescu--Guernsey--Lenard equation are described
by the Hamiltonian $H_{BGL}^{(\text{conjecture)}}$ that reads
\begin{eqnarray}
H_{BGL}^{(\text{conjecture)}}\left[f,p\right] & = & \int\text{d}\mathbf{r}\text{d}\mathbf{v}_{1}f\left\{ \mathbf{b}\left[f\right].\frac{\partial p}{\partial\mathbf{v}_{1}}+\frac{\partial}{\partial\mathbf{v}_{1}}\left(\mathbf{D}\left[f\right]\frac{\partial p}{\partial\mathbf{v}_{1}}\right)+\mathbf{D}\left[f\right]:\frac{\partial p}{\partial\mathbf{v}_{1}}\frac{\partial p}{\partial\mathbf{v}_{1}}\right\} \ldots\label{eq:conject_BGL}\\
 & - & \int\text{d}\mathbf{r}\text{d\ensuremath{\mathbf{v}_{1}}}\text{d}\mathbf{v}_{2}f(\mathbf{v}_{1})f(\mathbf{v}_{2})\frac{\partial p}{\partial\mathbf{v}_{1}}\frac{\partial p}{\partial\mathbf{v}_{2}}:\mathbf{\mathbf{B}}\left[f\right]\left(\mathbf{v}_{1},\mathbf{v}_{2}\right).\nonumber 
\end{eqnarray}
One can check that this large deviation Hamiltonian has all the expected
properties: it has the conservation law symmetries and the negative
of the entropy $-S$ (see \eqref{eq:Entropy}) solves the stationary
Hamilton--Jacobi equation. However, we will prove in section
\ref{sec:LD-from-microscopic-dynamics} that the correct Hamiltonian for the Balescu--Guernsey--Lenard
equation is not quadratic in $p$, and that a quadratic Hamiltonian
in $p$ is obtained only in the Landau limit $k\lambda_{D}\gg1$ (or
$k\gg1$ in our set of non-dimensional variables). 

We thus conclude that although very natural, $H_{BGL}^{(\text{conjecture)}}$
is not the Hamiltonian for the large deviations associated with the
Balescu--Guernsey--Lenard equation.

\section{Large deviations associated to the Landau kinetic theory from the
microscopic dynamics \label{sec:LD-from-microscopic-dynamics}}

In this section, we compute the Hamiltonian for the large deviations
of the empirical density for plasma directly from the dynamics (\ref{eq:coulomb_dynamics}).
We use the formalism of large deviations for slow-fast system presented
in section \ref{sec:Large_Deviations_Slow_Fast}. Our result is a
series representation of the large deviation Hamiltonian for the empirical
density of $N$ particles coupled through Coulomb interactions. We
compute explicitly the terms of this series only to order four. This
expansion can be truncated at order two, and then fluctuations are
Gaussian, either the limit of a large plasma parameter ($\Lambda\rightarrow\infty$)
when $L>\lambda_{D}$ or in the limit of large $N$, when $L<\lambda_{D}$. 

We then discuss the Landau approximation. The Landau approximation
is valid to describe the contributions to the kinetic equation or
to the large deviation Hamiltonian of very large wavevectors compared
to the inverse of the Debye length ($k\gg1$ in non-dimensional plasma
variables or equivalently $k\lambda_{D}\gg1$ for physical variables).
We show that within the Landau equation, this series expansion of
the Hamiltonian can be truncated at order two, with larger order terms
being negligible. Through this truncation and noting that some terms
of order two in $p$ are also negligible, we obtain the large deviation
Hamiltonian for the Landau equation. The Landau large deviation Hamiltonian
indeed coincides with the one computed in section \ref{sec:Large-dev-from-boltzmann}
from the Boltzmann equation, as expected. In this Landau limit, the
large deviation Hamiltonian describes locally Gaussian fluctuations.
We note however, that beyond the Landau limit, when one cannot assume
that $k\gg1$ in non-dimensional plasma variables (or equivalently
$k\lambda_{D}\gg1$ for physical variables), the full expression for
the Gaussian fluctuations does not coincide with the Landau Gaussian
fluctuations.\\

In section \ref{subsec:The-quasi-linear-model}, we introduce the
quasi-linear dynamics of the empirical density of $N$ particles coupled
with a Coulomb interaction, for which the law of large numbers is
the Balescu--Guernsey--Lenard kinetic theory.
We also explain that this quasi-linear dynamics of the empirical density
can be seen as a slow-fast system. We can use the slow-fast large
deviation formalism, as presented in section \ref{sec:Large_Deviations_Slow_Fast}.
In section \ref{subsec:The-quasi-stationary-Gaussian}, we characterize
the stationary process of the fast variables, which is the fluctuating
part of the empirical density. We also perform the computation of
the two first terms of its cumulant series expansion. In section \ref{subsec:Hierarchization-of-the}
we show that the terms of this cumulant series expansion are naturally
ordered as powers of the wavevectors. As a consequence, from the two
first cumulants, we can deduce the expression of the large deviation
Hamiltonian for the Landau equation. In section \ref{subsec:Large-deviations-for-the-landau-eq},
we show that the large deviation Hamiltonian for the Landau equation,
obtained either from the microscopic dynamics or from the Boltzmann
equation, are the same. In section \ref{subsec:Large-deviations-forL<Lambda},
we discuss the large deviation result for the case where the size
of the domain is smaller than the Debye length: $L<\lambda_{D}$.
In section \ref{subsec:phys-var}, we switch back to dimensional variables
and we express the large deviation principle associated with the Landau
equation in physical units.

\subsection{The Klimontovich approach, quasilinear and slow-fast dynamics\label{subsec:The-quasi-linear-model}}

We consider the empirical density
\[
g_{\Lambda}\left(\mathbf{r},\mathbf{v},t\right)=\frac{1}{\Lambda}\sum_{n=1}^{N}\delta\left(\mathbf{v}-\mathbf{v}_{n}\left(t\right)\right)\delta\left(\mathbf{r}-\mathbf{r}_{n}\left(t\right)\right),
\]
rescaled by the plasma parameter $\Lambda$, of $N$ particles interacting
via a Coulomb potential according to the dynamics \eqref{eq:coulomb_dynamics}.
From these equations of motion, we can deduce the Klimontovich equation
\begin{equation}
\frac{\partial g_{\Lambda}}{\partial t}+\mathbf{v}\cdot\frac{\partial g_{\Lambda}}{\partial\mathbf{r}}-\frac{\partial V\left[g_{\Lambda}\right]}{\partial\mathbf{r}}\cdot\frac{\partial g_{\Lambda}}{\partial\mathbf{v}}=0.\label{eq:Klimontovich}
\end{equation}

We consider the decomposition
\[
g_{\Lambda}\left(\mathbf{r},\mathbf{v},t\right)=f_{\Lambda}\left(\mathbf{v}\right)+\frac{1}{\sqrt{\Lambda}}\delta g_{\Lambda}\left(\mathbf{r},\mathbf{v},t\right),
\]
where $f_{\Lambda}\left(\mathbf{v},t\right)=\left(\frac{\lambda_{D}}{L}\right)^{3}\int_{\left[0,L/\lambda_{D}\right]^{3}}\text{d}\mathbf{r}\,g_{\Lambda}\left(\mathbf{r},\mathbf{v},t\right)$
is the projection of $g_{\Lambda}$ on homogeneous distributions (distributions
that only depend on the velocity). From the Klimontovich equation
\eqref{eq:Klimontovich}, we straightforwardly write 

\begin{eqnarray}
\frac{\partial f_{\Lambda}}{\partial t} & = & \frac{1}{\Lambda}\left(\frac{\lambda_{D}}{L}\right)^{3}\int\text{d}\mathbf{r}\,\left(\frac{\partial V\left[\delta g_{\Lambda}\right]}{\partial\mathbf{r}}.\frac{\partial\delta g_{\Lambda}}{\partial\mathbf{v}}\right),\label{eq:slow}\\
\frac{\partial\delta g_{\Lambda}}{\partial t} & = & -\mathbf{v}.\frac{\partial\delta g_{\Lambda}}{\partial\mathbf{r}}+\frac{\partial V\left[\delta g_{\Lambda}\right]}{\partial\mathbf{r}}.\frac{\partial f_{\Lambda}}{\partial\mathbf{v}} \\
& & + \frac{1}{\sqrt{\Lambda}}\left[\frac{\partial V\left[\delta g_{\Lambda}\right]}{\partial\mathbf{r}}.\frac{\partial\delta g_{\Lambda}}{\partial\mathbf{v}}-\frac{\lambda_{D}^{3}}{\Lambda L^{3}}\int\text{d}\mathbf{r}\,\left(\frac{\partial V\left[\delta g_{\Lambda}\right]}{\partial\mathbf{r}}.\frac{\partial\delta g_{\Lambda}}{\partial\mathbf{v}}\right)\right].  \nonumber \label{eq:fast}
\end{eqnarray}

Those equations are similar to (\ref{eq:decomp}-\ref{eq:decomp2}),
but we do not take statistical averages. We will study in this section
the complete statistics of the right hand side of \eqref{eq:slow}
and not just its average as in (\ref{eq:decomp}-\ref{eq:decomp2}). 

We now assume the validity of the quasi-linear approximation, which
amounts to neglecting the terms of order $\Lambda^{-1/2}$ in the evolution
equation for $\delta g_{\Lambda}$. We also change the timescale $\tau=t/\Lambda$
and obtain the quasilinear dynamics

\begin{eqnarray}
\frac{\partial f_{\Lambda}}{\partial\tau} & = & \left(\frac{\lambda_{D}}{L}\right)^{3}\int\text{d}\mathbf{r}\,\left(\frac{\partial V\left[\delta g_{\Lambda}\right]}{\partial\mathbf{r}}.\frac{\partial\delta g_{\Lambda}}{\partial\mathbf{v}}\right),\label{eq:slow-1}\\
\frac{\partial\delta g_{\Lambda}}{\partial\tau} & =\Lambda & \left\{ -\mathbf{v}.\frac{\partial\delta g_{\Lambda}}{\partial\mathbf{r}}+\frac{\partial V\left[\delta g_{\Lambda}\right]}{\partial\mathbf{r}}.\frac{\partial f_{\Lambda}}{\partial\mathbf{v}}\right\} .\label{eq:fast-1}
\end{eqnarray}
When $\Lambda$ goes to infinity, we observe that the equation for
$\delta g_{\Lambda}$ is a fast process, with timescales for $\tau$
of order $1/\Lambda$, while the equation for $f_{\Lambda}$ is a
slow one with timescales for $\tau$ of order $1$. For such slow-fast
dynamics, it is natural to consider $f_{\Lambda}$ fixed (frozen)
in equation \eqref{eq:fast-1} on time scales for $\tau$ of order
$\tau=1/\Lambda.$ For fixed $f_{\Lambda}$, the dynamics for $\delta g_{\Lambda}$
is linear and can be solved. Computing the average of the term $\int\text{d}\mathbf{r}\,\frac{\partial V\left[\delta g_{\Lambda}\right]}{\partial\mathbf{r}}.\frac{\partial\delta g_{\Lambda}}{\partial\mathbf{v}}$,
for the asymptotic process for $\delta g_{\Lambda}$ for fixed $f_{\Lambda}$
leads to the Balescu--Guernsey--Lenard equation,
as explained in section \ref{sec:Dynamics-of-plasma}, or its Landau
approximation whenever small length scales dominate the collision
kernel of the Balescu--Guernsey--Lenard equation.
Those computations can be found in classical textbooks \cite{Nicholson_1991,Lifshitz_Pitaevskii_1981_Physical_Kinetics,schram2012kinetic}. 

In the following we want to go beyond these classical computations,
by estimating not just the average of the right hand side in \eqref{eq:slow-1}
$\int\text{d}\mathbf{r}\,\frac{\partial V\left[\delta g_{\Lambda}\right]}{\partial\mathbf{r}}.\frac{\partial\delta g_{\Lambda}}{\partial\mathbf{v}}$,
but all the cumulants of the time averages $\int_{0}^{T}\int\text{d}\mathbf{r}\,\frac{\partial V\left[\delta g_{\Lambda}\right]}{\partial\mathbf{r}}.\frac{\partial\delta g_{\Lambda}}{\partial\mathbf{v}}$
in order to describe the large deviations for the process $f_{\Lambda}$.
Using the classical result of large deviations for slow-fast dynamics,
as explained in section \ref{sec:Large_Deviations_Slow_Fast} (see
equations (\ref{eq:Large_Deviations_Slow_Fast}-\ref{eq:Hamiltonian_Slow_Fast})),
we conclude that 
\begin{equation}
\mathbf{P}(f_{\Lambda}=f)\underset{\Lambda\rightarrow\infty}{\asymp}\text{e}^{-\Lambda\text{Sup}_{p}\int_{0}^{T}\left\{ \int\text{d}\mathbf{r}\text{d}\mathbf{v}\,\dot{f}p-H[f,p]\right\} }.\label{eq:PGD_H_GE}
\end{equation}
with the prescription that $f_{\Lambda}(\tau=0)$ is in the neighborhood of $f(\tau=0)$, and where
\begin{equation}
H\left[f,p\right]=\underset{T\rightarrow\infty}{\lim}\frac{1}{T}\log\mathbb{E}_{f}\left[\text{exp\ensuremath{\left(\int_{0}^{T}\text{d}t\,\int\text{d}\mathbf{v}\,p\left(\mathbf{v}\right)\int\text{d}\mathbf{r}'\frac{\partial V\left[\delta g_{\Lambda}\right]}{\partial\mathbf{r}'}.\frac{\partial\delta g_{\Lambda}}{\partial\mathbf{v}}\right)}}\right]\label{eq:H_Gartner_ellis}
\end{equation}
and where $\mathbb{E}_{f}$ denotes the expectation on the process
for $\delta g_{\Lambda},$ where $\delta g_{\Lambda}$ evolves according
to 
\begin{equation}
\frac{\partial\delta g_{\Lambda}}{\partial t}= -\mathbf{v}.\frac{\partial\delta g_{\Lambda}}{\partial\mathbf{r}}+\frac{\partial V\left[\delta g_{\Lambda}\right]}{\partial\mathbf{r}}.\frac{\partial f}{\partial\mathbf{v}} .\label{eq:Linear_deltagN_fast}
\end{equation}
In this equation $f_{\Lambda}=f$ is frozen and time independent. 

We note that to obtain equation \eqref{eq:H_Gartner_ellis} from equation
\eqref{eq:Hamiltonian_Slow_Fast}, we have considered $f_{\Lambda}$
as a function of the $\mu$-space. Then the conjugated momentum $p\left(\mathbf{r},\mathbf{v}\right)$
should also be a function of the $\mu$-space and the scalar product
be the one of the $\mu$-space. However, recognizing that for homogeneous
$f$, $p$ should also be homogeneous ($p\left(\mathbf{r},\mathbf{v}\right)=p\left(\mathbf{v}\right)$),
and performing trivial integration over $\mathbf{r}$ leads to \eqref{eq:H_Gartner_ellis}.

The goal of the following subsections is to compute \eqref{eq:H_Gartner_ellis}.

\subsection{The quasi-stationary Gaussian process for $ \delta g_{\Lambda}$\label{subsec:The-quasi-stationary-Gaussian}}

In order to compute \eqref{eq:H_Gartner_ellis}, we first note that
for frozen $f$, equation \eqref{eq:Linear_deltagN_fast} is linear.
It thus describe a Gaussian process, for instance when the initial
conditions are distributed according to a Gaussian. Moreover, as explained
in \S 51 of \cite{Lifshitz_Pitaevskii_1981_Physical_Kinetics} such
a process is expected to converge to a stationary Gaussian process
irrespective of the initial condition. The properties of this process
are determined by the fact that we are dealing with a dynamics with
discrete particles. In the following we will thus consider averages
in equation \eqref{eq:H_Gartner_ellis} as averages over this stationary
Gaussian process. Such stationary averages are denoted $\mathbb{E}_{S}$.

We do not reproduce the classical computations of the correlation
functions of this stationary process, but just report the formulas
which can be found for instance in \S 51 of \cite{Lifshitz_Pitaevskii_1981_Physical_Kinetics}.
The potential autocorrelation function is homogeneous because of the
space translation symmetry. Then 
\[
\mathbb{E}_{S}\left(V\left[\delta g_{\Lambda}\right]\left(\mathbf{r}_{1},t_{1}\right)V\left[\delta g_{\Lambda}\right]\left(\mathbf{r}_{2},t_{2}\right)\right)=\mathcal{C}_{VV}\left(\mathbf{r}_{1}-\mathbf{r}_{2},t_{1}-t_{2}\right),
\]
We define $\tilde{\varphi}$ the Fourier-Laplace transform of a function
$\varphi$ as 
\begin{equation}
\tilde{\varphi}\left(\mathbf{k},\omega\right)=\int_{\left[0,L/\lambda_{D}\right]^{3}}\text{d}\mathbf{r}\int_{0}^{\infty}\text{d}t\,\text{e}^{-i\left(\mathbf{k}.\mathbf{r}-\omega t\right)}\varphi\left(\mathbf{r},t\right),\label{eq:FL_Transf}
\end{equation}
following the same convention as in \cite{Lifshitz_Pitaevskii_1981_Physical_Kinetics}.
The autocorrelation function of the Fourier-Laplace transform of the
potential then reads 
\begin{equation}
\mathbb{E}_{S}\left(V\left[\widetilde{\delta g_{\Lambda}}\right]\left(\mathbf{k}_{1},\omega_{1}\right)V\left[\widetilde{\delta g_{\Lambda}}\right]\left(\mathbf{k}_{2},\omega_{2}\right)\right)=2\pi\left(\frac{L}{\lambda_{D}}\right)^{3}\delta_{\mathbf{k}_{1},-\mathbf{k}_{2}}\delta\left(\omega_{1}+\omega_{2}\right)\widetilde{\mathcal{C}_{VV}}\left(\mathbf{k}_{1},\omega_{1}\right),\label{eq:Autocorrelation_VV}
\end{equation}
where $\widetilde{\mathcal{C}_{VV}}$ is the space-time Fourier transform
of $\mathcal{C}_{VV}$. Equation (51.20), \S 51 of \cite{Lifshitz_Pitaevskii_1981_Physical_Kinetics},
with the identifications $V=\varphi$, $\hat{W}\left(\mathbf{k}\right)=1/k^{2}$,
and with the dimensionless variables defined in section \ref{subsec:plasma},
gives 
\begin{equation}
\widetilde{\mathcal{C}_{VV}}\left(\mathbf{k},\omega\right)=2\pi\left[\int\text{d}\mathbf{v}'\,f\left(\mathbf{v}'\right)\delta\left(\omega-\mathbf{k}.\mathbf{v}'\right)\right]\frac{\hat{W}\left(\mathbf{k}\right)^{2}}{\left|\varepsilon\left[f\right]\left(\mathbf{k},\omega\right)\right|^{2}},\label{eq:Autocorrelation_Potentiel_Fourier_Laplace}
\end{equation}

Similarly the time stationary correlation functions between the potential
and the distribution fluctuations is space-time homogeneous $\mathbb{E}_{S}\left(V\left[\delta g_{N}\right]\left(\mathbf{r}_{1},t_{1}\right)\delta g_{N}\left(\mathbf{r}_{2},\text{\ensuremath{\mathbf{v}}},t_{2}\right)\right)=\mathcal{C}_{VG}\left(\mathbf{r}_{1}-\mathbf{r}_{2},t_{1}-t_{2},\mathbf{v}\right),$
with space-time Fourier transform 
\begin{equation}
\mathbb{E}_{S}\left(V\left[\widetilde{\delta g_{\Lambda}}\right]\left(\mathbf{k}_{1},\omega_{1}\right)\widetilde{\delta g_{\Lambda}}\left(\mathbf{k}_{2},\omega_{2}\right)\right)=2\pi\left(\frac{L}{\lambda_{D}}\right)^{3}\delta_{\mathbf{k}_{1},-\mathbf{k}_{2}}\delta\left(\omega_{1}+\omega_{2}\right)\widetilde{\mathcal{C}_{VG}}\left(\mathbf{k}_{1},\omega_{1},\mathbf{v}\right).\label{eq:Autocorrelation_VG}
\end{equation}
Similarly $\mathbb{E}_{S}\left(\delta g_{\Lambda}\left(\mathbf{r}_{1},\mathbf{v}_{1},t_{1}\right)\delta g_{\Lambda}\left(\mathbf{r}_{2},\mathbf{v}_{2},t_{2}\right)\right)=\mathcal{C}_{GG}\left(\mathbf{r}_{1}-\mathbf{r}_{2},t_{1}-t_{2},\mathbf{v}_{1},\mathbf{v}_{2}\right),$
with 
\begin{equation}
\mathbb{E}_{S}\left(\widetilde{\delta g_{\Lambda}}\left(\mathbf{k}_{1},\omega_{1}\right)\widetilde{\delta g_{\Lambda}}\left(\mathbf{k}_{2},\omega_{2}\right)\right)=2\pi\left(\frac{L}{\lambda_{D}}\right)^{3}\delta_{\mathbf{k}_{1},-\mathbf{k}_{2}}\delta\left(\omega_{1}+\omega_{2}\right)\widetilde{\mathcal{C}_{GG}}\left(\mathbf{k}_{1},\omega_{1},\mathbf{v}_{1},\mathbf{v}_{2}\right).\label{eq:Autocorrelation_GG}
\end{equation}
The formulas for $\widetilde{\mathcal{C}_{VG}}$ are given by equation
(51.21) and (51.23) respectively, in \cite{Lifshitz_Pitaevskii_1981_Physical_Kinetics},
with the identifications $V=\varphi$, $\hat{W}\left(\mathbf{k}\right)=1/k^{2}$.
They are 

\begin{equation}
\widetilde{\mathcal{C}_{VG}}\left(\mathbf{k},\omega,\mathbf{v}\right)=-\frac{\mathbf{k}}{\omega-\mathbf{k}.\mathbf{v}-i\tilde{\epsilon}}.\frac{\partial f}{\partial\mathbf{v}}\left(\mathbf{v}\right)\widetilde{\mathcal{C}_{VV}}\left(\mathbf{k},\omega\right)+\frac{2\pi\hat{W}\left(\mathbf{k}\right)}{\varepsilon\left[f\right]\left(\mathbf{k},\omega\right)}f\left(\mathbf{v}\right)\delta\left(\omega-\mathbf{k}.\mathbf{v}\right),\label{eq:Autocorrelation_VG_Fourier_Laplace}
\end{equation}
and
\begin{eqnarray}
\widetilde{\mathcal{C}_{GG}}\left(\mathbf{k},\omega,\mathbf{v}_{1},\mathbf{v}_{2}\right) & = & 2\pi\delta\left(\mathbf{v}_{1}-\mathbf{v}_{2}\right)f\left(\mathbf{v}_{1}\right)\delta\left(\omega-\mathbf{k}.\mathbf{v}_{1}\right)\label{eq:Autocorrelation_GG_Fourier_Laplace}\\
 & + & \frac{\widetilde{\mathcal{C}_{VV}}\left(\mathbf{k},\omega\right)}{\left(\omega-\mathbf{k}.\mathbf{v}_{1}+i\tilde{\epsilon}\right)\left(\omega-\mathbf{k}.\mathbf{v}_{2}-i\tilde{\epsilon}\right)}\mathbf{k}.\frac{\partial f}{\partial\mathbf{v}}\left(\mathbf{v}_{1}\right)\mathbf{k}.\frac{\partial f}{\partial\mathbf{v}}\left(\mathbf{v}_{2}\right)\nonumber \\
 & - & 2\pi\hat{W}\left(\mathbf{k}\right)\mathbf{k}.\frac{\partial f}{\partial\mathbf{v}}\left(\mathbf{v}_{1}\right)\frac{f(\mathbf{v}_{2})\delta\left(\omega-\mathbf{k}.\mathbf{v}_{2}\right)}{\varepsilon\left(\mathbf{k},\omega\right)\left(\omega-\mathbf{k}.\mathbf{v}_{1}+i\tilde{\epsilon}\right)}\nonumber \\
 & - & 2\pi\hat{W}\left(\mathbf{k}\right)\mathbf{k}.\frac{\partial f}{\partial\mathbf{v}}\left(\mathbf{v}_{2}\right)\frac{f(\mathbf{v}_{1})\delta\left(\omega-\mathbf{k}.\mathbf{v}_{1}\right)}{\varepsilon^{*}\left(\mathbf{k},\omega\right)\left(\omega-\mathbf{k}.\mathbf{v}_{2}-i\tilde{\epsilon}\right)}.\nonumber 
\end{eqnarray}
We note that the order in the correlation functions for $V$ and $\delta g_{\Lambda}$
matters. We have $\mathbb{E}_{S}\left(\delta g_{\Lambda}\left(\mathbf{r}_{1},\text{\ensuremath{\mathbf{v}}},t_{1}\right)V\left[\delta g_{\Lambda}\right]\left(\mathbf{r}_{2},t_{2}\right)\right)=\mathcal{C}_{GV}\left(\mathbf{r}_{1}-\mathbf{r}_{2},t_{1}-t_{2},\mathbf{v}\right)$. Then 
$\widetilde{\mathcal{C}_{VG}}\left(\mathbf{k},\omega,\mathbf{v}\right)$ $=\widetilde{\mathcal{C}_{GV}}\left(-\mathbf{k},-\omega,\mathbf{v}\right)$ $=\widetilde{\mathcal{C}_{GV}}^{*}\left(\mathbf{k},\omega,\mathbf{v}\right).$
We also note the symmetry property of $\widetilde{\mathcal{C}_{GG}}$:
$\widetilde{\mathcal{C}_{GG}}\left(\mathbf{k},\omega,\mathbf{v}_{1},\mathbf{v}_{2}\right)=\widetilde{\mathcal{C}_{GG}}\left(-\mathbf{k},-\omega,\mathbf{v}_{2},\mathbf{v}_{1}\right)$,
which is a consequence of the symmetry $\mathcal{C}_{GG}\left(\mathbf{r},t,\mathbf{v}_{1},\mathbf{v}_{2}\right)=\mathcal{C}_{GG}\left(-\mathbf{r},-t,\mathbf{v}_{2},\mathbf{v}_{1}\right)$.
\\

From this stationary Gaussian process, we are now ready to compute
the large deviation Hamiltonian through a cumulant expansion in the
two following sections. 

\subsection{Computation of a series expansion of large deviation Hamiltonian\label{subsec:LDH-as-a-series-exp}}

In order to have explicit formulas for \eqref{eq:H_Gartner_ellis},
in this section we first compute the two first cumulants for 
\begin{equation}
X\left[f\right]=-\int_{0}^{T}\text{d}t\,\int\text{d}\mathbf{v}\,\frac{\partial p}{\partial\mathbf{v}}\int\text{d}\mathbf{r}\frac{\partial V\left[\delta g_{\Lambda}\right]}{\partial\mathbf{r}}.\delta g_{\Lambda}.\label{eq:Xf}
\end{equation}
If we expand a cumulant generating function for a random variable
$X$, we obtain $\log\mathbb{E}\exp(X)=\mathbb{E}(X)+\mathbb{E}\mathbb{E}\left(X^{2}\right)/2+H^{>2}$,
where for the second order cumulant we use the notation $\mathbb{E}\mathbb{E}\left(X^{2}\right)=\mathbb{E}\left(X^{2}\right)-\left[\mathbb{E}\left(X\right)\right]^{2}$,
and where $H^{>2}$ is the contribution of all cumulants of order
larger than $2$. We thus have 
\begin{equation}
H=H^{(1)}+H^{(2)}+H^{>2}.\label{eq:H_cumul_decomp-1}
\end{equation}
If $X$ is given by (\ref{eq:Xf}), we have 

\begin{equation}
H^{(1)}=\int\text{d}\mathbf{r}\int\text{d}\mathbf{v}\,\frac{\partial p}{\partial\mathbf{v}} \mathbf{C}^{(1)}(\mathbf{v}),\,\,\,\text{where}\,\,\,\mathbf{C}^{(1)}(\mathbf{v})=-\mathbb{E}_{S}\left(\frac{\partial V\left[\delta g_{\Lambda}\right]}{\partial\mathbf{r}}\delta g_{\Lambda}\right).\label{eq:firstcumul}
\end{equation}
and
\begin{equation}
H^{(2)}=\int\text{d}\mathbf{r}\text{d}\mathbf{v}_{1}\text{d}\mathbf{v}_{2}\,\frac{\partial p}{\partial\mathbf{v}_1}\frac{\partial p}{\partial\mathbf{v}_2}:\mathbf{C}(\mathbf{v}_{1},\mathbf{v}_{2}),\label{eq:H^2}
\end{equation}
where
\small
\begin{equation}
\mathbf{C}(\mathbf{v}_{1},\mathbf{v}_{2})=\lim_{T\rightarrow\infty}\frac{1}{2T}\int_{0}^{T}\text{d}t_{1}\int_{0}^{T}\text{d}t_{2}\int\text{d}\mathbf{r}_{1}\int\text{d}\mathbf{r}_{2}\,\mathbb{E}\mathbb{E}\left[\frac{\partial V\left[\delta g_{\Lambda}\right]^{(1)}}{\partial\mathbf{r}}\delta g_{\Lambda}^{(1)}\frac{\partial V\left[\delta g_{\Lambda}\right]^{(2)}}{\partial\mathbf{r}}\delta g_{\Lambda}^{(2)}\right].\label{eq:Definition_C}
\end{equation}
\normalsize
We note that $\mathbf{C}$ is a second order tensor and that in the
formula for $H^{(2)}$ , the symbol ``$:$'' means the contraction
of two second order tensors. In the formula for $\mathbf{C}$,
the superscripts $(1)$ or $(2)$ mean that the quantities are evaluated
at either $\left(\mathbf{r}_{1},t_{1}\right)$ and $\left(\mathbf{r}_{2},t_{2}\right)$,
respectively, or $\left(\mathbf{r}_{1},\mathbf{v}_{1},t_{1}\right)$
and $\left(\mathbf{r}_{2},\mathbf{v}_{2},t_{2}\right)$, respectively.

We note that a truncation at second order of the cumulant expansion
gives a Hamiltonian which is quadratic in $p$.

\paragraph{Computation of the first cumulant}

Using (\ref{eq:Xf}) and (\ref{eq:Autocorrelation_VG_Fourier_Laplace})
one can compute $\mathbf{C}^{(1)}$. The computations can be found
in appendix \ref{subsec:Comput-linear}. The computations are not
exactly the same, but really similar to the one in \S 51 of \cite{Lifshitz_Pitaevskii_1981_Physical_Kinetics}.
One obtains 
\begin{eqnarray*}
\mathbf{C}^{(1)}(\mathbf{v})&=&\int\text{d}\mathbf{v}_{2}\,\mathbf{B}\left[f\right](\mathbf{v},\mathbf{v}_{2})\left(-\frac{\partial f}{\partial\mathbf{v}_{2}}f(\mathbf{v})+f(\mathbf{v}_{2})\frac{\partial f}{\partial\mathbf{v}}\right)\\&=&\mathbf{b}\left[f\right]\left(\mathbf{v}\right)f\left(\mathbf{v}\right)-\mathbf{D}\left[f\right]\left(\mathbf{v}\right).\frac{\partial f}{\partial\mathbf{v}},
\end{eqnarray*}
where $\mathbf{B}$ is the tensor defined in equation
(\ref{eq:B_GLB}). Integrating over $\mathbf{r}$ in equation (\ref{eq:firstcumul}),
we find that $H^{(1)}$ is then given by 
\begin{equation}
H^{(1)}=\int\text{d}\mathbf{r}\text{d}\mathbf{v}\,f\left(\mathbf{v}\right)\left\{ \mathbf{b}\left[f\right].\frac{\partial p}{\partial\mathbf{v}}+\frac{\partial}{\partial\mathbf{v}}\left(\mathbf{D}\left[f\right]\frac{\partial p}{\partial\mathbf{v}}\right)\right\} ,\label{eq:H^1_fin}
\end{equation}
where $\mathbf{b}\left[f\right]$ and $\mathbf{D}\left[f\right]$
are defined in equation (\ref{eq:coeffLandau}). $H^{(1)}$, which
is the linear part with respect to $p$, gives the formula that corresponds
to the Balescu--Guernsey--Lenard operator, as
expected.

\paragraph{Computation of the second cumulant}

Now, the more challenging and new part is to compute $H^{(2)}$ the
second cumulant. In order to compute (\ref{eq:Definition_C}) using
(\ref{eq:Xf}), we see that we will have to evaluate four-point correlation
functions. As the fluctuations are locally Gaussian, we can use Wick's
theorem in order to express the four-points correlation functions
$\mathbb{E}_{S}\left(\frac{\partial V\left[\delta g_{N}\right]^{(1)}}{\partial\mathbf{r}}\frac{\partial V\left[\delta g_{N}\right]^{(2)}}{\partial\mathbf{r}}\delta g_{N}^{(1)}\delta g_{N}^{(2)}\right)$
as a sum of products of two-points correlation functions, and use
the formulas for the two point correlation functions. After some lengthy
computations reported in appendix \ref{subsec:Appendice_Second_Cumulant},
we obtain the result
\begin{multline}
H^{(2)}=\int\text{d}\mathbf{r}\text{d}\mathbf{v}_{1}\,\frac{\partial p}{\partial\mathbf{v}_{1}}\frac{\partial p}{\partial\mathbf{v}_{1}}:\mathbf{D}(\mathbf{v}_{1})f\left(\mathbf{v}_{1}\right)\\
-\int\text{d}\mathbf{r}\text{d}\mathbf{v}_{1}\text{d}\mathbf{v}_{2}\,\frac{\partial p}{\partial\mathbf{v}_{1}}\frac{\partial p}{\partial\mathbf{v}_{2}}:\mathbf{B}[f](\mathbf{v}_{1},\mathbf{v}_{2})f\left(\mathbf{v}_{1}\right)f\left(\mathbf{v}_{2}\right)\\
+\int\text{d}\mathbf{r}\text{d\ensuremath{\mathbf{v}_{1}}}\text{d}\mathbf{v}_{2}\text{d\ensuremath{\mathbf{v}_{3}}}\text{d}\mathbf{v}_{4}\,\frac{\partial p}{\partial\mathbf{v}_{1}}\frac{\partial p}{\partial\mathbf{v}_{2}}\mathbf{B}^{(2)}\left(\mathbf{v}_{1},\mathbf{v}_{2},\mathbf{v}_{3},\mathbf{v}_{4}\right)\left\{ f(\mathbf{v}_{1})f(\mathbf{v}_{2})\frac{\partial f}{\partial\mathbf{v}_{3}}\frac{\partial f}{\partial\mathbf{v}_{4}}\right.\\
\left.-2f(\mathbf{v}_{1})\frac{\partial f}{\partial\mathbf{v}_{2}}f(\mathbf{v}_{3})\frac{\partial f}{\partial\mathbf{v}_{4}}+\frac{\partial f}{\partial\mathbf{v}_{1}}\frac{\partial f}{\partial\mathbf{v}_{2}}f(\mathbf{v}_{3})f(\mathbf{v}_{4})\right\} ,\label{eq:H^2_rec-1}
\end{multline}
with
\begin{equation}
\mathbf{B}^{(2)}\left(\mathbf{v}_{1},\mathbf{v}_{2},\mathbf{v}_{3},\mathbf{v}_{4}\right)=2\pi^{3}\left(\frac{\lambda_{D}}{L}\right)^{3}\sum_{\mathbf{k}}\int\text{d}\omega\,\mathbf{k}\mathbf{k}\mathbf{k}\mathbf{k}\frac{\hat{W}\left(\mathbf{k}\right)^{4}}{\left|\varepsilon\left(\mathbf{k},\omega\right)\right|^{4}}\prod_{i=1}^{4}\delta\left(\omega-\mathbf{k}.\mathbf{v}_{i}\right),\label{eq:B^2-1}
\end{equation}
being a fully symmetric order-4 tensor.

\subsection{Computation of higher order cumulants}

Using this same method, it is possible to compute, by induction, every
terms of the cumulant expansion. However, there is a infinity of them,
and a priori they are non-zero and, in general, they are of the same
order of magnitude as the second order one. Nevertheless, we can recognize
a pattern in the cumulant expansion. To understand it better, let
us compute the following term in the cumulant expansion of the large
deviation Hamiltonian .

In this subsection, we compute the third cumulant, but the procedure
would be exactly the same if we were to compute the $n$-th cumulant.
We already computed the first two cumulants $H^{(1)}$ and $H^{(2)}$
associated to this cumulant generating function. Now let us compute
the third cumulant $H^{(3)}$ which can be expressed as a combination
of moments of the random variable $X$ (\ref{eq:Xf}): 
\[H^{(3)}=\underset{T\rightarrow\infty}{\lim}\frac{1}{T}\left(\mathbb{E}\left(X^{3}\right)-3\mathbb{E}\left(X^{2}\right)\mathbb{E}\left(X\right)+2\mathbb{E}\left(X\right)^{3}\right).
\]
Similarly, we denote $H^{(n)}$ the term of the large deviation Hamiltonian
\eqref{eq:H_Gartner_ellis} accounting for the contribution of the
$n$-th cumulant of the random variable $X$ (\ref{eq:Xf}).

Because the process for $\delta g_{\Lambda}$ is Gaussian, we can
compute all the moments of $X$ from the two-points correlation functions
(\ref{eq:Autocorrelation_Potentiel_Fourier_Laplace}, \ref{eq:Autocorrelation_VG_Fourier_Laplace},
\ref{eq:Autocorrelation_GG_Fourier_Laplace}) and the Wick theorem.
To express the result, let us introduce the fully symmetric order-2n
tensor $B^{(n)}$ defined as 
\begin{equation}
\mathbf{B}^{(n)}\left(\mathbf{v}_{1},\ldots,\mathbf{v}_{2n}\right)=\frac{\left(2\pi\right)^{2n}}{4\pi n}\left(\frac{\lambda_{D}}{L}\right)^{3}\sum_{\mathbf{k}}\int_{\Gamma}\text{d}\omega\,\frac{\mathbf{k}^{\otimes2n}\hat{W}\left(\mathbf{k}\right)^{2n}}{\left|\varepsilon\left(\mathbf{k},\omega\right)\right|^{2n}}\prod_{i=1}^{2n}\delta\left(\omega-\mathbf{k}.\mathbf{v}_{i}\right),\label{eq:tensor_B_Generalized}
\end{equation}
where $\mathbf{k}^{\otimes2n}$ is the tensor $\underset{2n\,\text{times}}{\mathbf{k}\otimes...\otimes\mathbf{k}}$,
such as $\mathbf{B}^{(1)}=\mathbf{B}$, and it is consistent with
the definition of $\mathbf{B}^{(2)}$ (\ref{eq:B^2}). Then, the third
cumulant reads
\small
\begin{multline}
H^{(3)}=2\int\text{d}\mathbf{r}\text{d\ensuremath{\mathbf{v}_{1}}}\text{d}\mathbf{v}_{2}\text{d\ensuremath{\mathbf{v}_{3}}}\text{d}\mathbf{v}_{4}\,\frac{\partial p}{\partial\mathbf{v}_{1}}\frac{\partial p}{\partial\mathbf{v}_{2}}\left\{ \frac{\partial p}{\partial\mathbf{v}_{2}}-\frac{\partial p}{\partial\mathbf{v}_{3}}\right\}\mathbf{B}^{(2)}f(\mathbf{v}_{2})f(\mathbf{v}_{3})\left(f(\mathbf{v}_{1})\frac{\partial f}{\partial\mathbf{v}_{4}}-\frac{\partial f}{\partial\mathbf{v}_{1}}f(\mathbf{v}_{4})\right) \\
+\int\text{d}\mathbf{r}\text{d\ensuremath{\mathbf{v}_{1}}}\ldots\text{d}\mathbf{v}_{6}\,\frac{\partial p}{\partial\mathbf{v}_{1}}\frac{\partial p}{\partial\mathbf{v}_{2}}\frac{\partial p}{\partial\mathbf{v}_{3}}\mathbf{B}^{(3)}\left\{ f(\mathbf{v}_{1})f(\mathbf{v}_{2})f(\mathbf{v}_{3})\frac{\partial f}{\partial\mathbf{v}_{4}}\frac{\partial f}{\partial\mathbf{v}_{5}}\frac{\partial f}{\partial\mathbf{v}_{6}}\right.\\
-3\frac{\partial f}{\partial\mathbf{v}_{1}}f(\mathbf{v}_{2})f(\mathbf{v}_{3})f\left(\mathbf{v}_{4}\right)\frac{\partial f}{\partial\mathbf{v}_{5}}\frac{\partial f}{\partial\mathbf{v}_{6}}\\
\left.+3f(\mathbf{v}_{1})\frac{\partial f}{\partial\mathbf{v}_{2}}\frac{\partial f}{\partial\mathbf{v}_{3}}f\left(\mathbf{v}_{4}\right)\frac{\partial f}{\partial\mathbf{v}_{5}}\frac{\partial f}{\partial\mathbf{v}_{6}}-\frac{\partial f}{\partial\mathbf{v}_{1}}\frac{\partial f}{\partial\mathbf{v}_{2}}\frac{\partial f}{\partial\mathbf{v}_{3}}f\left(\mathbf{v}_{4}\right)f\left(\mathbf{v}_{5}\right)f\left(\mathbf{v}_{6}\right)\right\} .\label{eq:H^(3)}
\end{multline}
\normalsize
We note that $H^{(3)}$ involves a term which is proportional to $\mathbf{B}^{(3)}$,
but also a term which is proportional to $\mathbf{B}^{(2)}$.

We do not report the detailed computation, but we have explicitly
computed $H^{(4)}$ (see appendix \ref{sec:Computation-of-higher}).
We observe that it involves terms which are proportional to $\mathbf{B}^{(4)}$, $\mathbf{B}^{(3)}$, and $\mathbf{B}^{(2)}$  but no terms which are proportional to $\mathbf{B}^{(1)}$. Based on this remark,
we conjecture that $H^{(n)}$ contains only terms which are proportional
to the tensors $\mathbf{B}^{(k)}$  with $k \geq n/2$. As a consequence of this conjecture, only the two first cumulants $H^{(1)}$ and $H^{(2)}$ involve the tensor $\mathbf{B}=\mathbf{B}^{(1)}$ whereas all the other cumulants $H^{(n)}$ for $n>2$  only involve the tensors $\mathbf{B}^{(k)}$ with $k \geq 2$.

As we will explain in the next subsection, in the context of the Landau
approximation, there is a natural hierarchy between the tensors $\mathbf{B}^{(n)}$
and the cumulant expansion can be simply truncated.

\subsection{Hierarchy of the series expansion within the Landau approximation\label{subsec:Hierarchization-of-the}}

Let us first recall that we can obtain the Landau equation from the
Balescu--Guernsey-- Lenard equation. The collision
kernel for the Balescu--Guernsey--Lenard equation
converges to the Landau collision kernel in the limit where all the
wavevectors in \eqref{eq:B_GLB} satisfy $k\lambda_{D}\gg1$. In our
system of plasma unit, where the length unit is renormalized by the
Debye length, this means that the Balescu--Guernsey--Lenard
collision kernel converges toward the Landau collision kernel in the
limit of infinitely large wavevectors. In a similar way, we obtain
the large deviation Hamiltonian for the Landau equation $H_{\text{Landau}}$
from the large deviation Hamiltonian $H$ \eqref{eq:H_cumul_decomp-1}
of the empirical density of $N$ Coulomb interacting particles using
the same limit.

In the expression of the tensor $\mathbf{B}^{(1)}=\mathbf{B}$
\[
\mathbf{B}^{(1)}=\mathbf{B}(\mathbf{v}_{1},\mathbf{v}_{2})=\pi\left(\frac{\lambda_{D}}{L}\right)^{3}\int_{-\infty}^{+\infty}\text{d}\omega\,\sum_{\mathbf{k}}\frac{\mathbf{k}\mathbf{k}}{k^{4n}\left|\varepsilon[f]\left(\omega,\mathbf{k}\right)\right|^{2}}\delta\left(\omega-\mathbf{k}.\mathbf{v}_{1}\right)\delta\left(\omega-\mathbf{k}.\mathbf{v}_{2}\right),
\]
the Landau approximation implies that $k\gg1$. In this context, we
can consider that the dielectric function $\varepsilon$ is equal
to one. From there, a clear hierarchy appears in the cumulant series
expansion (\ref{eq:H_cumul_decomp-1}). For $n\geq2$, the terms involving
\[
\mathbf{B}^{(n)}\left(\mathbf{v}_{1},\ldots,\mathbf{v}_{2n}\right)=\frac{\left(2\pi\right)^{2n}}{4\pi n}\left(\frac{\lambda_{D}}{L}\right)^{3}\sum_{\mathbf{k}}\int_{\Gamma}\text{d}\omega\,\frac{\mathbf{k}^{\otimes2n}}{k^{4n}\left|\varepsilon\left(\mathbf{k},\omega\right)\right|^{2n}}\prod_{i=1}^{2n}\delta\left(\omega-\mathbf{k}.\mathbf{v}_{i}\right)
\]
will be negligible with respect to the terms involving $\mathbf{B}^{(1)}=\mathbf{B}$.

Let us define 
\[
\mathbf{B}_{\mathbf{k}}^{(n)}\left(\mathbf{v}_{1},\ldots,\mathbf{v}_{2n}\right)=\int_{\Gamma}\text{d}\omega\,\frac{\mathbf{k}^{\otimes2n}}{k^{4n}\left|\varepsilon\left(\mathbf{k},\omega\right)\right|^{2n}}\prod_{i=1}^{2n}\delta\left(\omega-\mathbf{k}.\mathbf{v}_{i}\right),
\]
such that
\[
\mathbf{B}^{(n)}\left(\mathbf{v}_{1},\ldots,\mathbf{v}_{2n}\right)=\frac{\left(2\pi\right)^{2n}}{4\pi n}\sum_{\mathbf{k}}\left(\frac{\lambda_{D}}{L}\right)^{3}\mathbf{B}_{\mathbf{k}}^{(n)}\left(\mathbf{v}_{1},\ldots,\mathbf{v}_{2n}\right).
\]
Let us evaluate the size of $\mathbf{B}_{\mathbf{k}}^{(n)}$ in terms
of the wavevectors $\mathbf{k}$. We have,

\[
\mathbf{B}_{\mathbf{k}}^{(n)}\left(\mathbf{v}_{1},\ldots,\mathbf{v}_{2n}\right)=k^{1-4n}\frac{\mathbf{m}^{\otimes2n}}{\left|\varepsilon\left(\mathbf{k},\omega\right)\right|^{2n}}\prod_{i=2}^{2n}\delta\left(\mathbf{m}.\left(\mathbf{v}_{1}-\mathbf{v}_{i}\right)\right),
\]
where $\mathbf{m}=\mathbf{k}/k$. Then,

\begin{equation}
\left(\frac{\lambda_{D}}{L}\right)^{3}\mathbf{B}_{\mathbf{k}}^{(n)}\underset{k\gg1}{=}\mathcal{O}\left(\left(\frac{\lambda_{D}}{Lk}\right)^{3}\left(\frac{1}{k}\right)^{4n-4}\right),\label{eq:B_k_ordre}
\end{equation}
where $\mathcal{O}\left(k^{m}\right)$ means that the term is of order
$k^{m}$. 

Furthermore, we note the wavevectors \textbf{$\mathbf{k}$ }are of
the form $2\pi\left(\lambda_{D}/L\right)\text{\ensuremath{\mathbf{l}}}$
with $\mathbf{l}\in\mathbb{Z}^{3}$. Then $\left(\frac{\lambda_{D}}{Lk}\right)^{3}$
is of order one at most. Thus, we can conclude that within the Landau
approximation ($k\gg1$ in our non-dimensional plasma variables) all
the tensors $\mathbf{B}^{(n)}$ are negligible except for $\mathbf{B}^{(1)}=\mathbf{B}$.
We have presented all the computation and this estimation in a finite
box of length $L$. However similar reasoning generalize easily to
an infinite box. \\

As a conclusion, at leading order, we can just keep the terms involving
$\mathbf{B}^{(1)}$ in the cumulant series expansion, and the large
deviations Hamiltonian for the Landau equation reads
\begin{eqnarray}
H_{\text{Landau}}\left[f,p\right] & = & \int\text{d}\mathbf{r}\text{d}\mathbf{v}_{1}f\left\{ \mathbf{b}\left[f\right].\frac{\partial p}{\partial\mathbf{v}_{1}}+\frac{\partial}{\partial\mathbf{v}_{1}}\left(\mathbf{D}\left[f\right]\frac{\partial p}{\partial\mathbf{v}_{1}}\right)+\mathbf{D}\left[f\right]:\frac{\partial p}{\partial\mathbf{v}_{1}}\frac{\partial p}{\partial\mathbf{v}_{1}}\right\} \nonumber \\
 & - & \int\text{d}\mathbf{r}\text{d\ensuremath{\mathbf{v}_{1}}}\text{d}\mathbf{v}_{2}f(\mathbf{v}_{1})f(\mathbf{v}_{2})\frac{\partial p}{\partial\mathbf{v}_{1}}\frac{\partial p}{\partial\mathbf{v}_{2}}:\mathbf{B}\left(\mathbf{v}_{1},\mathbf{v}_{2}\right).\label{eq:H_Landau_from_dynamics}
\end{eqnarray}
This is exactly the Hamiltonian we derived from the Boltzmann equation
large deviation Hamiltonian in section \ref{subsec:Deriving-Landau's-large}.

\subsection{Large deviations for the Landau equation\label{subsec:Large-deviations-for-the-landau-eq}}

In the previous section \ref{subsec:Hierarchization-of-the}, we
have established a large deviation principle for the homogeneous projection
of the empirical density of $N$ particles submitted to pairwise Coulomb
interactions in the Landau approximation. It describes dynamical fluctuations
beyond the Landau equation. More precisely, if we consider $N$ particles
evolving according to the dynamics \eqref{eq:coulomb_dynamics}, in
a $3$-dimensional torus of size $\left(L/\lambda_{D}\right)^{3}$
where $\lambda_{D}$ is the Debye length, $f_{\Lambda}$ the homogeneous
projection of the empirical density
\[
f_{\Lambda}\left(\mathbf{v},t\right)=\frac{1}{\Lambda}\left(\frac{\lambda_{D}}{L}\right)^{3}\sum_{n=1}^{N}\delta\left(\mathbf{v}-\mathbf{v}_{n}\left(t\right)\right),
\]
follows the large deviation principle
\begin{equation}
\mathbf{P}(f_{\Lambda}=f)\underset{\Lambda\rightarrow\infty}{\asymp}\text{e}^{-\Lambda\text{Sup}_{p}\int_{0}^{T}\left\{ \int\text{d}\mathbf{r}\text{d}\mathbf{v}\,\dot{f}p-H_{\text{Landau}}[f,p]\right\} },\label{eq:PGD_L>=00005Clambda_D}
\end{equation}
with the prescription that $f_{\Lambda}(\tau=0)$ is in the neighborhood of $f(\tau=0)$, and where the large deviation Hamiltonian $H_{\text{Landau}}$ is given
by \eqref{eq:H_Landau_from_dynamics}. 

Although this Hamiltonian is exactly the one we derived in section
\ref{sec:Large-dev-from-boltzmann} from the large deviation Hamiltonian
associated to the Boltzmann equation, the large deviation principle
\eqref{eq:PGD_L>=00005Clambda_D} is slightly different. Indeed, the
large deviation principle \eqref{eq:LDP_LAndau_from_Boltz} describes
large deviations of the empirical density $g_{\Lambda}$, whereas
the large deviation principle \eqref{eq:PGD_L>=00005Clambda_D} only
describes the large deviations for $f_{\Lambda}$ which is the projection
of $g_{\Lambda}$ over homogeneous distributions. However, it is possible
to obtain \eqref{eq:PGD_L>=00005Clambda_D} from \eqref{eq:LDP_LAndau_from_Boltz}
through the use of the contraction principle. In large deviation theory,
the contraction principle states that if we know a large deviation
principle for a random variable $X$ with a large deviation function
$I\left(x\right)$ it is possible to obtain a large deviation principle
for any function $\varphi\left(X\right)$ of this random variable
and the associated large deviation function is $I_{\varphi}\left(y\right)=\inf_{\varphi\left(x\right)=y}I\left(x\right)$.
The two results are thus fully consistent. 

Based on the discussion of section \ref{subsec:Verifications-of-the},
the large deviation Hamiltonian $H_{\text{Landau}}$ satisfies all
the expected properties of the large deviation Hamiltonian for the
Landau equation: mass, momentum and energy conservation, as well as
entropy as the opposite of the quasipotential and time-reversal symmetry.

\subsection{Large deviations for the Landau equation when $L<\lambda_{D}$\label{subsec:Large-deviations-forL<Lambda}}

Whenever the size of the domain is smaller than the Debye length,
the relevant large deviation parameter is the number of particles
in a box of the size of the effective interaction length scale $\ell=L$;
i.e. the relevant large deviation parameter is $N$. We can then study
the asymptotics of the empirical density $g_{\Lambda}$ and its homogeneous
projection as $N$ goes to infinity. Because $\Lambda=\left(\lambda_{D}/L\right)^{3}N$,
when $L<\lambda_{D}$ the large $N$ limit implies the large $\Lambda$
limit, which is responsible for the kinetic behavior of the empirical
density. In order to make explicit that $N$ is the natural large
deviation rate, we perform the trivial integral on the positions in
the large deviation principle (\ref{eq:PGD_L>=00005Clambda_D}). It
is then possible to rephrase the large deviation principle \eqref{eq:PGD_L>=00005Clambda_D}
as following
\begin{equation}
\mathbf{P}(f_{\Lambda}=f)\underset{N\rightarrow\infty}{\asymp}\text{e}^{-N\text{Sup}_{p}\int_{0}^{T}\left\{ \int\text{d}\mathbf{v}\,\dot{f}p-H_{\text{Landau},h}[f,p]\right\} },\label{eq:PGD_L<=00005Clambda_D}
\end{equation}
with the prescription that $f_{\Lambda}(\tau=0)$ is in the neighborhood of $f(\tau=0)$, and by defining $H_{\text{Landau},h}$ as the large deviation Hamiltonian
divided by the volume of the domain, such that
\[
H_{\text{Landau}}=\int\text{d}\mathbf{r}H_{\text{Landau},h}=\left(\frac{L}{\lambda_{D}}\right)^{3}H_{\text{Landau},h},
\]
and
\begin{multline*}
H_{\text{Landau},h}\left[f,p\right]=\int\text{d}\mathbf{v}_{1}f\left\{ \mathbf{b}\left[f\right].\frac{\partial p}{\partial\mathbf{v}_{1}}+\frac{\partial}{\partial\mathbf{v}_{1}}.\left(\mathbf{D}\left[f\right]\frac{\partial p}{\partial\mathbf{v}_{1}}\right)+\mathbf{D}\left[f\right]:\frac{\partial p}{\partial\mathbf{v}_{1}}\frac{\partial p}{\partial\mathbf{v}_{1}}\right\} \\
-\int\text{d\ensuremath{\mathbf{v}_{1}}}\text{d}\mathbf{v}_{2}f(\mathbf{v}_{1})f(\mathbf{v}_{2})\frac{\partial p}{\partial\mathbf{v}_{1}}\frac{\partial p}{\partial\mathbf{v}_{2}}:\mathbf{B}\left(\mathbf{v}_{1},\mathbf{v}_{2}\right).
\end{multline*}
Using this same relation between $N$ and $\Lambda$, we already have
remarked that
\[
f_{\Lambda}\left(\mathbf{v},t\right)=\frac{1}{\Lambda}\left(\frac{\lambda_{D}}{L}\right)^{3}\sum_{n=1}^{N}\delta\left(\mathbf{v}-\mathbf{v}_{n}\left(t\right)\right)=\frac{1}{N}\sum_{n=1}^{N}\delta\left(\mathbf{v}-\mathbf{v}_{n}\left(t\right)\right)=h_{N}\left(\mathbf{v},t\right),
\]
where $h_{N}$ is the velocity empirical density rescaled by the number
of particles defined in section \ref{subsec:The-Balescu=002013Guernsey=002013Lenard-and}.
Then, we have the following large deviation principle for $h_{N}$
\[
\mathbf{P}(h_{N}=f)\underset{N\rightarrow\infty}{\asymp}\text{e}^{-N\text{Sup}_{p}\int_{0}^{T}\left\{ \int\text{d}\mathbf{v}\,\dot{f}p-H_{\text{Landau},h}[f,p]\right\} },
\]
with the prescription that $h_N(\tau=0)$ is in the neighborhood of $f(\tau=0)$. It is very similar to the large deviation principle \eqref{eq:PGD_mean_field_indep_diff}
we established for the velocities empirical distribution of $N$ diffusing
particles coupled in a mean field way, except that the large deviation
Hamiltonian $H_{\text{Landau},h}$ contains an additional term in
addition to $H_{MF,h}$ \eqref{eq:Mean_field_Hamiltonian-1}, accounting
for the weak interactions between the particles.

If in addition to $L<\lambda_{D}$ we have $L\ll\lambda_{D}$, then,
because the wavevectors $\mathbf{k}$ are in $2\pi\left(\lambda_{D}/L\right)\mathbb{Z}^{3}$
we have for all scales $k\gg1$. This amounts at saying that the Landau
approximation holds at all scales and that the large deviations described
by \eqref{eq:PGD_L<=00005Clambda_D} are Gaussian regardless of the
scale of the fluctuations.

\subsection{Large deviations for the Landau equation expressed in physical variables\label{subsec:phys-var}}

In section \ref{subsec:Deriving-Landau's-large}, we established a
large deviation principle (equations \eqref{eq:ham_from_boltz}-\eqref{eq:LDP_LAndau_from_Boltz})
that describes the large deviations of the probability of homogeneous
evolution paths for the empirical density $g_{\Lambda}\left(\mathbf{r},\mathbf{v},t\right)=\Lambda^{-1}\sum_{n=1}^{N}\delta(\mathbf{v}-\mathbf{v}_{n}(t))\delta\left(\mathbf{r}-\mathbf{r}_{n}(t)\right)$.
As discussed in section \ref{subsec:Large-deviations-for-the-landau-eq},
this result is consistent with the large deviation principle for the
projection of the empirical density on homogeneous paths $f_{\Lambda}\left(\mathbf{v},t\right)=\Lambda^{-1}\left(\lambda_{D}/L\right)^{3}\sum_{n=1}^{N}\delta\left(\mathbf{v}-\mathbf{v}_{n}\left(t\right)\right).$
So far, we expressed those results in a set of non-dimensional variables
adapted to Coulomb plasmas.

We can express this large deviation result in physical variables,
with the change of variables
\[
\mathbf{v}_{\varphi}=v_{T}\mathbf{v},\mathbf{k}_{\varphi}=\mathbf{k}/\lambda_{D},t_{\varphi}=\Lambda\tau/\omega_{pe},
\]
where $v_{T}$ the thermal velocity, $\lambda_{D}$ the Debye length,
and $\omega_{pe}$ the plasma electron frequency are defined in section
\ref{subsec:plasma}., and we denoted dimensional variables expressed
in physical units with a subscript $\varphi$.

In the following we omit the subscript $\varphi$. The result is a
large deviation principle for the empirical density in physical units
\[
g_{\Lambda}\left(\mathbf{r},\mathbf{v},t\right)=\frac{1}{\Lambda}\sum_{n=1}^{N}\delta(\mathbf{v}-\mathbf{v}_{n}(t))\delta\left(\mathbf{r}-\mathbf{r}_{n}(t)\right)
\]
which reads
\[
\mathbf{P}\left(\left\{ g_{\Lambda}\right\} _{0\leq t\leq T}=\left\{ f\right\} _{0\leq t\leq T}\right)\underset{\Lambda\rightarrow\infty}{\asymp}\text{e}^{-\Lambda\text{Sup}_{p}\int_{0}^{T}\text{d}t\left\{ \int\text{d}\mathbf{r}\text{d}\mathbf{v}\dot{f}p-H_{\text{Landau}}\left[f,p\right]\right\} },
\]
with the prescription that $g_{\Lambda}(t=0)$ is in the neighborhood of $f(t=0)$,
and where
\begin{multline*}
H_{\text{Landau}}\left[f,p\right]=\int\text{d}\mathbf{r}\text{d}\mathbf{v}_{1}f\left\{ \mathbf{b}\left[f\right].\frac{\partial p}{\partial\mathbf{v}_{1}}+\frac{\partial}{\partial\mathbf{v}_{1}}.\left(\mathbf{D}\left[f\right]\frac{\partial p}{\partial\mathbf{v}_{1}}\right)+\mathbf{D}\left[f\right]:\frac{\partial p}{\partial\mathbf{v}_{1}}\frac{\partial p}{\partial\mathbf{v}_{1}}\right\} \\
-\int\text{d}\mathbf{r}\text{d\ensuremath{\mathbf{v}_{1}}}\text{d}\mathbf{v}_{2}f(\mathbf{v}_{1})f(\mathbf{v}_{2})\frac{\partial p}{\partial\mathbf{v}_{1}}\frac{\partial p}{\partial\mathbf{v}_{2}}:\mathbf{B}\left(\mathbf{v}_{1},\mathbf{v}_{2}\right).
\end{multline*}
with
\begin{equation}
\begin{cases}
\mathbf{b}\left[f\right](\mathbf{v}) & =\int\text{d}\mathbf{v}_{2}\mathbf{B}(\mathbf{v},\mathbf{v}_{2})\frac{\partial f}{\partial\mathbf{v}_{2}}\\
\mathbf{D}\left[f\right](\mathbf{v}) & =\int\text{d}\mathbf{v}_{2}\mathbf{B}(\mathbf{v},\mathbf{v}_{2})f(\mathbf{v}_{2}),
\end{cases}\label{eq:b_D_phys}
\end{equation}
and
\begin{equation}
\mathbf{B}\left(\mathbf{v}_{1},\mathbf{v}_{2}\right)=\frac{\Lambda q^{4}}{m^{2}\epsilon_{0}^{2}}\frac{\pi}{L^{3}}\sum_{\mathbf{k}\in\left(2\pi/L\right)\mathbb{Z^{*}}^{3}}\left(\hat{W}\left(\mathbf{k}\right)\right)^{2}\mathbf{k}\mathbf{k}\delta\left(\mathbf{k}.\mathbf{v}_{2}-\mathbf{k}.\mathbf{v}_{1}\right).\label{eq:B_phys}
\end{equation}
And the associated Landau equation reads
\begin{equation}
\frac{\partial f}{\partial t}=\frac{\partial}{\partial\mathbf{v}}.\int\text{d}\mathbf{v}_{2}\,\mathbf{B}\left(\mathbf{v}_{1},\mathbf{v}_{2}\right)\left(-\frac{\partial f}{\partial\mathbf{v}_{2}}f(\mathbf{v})+f(\mathbf{v}_{2})\frac{\partial f}{\partial\mathbf{v}}\right).\label{eq:Landau_phys_presque}
\end{equation}
This differs slightly with the Landau equation one can found in the
plasma literature \cite{schram2012kinetic,Lifshitz_Pitaevskii_1981_Physical_Kinetics,Nicholson_1991}
by a factor $\Lambda$ in the tensor $\mathbf{B}$ (\ref{eq:B_phys}).
Typically, in those references, the Landau equation is an evolution
equation for the average of the non-rescaled empirical density. Here,
we rescaled the empirical density by the plasma parameter $\Lambda$.
In order to recover the Landau equation of \cite{schram2012kinetic,Lifshitz_Pitaevskii_1981_Physical_Kinetics,Nicholson_1991},
one should replace $f$ in equation (\ref{eq:Landau_phys_presque})
by $f_{0}/\Lambda$. The resulting evolution equation for $f_{0}$
would be the usual Landau equation, where $f_{0}=\mathbb{E}\left(\Lambda g_{\Lambda}\right)$
is the distribution function typically used in plasma textbooks.

\section*{Conclusions}

The main result of this paper is the large deviation principle for
the dynamics of the empirical density of a homogeneous Coulomb plasma
of $N$ equal charges particles. More precisely, we have shown that
the probability of a homogeneous evolution path $\left\{ f(\tau)\right\} _{0\leq \tau\leq T}$
for the empirical density $g_{\Lambda}\left(\mathbf{r},\mathbf{v},\tau\right)=\Lambda^{-1}\sum_{n=1}^{N}\delta(\mathbf{v}-\mathbf{v}_{n}(\tau))\delta\left(\mathbf{r}-\mathbf{r}_{n}(\tau)\right)$
follows a large deviation principle
\[
\mathbf{P}\left(\left\{ g_{\Lambda}(\tau)\right\} _{0\leq \tau\leq T}=\left\{ f(\tau)\right\} _{0\leq \tau\leq T}\right)\underset{\Lambda\rightarrow\infty}{\asymp}\text{e}^{-\Lambda\int_{0}^{T}\text{d}\tau\,\text{Sup}_{p}\left\{ \int\text{d}\mathbf{r}\text{d}\mathbf{v}\,\dot{f}p-H_{\text{Landau}}[f,p]\right\} },
\]
with the prescription that $g_{\Lambda}(\tau=0)$ is in the neighborhood of $f(\tau=0)$, and 
where the large deviation Hamiltonian $H_{\text{Landau}}[f,p]$ is
\[
H_{\text{Landau}}[f,p]=H_{MF}\left[f,p\right]+H_{I}\left[f,p\right],
\]
with 
\[
H_{MF}\left[f,p\right]=\int\text{d}\mathbf{r}\text{d}\mathbf{v}f\left\{ \mathbf{b}\left[f\right].\frac{\partial p}{\partial\mathbf{v}}+\frac{\partial}{\partial\mathbf{v}}.\left(\mathbf{D}\left[f\right].\frac{\partial p}{\partial\mathbf{v}}\right)+\mathbf{D}\left[f\right]:\frac{\partial p}{\partial\mathbf{v}}\frac{\partial p}{\partial\mathbf{v}}\right\} ,
\]
and 
\[
H_{I}\left[f,p\right]=-\int\text{d}\mathbf{r}\text{d\ensuremath{\mathbf{v}_{1}}}\text{d}\mathbf{v}_{2}f(\mathbf{v}_{1})f(\mathbf{v}_{2})\frac{\partial p}{\partial\mathbf{v}_{1}}\frac{\partial p}{\partial\mathbf{v}_{2}}:\mathbf{B}\left(\mathbf{v}_{1},\mathbf{v}_{2}\right).
\]
where $\mathbf{D}\left[f\right]$, $\mathbf{b}\left[f\right]$ are
defined in equation (\ref{eq:b_D_phys}), and $\mathbf{B}$ is defined
in equation (\ref{eq:B_phys}). This result has been obtained both
from the large deviation Hamiltonian associated with the Boltzmann
equation, and directly from the dynamics. This result is expressed
in physical variables, but throughout the paper we worked with a non-dimensional
set of variables adapted to plasmas. The connection is made between
these two sets of variable in section \ref{subsec:phys-var}.

This result is valid only for fluctuations at wavenumbers $k$ such
that $k\lambda_{D}\gg1$ in physical units. This large deviation Hamiltonian
is quadratic in its conjugate momentum meaning that large deviations
are Gaussian. It also satisfies all the expected properties: conservation
laws, time-reversal symmetry and consistency with equilibrium thermodynamics.\\

This paper also contains a set of complementary results. It contains
the expression for the Hamiltonian for the path large deviations of
the empirical density of $N$ independent Markov processes \eqref{eq:Hamiltonien_N_Markov_Processes}
, of $N$ independent diffusions \eqref{eq:N_Diff_Hamiltonian}, and
of $N$ diffusions coupled in a mean field way \eqref{eq:Mean_field_Hamiltonian-1}.
It also contains an explicit gradient flow structure for the Landau
equation \eqref{eq:Gradient-flow-landau}, deduced from the large
deviation Hamiltonian. We also obtained results for the empirical
density of $N$ particles with long-range interactions without the
Landau approximation. In this general case, we established a cumulant
generating function representation of the large deviation Hamiltonian
for the empirical density \eqref{eq:H_Gartner_ellis}. We computed
a cumulant expansion of this cumulant generating function up to order
four.\\

Our results are exact computations, once natural hypothesis are made.
The first main hypothesis is the validity of the quasilinear approximation.
The second one is convergence of the Gaussian process of fluctuation
to a stationary process. The third one is the validity of the classical
expression for the large deviation Hamiltonian, in this context. The
quasilinear approximation is very natural and is obtained naturally
as the leading order contribution in a series expansion. The second
hypothesis is partly justified in classical textbooks, although a rigorous
proof is missing. Actually, from a mathematical point of view, the
type of convergence to consider is not clear. About the third one,
we note that classical theorems for large deviations for slow-fast
systems use sufficient ergodicity hypothesis which are probably wrong
for this problem. A mathematical proof would thus require interesting
mathematical developments. Actually the second and third hypothesis
are strongly connected. In order to obtain a theorem, these three
hypothesis should be proven. As far as we understand such a task seems
out of reach of the best mathematicians, currently. However it might
be achievable in the future, which would be a fascinating perspective. 

A natural extension of this work would be to compute the large deviation
Hamiltonian associated with the Balescu--Guernsey--Lenard
equation. This would be a large deviation principle for the empirical
density of $N$ particles which interact through long-range interactions,
for instance through Coulomb interactions, but without the Landau
approximation. This will be the subject of an upcoming paper. 

In this paper, we obtained results quantifying the dynamical fluctuations
of the empirical density of a Coulomb plasma in the large plasma parameter
limit. A series of mathematical papers \cite{serfaty2017microscopic,serfaty2020gaussian,padillagarza2020large,leble2017large,leble2018fluctuations}
focus on fluctuations of stationary observables for Coulomb gases
without using the large plasma parameter limit. This raises the question
of whether it would be possible to obtain results about the dynamical
fluctuations of the empirical density without the hypothesis of a
large plasma parameter limit. This is an interesting perspective that
would extend our present work, and at the same time would extend the
static picture discussed in \cite{serfaty2017microscopic,serfaty2020gaussian,padillagarza2020large,leble2017large,leble2018fluctuations}.

Another perspective and extension would be to obtain a large deviation
principle and Hamiltonian for the evolution of inhomogeneous distribution
functions. This would be particularly relevant for systems that are
naturally inhomogeneous, for instance self-gravitating systems with
application in galactic and globular cluster dynamics. 

Finally, a large part of the computations and reasonings of this paper
can be formulated beyond the framework of Coulomb plasmas. One of
the first generalization we think about is to investigate the large
deviations for the empirical density of particles which interact through
long range potential and are stochastically forced out-of-equilibrium.
Beyond interacting particles system, it will be interesting to use
this tool to investigate two-dimensional and geostrophic turbulence.
The dynamics of those hydrodynamical systems has deep analogies with
systems with long-range interactions.

\appendix

\section{The relative entropy for $N$ independent diffusions solves the stationary
Hamilton--Jacobi equation\label{sec:Quasipotential-and-relative}}

We consider the relative entropy
\[
S_{\text{rel}}\left[h\right]=-\int\text{d}\mathbf{v}\,h\log\left(\frac{h}{h_{\text{eq}}}\right),
\]
where $h_{\text{eq}}$ is the equilibrium distribution. In this appendix,
we shows that $-S_{\text{rel}}$ solves the stationary Hamilton-Jacobi
equation ($H_{MF,h}\left[h,-\frac{\delta S_{\text{rel}}}{\delta h}\right]=0$),
for the case of $N$ independent diffusions \eqref{eq:Diffusion_Particules_Independantes}.
We recall that $H_{MF,h}\left[h,-\frac{\delta S_{\text{rel}}}{\delta h}\right]=0$
is a necessary condition for $-S_{\text{rel}}$ to be the quasipotential.
By contrast, when those $N$ diffusions are coupled in a mean field
way (in \eqref{eq:N_Diffusions-1}) and the drift and diffusion coefficients
depend actually on $h$, we are no more able to conclude that $H_{MF,h}\left[h,-\frac{\delta S_{\text{rel}}}{\delta h}\right]=0$
and we believe this is actually wrong in general. 

In both cases, the large deviation Hamiltonian for the empirical density
$h_{N}$ reads
\begin{equation}
H_{MF,h}\left[h,p\right]=\int\text{d}\mathbf{v}h\left\{ \mathbf{b}\left[h\right].\frac{\partial p}{\partial\mathbf{v}}+\frac{\partial}{\partial\mathbf{v}}\left(\mathbf{D}\left[h\right]\frac{\partial p}{\partial\mathbf{v}}\right)+\mathbf{D}\left[h\right]:\frac{\partial p}{\partial\mathbf{v}}\frac{\partial p}{\partial\mathbf{v}}\right\} .\label{eq:H_h}
\end{equation}
In the simple case where the $N$ diffusions are independent, the
drift and the diffusion coefficients do not depend on the actual distribution
$h$: $\mathbf{b}\left[h\right]=\mathbf{b}$ and $\mathbf{D}\left[h\right]=\mathbf{D}$.
In order to check that the relative entropy $S_{\text{rel}}$ is the
opposite of the quasipotential, according to property 11 from section
\ref{subsec:LD_Kinetic_theories}, we shall check that it solves the
stationary Hamilton--Jacobi equation
\begin{equation}
H_{MF,h}\left[h,-\frac{\delta S_{\text{rel}}}{\delta h}\right]=0.\label{eq:HJ_S_Relative}
\end{equation}

We have
\begin{equation}
-\frac{\partial}{\partial\mathbf{v}}\left(\frac{\delta S_{\text{rel}}}{\delta h}\right)=\frac{1}{h}\frac{\partial h}{\partial\mathbf{v}}-\frac{1}{h_{\text{eq}}}\frac{\partial h_{\text{eq}}}{\partial\mathbf{v}},\label{eq:partial_derivative_relative_S}
\end{equation}
and $h_{\text{eq}}$ solves the stationary Fokker--Planck
equation 
\begin{equation}
\frac{\partial}{\partial\mathbf{v}}\left(\mathbf{D}\left[h_{\text{eq}}\right]\frac{\partial h_{\text{eq}}}{\partial\mathbf{v}}-\mathbf{b}\left[h_{\text{eq}}\right]h_{\text{eq}}\right)=0.\label{eq:Stationary-FP}
\end{equation}
Using \eqref{eq:partial_derivative_relative_S} we have
\[
H_{MF,h}\left[h,-\frac{\delta S_{\text{rel}}}{\delta h}\right]=\int\text{d}\mathbf{v}\,\left\{ \mathbf{b}\left[h\right]\frac{\partial h}{\partial\mathbf{v}}-\frac{h}{h_{\text{eq}}}\frac{\partial h_{\text{eq}}}{\partial\mathbf{v}}\mathbf{b}\left[h\right]+\mathbf{D}\left[h\right]\frac{\partial h_{\text{eq}}}{\partial\mathbf{v}}\frac{1}{h_{\text{eq}^{2}}}\left(\frac{\partial h}{\partial\mathbf{v}}h_{\text{eq}}-\frac{\partial h_{\text{eq}}}{\partial\mathbf{v}}h\right)\right\} .
\]
Now, we integrate by parts the first and the last term of the expression
above, noting that
\[
\frac{1}{h_{\text{eq}^{2}}}\left(\frac{\partial h}{\partial\mathbf{v}}h_{\text{eq}}-\frac{\partial h_{\text{eq}}}{\partial\mathbf{v}}h\right)=\frac{\partial}{\partial\mathbf{v}}\left(\frac{h}{h_{\text{eq}}}\right),
\]
and
\[
-h\frac{\partial\mathbf{b}\left[h\right]}{\partial\mathbf{v}}-\frac{h}{h_{\text{eq}}}\frac{\partial h_{\text{eq}}}{\partial\mathbf{v}}\mathbf{b}\left[h\right]=-\frac{h}{h_{\text{eq}}}\frac{\partial}{\partial\mathbf{v}}\left(\mathbf{b}\left[h\right]h_{\text{eq}}\right).
\]
We obtain
\[
H_{MF,h}\left[h,-\frac{\delta S_{\text{rel}}}{\delta h}\right]=\int\text{d}\mathbf{v}\,\frac{h}{h_{\text{eq}}}\frac{\partial}{\partial\mathbf{v}}\left(\mathbf{D}\left[h\right]\frac{\partial h_{\text{eq}}}{\partial\mathbf{v}}-\mathbf{b}\left[h\right]h_{\text{eq}}\right).
\]
We see that if for any $h$
\begin{equation}
\frac{\partial}{\partial\mathbf{v}}\left(\mathbf{D}\left[h\right]\frac{\partial h_{\text{eq}}}{\partial\mathbf{v}}-\mathbf{b}\left[h\right]h_{\text{eq}}\right)=0,\label{eq:Identite}
\end{equation}
then $H_{MF,h}\left[h,-\frac{\delta S_{\text{rel}}}{\delta h}\right]=0$
for any $h$. When $\mathbf{b}\left[h\right]=\mathbf{b}$ and $\mathbf{D}\left[h\right]=\mathbf{D}$
do not depend of $f$, i.e. when the $N$ diffusions are independent,
this identity is equivalent to the stationary Fokker--Planck
equation \eqref{eq:Stationary-FP}. It thus holds. It follows that
the Hamilton--Jacobi equation \eqref{eq:HJ_S_Relative}
is verified and that the negative of the relative entropy solves the
stationary Hamilton--Jacobi equation, for the case of $N$
independent diffusions.

However, when the drift and the diffusion coefficient do depend on
the distribution, (\ref{eq:Identite}) is no more true for any $h$.
Then, we cannot conclude anymore that the relative entropy solves
the stationary Hamilton--Jacobi equation.

\section{Consistence of the two definitions of the tensor $\mathbf{B}$\label{sec:Lien_tenseur_B_boltz_LB}}

We prove that for Coulomb interaction the two expressions for $\mathbf{B}$,
\eqref{eq:tensor_B} and \eqref{eq:B_Landau} are equal.\textbf{ }

The first expression for $\mathbf{B}$, \eqref{eq:tensor_B}, is 

\[
\mathbf{B}(\mathbf{v}_{1},\mathbf{v}_{2})=\frac{1}{2}\Lambda\int\text{d}\mathbf{q}\,w(\mathbf{v}_{1},\mathbf{v}_{2};\mathbf{q})\mathbf{q}\otimes\mathbf{\mathbf{q}},
\]
Expressing $w$ in terms of the cross-section $\sigma_{0}$ through
\eqref{eq:Nouvelles_Notations} with $\gamma=\left(\lambda_{D}/L\right)^{3}$,
using \eqref{eq:cross_section-1}, and choosing for $\sigma_{0}$
the Rutherford diffusion cross-section 
\[
\sigma_{0}(\mathbf{v}_{1}+\mathbf{q},\mathbf{v}_{2}-\mathbf{q};\mathbf{v}_{1},\mathbf{v}_{2})=\frac{1}{4\pi^{2}\Lambda^{2}q^{4}},
\]
for two-body collisions of particles with electrostatic interactions
\cite{schram2012kinetic}, we obtain

\begin{equation}
\mathbf{B}(\mathbf{v}_{1},\mathbf{v}_{2})=\int\text{d}\mathbf{q}\,\frac{\mathbf{q\otimes q}}{8\pi^{2}q^{4}}\delta\left(2\mathbf{q}.\left(\mathbf{v}_{2}-\mathbf{v}_{1}\right)\right).\label{eq:link_B_intermediaire}
\end{equation}
We perform the integration over $\mathbf{q}$ angle in \eqref{eq:link_B_intermediaire}
to get 
\[
\mathbf{B}(\mathbf{v}_{1},\mathbf{v}_{2})=C\frac{g^{2}\text{Id}-\mathbf{g}\mathbf{g}}{g^{3}},
\]
with $C=\left(8\pi\right)^{-1}\int_{0}^{\infty}q^{-1}\text{d}q$,
$\mathbf{g}=\mathbf{v}_{2}-\mathbf{v}_{1}$, and where $\text{Id}$
is the identity matrix in three-dimension. We note that $\mathbf{B}(\mathbf{v}_{1},\mathbf{v}_{2})$
is proportional to $g^{2}\text{Id}-\mathbf{g}\otimes\mathbf{g}$,
which is the projector on the plane orthogonal to $\mathbf{v}_{2}-\mathbf{v}_{1}$.
This should have been expected as a consequence of symmetries. 

In order to obtain the proportionality coefficient $C$ we follow
equations (6.3.15-6.3.21) in chapter 6.3 of Schram's textbook \cite{schram2012kinetic}.
This chapter explains how one can deal with the logarithmic divergence
arising in the computation of $C$. Briefly, one has to regularize
the Coulomb interaction at large and small scales by introducing cut-offs,
justified by the geometry of grazing collision at small scales, and
by the Debye shielding at large scales. The final result reads
\begin{equation}
\mathbf{B}(\mathbf{v}_{1},\mathbf{v}_{2})=\frac{1}{8\pi}\ln\Lambda\frac{g^{2}\text{Id}-\mathbf{g}\mathbf{g}}{g^{3}}.\label{eq:B_q}
\end{equation}

Following the computations in chapter 8.4 of Schram's textbook \cite{schram2012kinetic},
we can show in a similar way that the definition of $\mathbf{B}$
given by \eqref{eq:B_Landau} is also equal to \eqref{eq:B_q}. We
have thus conclude that\textbf{ }the two expression for $\mathbf{B}$,
\eqref{eq:tensor_B} and \eqref{eq:B_Landau} are equal.

\section{Symmetries and conservation laws associated with the collision kernels\label{sec:Symmetries-and-conservation}}

\subsection{The Boltzmann collision kernel}

The time reversal symmetry of the microscopic Hamiltonian dynamics
imposes that 
\begin{equation}
w_{0}(\mathbf{v}'_{1},\mathbf{v}'_{2};\mathbf{v}_{1},\mathbf{v}_{2})=w_{0}(-\mathbf{v}{}_{1},-\mathbf{v}{}_{2};-\mathbf{v}'_{1},-\mathbf{v}'_{2}).\label{eq:time_reversibility}
\end{equation}
The space rotation symmetry imposes that for any rotation $\mathbf{R}$
that belongs to the orthogonal group $SO(3)$
\[
w_{0}(\mathbf{v}'_{1},\mathbf{v}'_{2};\mathbf{v}_{1},\mathbf{v}_{2})=w_{0}(\mathbf{R}\mathbf{v}{}_{1},\mathbf{R}\mathbf{v}{}_{2};\mathbf{R}\mathbf{v}'_{1},\mathbf{R}\mathbf{v}'_{2}).
\]
The combination of the time reversal symmetry and of the space rotation
symmetry for $\mathbf{R}=-\mathbf{I}$, where $\mathbf{I}$ is the
identity operator, implies the inversion symmetry 
\begin{equation}
w_{0}(\mathbf{v}'_{1},\mathbf{v}'_{2};\mathbf{v}_{1},\mathbf{v}_{2})=w_{0}(\mathbf{v}{}_{1},\mathbf{v}{}_{2};\mathbf{v}'_{1},\mathbf{v}'_{2}).\label{eq:inversion}
\end{equation}
The local conservation of momentum and energy implies that
\begin{equation}
w_{0}(\mathbf{v}'_{1},\mathbf{v}'_{2};\mathbf{v}_{1},\mathbf{v}_{2})=\sigma(\mathbf{v}'_{1},\mathbf{v}'_{2};\mathbf{v}_{1},\mathbf{v}_{2})\delta\left(\mathbf{v}_{1}+\mathbf{v}_{2}-\mathbf{v}'_{1}-\mathbf{v}'_{2}\right)\delta\left(\mathbf{v}_{1}^{2}+\mathbf{v}_{2}^{2}-\mathbf{v'}_{1}^{2}-\mathbf{v'}{}_{2}^{2}\right),\label{eq:cross_section}
\end{equation}
where $\sigma$ is the diffusion cross-section. $\sigma$ is of the
order of $a^{2}$ where $a$ is a typical atom size.

\subsection{The Landau collision kernel\label{subsec:The-Landau-collision}}

The tensor $\mathbf{B}$ defined by 
\begin{equation}
\mathbf{B}(\mathbf{v}_{1},\mathbf{v}_{2})=\frac{\Lambda}{2}\int\text{d}\mathbf{q}\,w(\mathbf{v}_{1},\mathbf{v}_{2};\mathbf{q})\mathbf{q}\otimes\mathbf{\mathbf{q}},\label{annex_B}
\end{equation}
involved in the Landau equation (\ref{eq:Landau}) has properties
related to the symmetry and conservation properties of the collision
process. In equation (\ref{annex_B}), $w(\mathbf{v}_{1},\mathbf{v}_{2};\mathbf{q})$
is an approximation at order zero of the collision kernel $w(\mathbf{v}_{1}+\mathbf{q}/2,\mathbf{v}_{2}-\mathbf{q}/2;\mathbf{q})$
associated with the collision of two particles with momenta $\left(\mathbf{v}_{1},\mathbf{v}_{2}\right)$
that exchange a momentum $\mathbf{q}$. We have:
\begin{enumerate}
\item $w(\mathbf{v}_{1},\mathbf{v}_{2};\mathbf{q})=w(\mathbf{v}_{2},\mathbf{v}_{1};\mathbf{q})$
because the incident particles are indiscernible,
\item $\mathbf{q}.\left(\mathbf{v}_{1}-\mathbf{v}_{2}\right)=0$ at leading
order in $\mathbf{q}$ because of the energy conservation condition
$\mathbf{v}_{1}^{2}+\mathbf{v}_{2}^{2}=\left(\mathbf{v}_{1}+\mathbf{q}\right)^{2}+\left(\mathbf{v}_{2}-\mathbf{q}\right)^{2}$,
\item $w(\mathbf{v}_{1},\mathbf{v}_{2};\mathbf{q})=w(\mathbf{v}_{1},\mathbf{v}_{2};-\mathbf{q})$,
which is a direct consequence of \eqref{eq:inversion} and the definition
of $w$ \eqref{eq:Nouvelles_Notations}. 
\end{enumerate}
We notice that the momentum conservation is already built-in in the
definition of $w$. The first property implies $\mathbf{B}(\mathbf{v}_{1},\mathbf{v}_{2})=\mathbf{B}(\mathbf{v}_{2},\mathbf{v}_{1})$.
The second property implies $\mathbf{B}(\mathbf{v}_{1},\mathbf{v}_{2}).\left(\mathbf{v}_{1}-\mathbf{v}_{2}\right)=0$.
In addition to that, $\mathbf{B}(\mathbf{v}_{1},\mathbf{v}_{2})$
is by construction a symmetric tensor for every pair $(\mathbf{v}_{1},\mathbf{v}_{2})$.

\section{Asymptotic expansions leading to the Landau equation and its large
deviation Hamiltonian}

\subsection{Asymptotic expansions leading to the Landau equation\label{subsec:Asymptotic-expansions-Landau}}

In this appendix, we start from the collision operator of the Boltzmann
equation (the right hand side of equation (\ref{eq:definition_I})),
we develop it at order 2 in $\mathbf{q}$ , and we prove that we recover
the collision term of the Landau equation (\ref{eq:Landau}).

We start from the expression of $I$ in equation (\ref{eq:definition_I}).
Noting that $\left[f(\mathbf{v}+\mathbf{q})f(\mathbf{v}_{2}-\mathbf{q})-f(\mathbf{v})f(\mathbf{v}_{2})\right]$
has no term of order zero, in order to compute an expansion at order
2 in $q=\left|\mathbf{q}\right|$, it will be sufficient to work with
the expansions:
\[
\begin{cases}
w(\mathbf{v}+\frac{1}{2}\mathbf{q},\mathbf{v}_{2}-\frac{1}{2}\mathbf{q};\mathbf{q}) & =w(\mathbf{v},\mathbf{v}_{2};\mathbf{q})+\frac{1}{2}\left(\frac{\partial w}{\partial\mathbf{v}}-\frac{\partial w}{\partial\mathbf{v}_{2}}\right).\mathbf{q}+\mathcal{O}(q^{2}),\,\,\,\text{and}\\
f(\mathbf{v}+\mathbf{q})f(\mathbf{v}_{2}-\mathbf{q})-f(\mathbf{v})f(\mathbf{v}_{2}) & =\left(\frac{\partial f}{\partial\mathbf{v}}f(\mathbf{v}_{2})-\frac{\partial f}{\partial\mathbf{v}_{2}}f(\mathbf{v})\right).\mathbf{q}+\\
& + \left(\frac{\partial^{2}f}{\partial\mathbf{v}\partial\mathbf{v}}f(\mathbf{v}_{2})+\frac{\partial^{2}f}{\partial\mathbf{v}_{2}\partial\mathbf{v}_{2}}f(\mathbf{v})-2\frac{\partial f}{\partial\mathbf{v}}\frac{\partial f}{\partial\mathbf{v}_{2}}\right):\mathbf{q}\mathbf{q}+\mathcal{O}(q^{3}).
\end{cases}
\]
Let us now compute the collision integral $I(\mathbf{v})$ order by
order. We directly notice that there is no term of order zero in $\mathbf{q}$.
Let us compute $I^{(1)}(\mathbf{v})$ the term of order 1 of the collision
integral
\[
I^{(1)}(\mathbf{v})=\Lambda\int\text{d}\mathbf{v}_{2}\text{d}\mathbf{q}\,w(\mathbf{v},\mathbf{v}_{2};\mathbf{q})\left(\frac{\partial f}{\partial\mathbf{v}}f(\mathbf{v}_{2})-\frac{\partial f}{\partial\mathbf{v}_{2}}f(\mathbf{v})\right).\mathbf{q}.
\]
We use that $w(\mathbf{v},\mathbf{v}_{2};\mathbf{q})$ is an even
function of $\mathbf{q}$ (point 3 of appendix \eqref{subsec:The-Landau-collision}).
This makes the integrand an odd function of $\mathbf{q}$, and implies
that $I^{(1)}(\mathbf{v})=0$.

At order 2 in $\mathbf{q}$ we have
\footnotesize
\[
I(\mathbf{v})=\frac{\Lambda}{2}\int\text{d}\mathbf{v}_{2}\text{d}\mathbf{q}\left\{ \left(\frac{\partial w}{\partial\mathbf{v}}-\frac{\partial w}{\partial\mathbf{v}_{2}}\right)\left(\frac{\partial f}{\partial\mathbf{v}}f(\mathbf{v}_{2})-\frac{\partial f}{\partial\mathbf{v}_{2}}f(\mathbf{v})\right)+w\left(\frac{\partial^{2}f}{\partial\mathbf{v}\partial\mathbf{v}}f(\mathbf{v}_{2})+\frac{\partial^{2}f}{\partial\mathbf{v}_{2}\partial\mathbf{v}_{2}}f(\mathbf{v})-2\frac{\partial f}{\partial\mathbf{v}}\frac{\partial f}{\partial\mathbf{v}_{2}}\right)\right\} :\mathbf{q}\mathbf{q}.
\]
\small
To obtain the Landau equation, we have to write $I(\mathbf{v})$ as
a divergence involving the tensor $\mathbf{B}$. In order to do so,
we integrate by parts the term involving $\frac{\partial w}{\partial\mathbf{v}_{2}}$
while keeping the terms involving $\frac{\partial w}{\partial\mathbf{v}}$
. This gives

\[
I(\mathbf{v})=\frac{\Lambda}{2}\int\text{d}\mathbf{v}_{2}\text{d}\mathbf{q}\left\{ \frac{\partial w}{\partial\mathbf{v}}\left(\frac{\partial f}{\partial\mathbf{v}}f(\mathbf{v}_{2})-\frac{\partial f}{\partial\mathbf{v}_{2}}f(\mathbf{v})\right)+w\frac{\partial}{\partial\mathbf{v}}\left(\frac{\partial f}{\partial\mathbf{v}}f(\mathbf{v}_{2})-\frac{\partial f}{\partial\mathbf{v}_{2}}f(\mathbf{v})\right)\right\} :\mathbf{q}\mathbf{q}.
\]
Now, by noting that $I(\mathbf{v})$ can be written as a total divergence
with respect to $\mathbf{v}$ and using equation \ref{eq:tensor_B}
we obtain 
\begin{equation}
I(\mathbf{v})=\frac{\partial}{\partial\mathbf{v}}\int\text{d}\mathbf{v}_{2}\mathbf{\mathbf{B}}(\mathbf{v},\mathbf{v}_{2})\left(-\frac{\partial f}{\partial\mathbf{v}_{2}}f(\mathbf{v})+\frac{\partial f}{\partial\mathbf{v}}f(\mathbf{v}_{2})\right)+o\left(q^{2}\right),\label{eq:Landau-2}
\end{equation}
with $\mathbf{\mathbf{B}}(\mathbf{v},\mathbf{v}_{2})=\Lambda\int\text{d}\mathbf{q}\,w(\mathbf{v},\mathbf{v}_{2};\mathbf{q})\mathbf{q}\otimes\mathbf{\mathbf{q}}/2$
(see equation \eqref{eq:tensor_B}), and $o\left(q^{2}\right)$ means
that we omitted terms of order larger than 2. The term of order 2
is the collision operator of the Landau equation (\ref{eq:Landau}).

\subsection{Asymptotic expansions leading to the large deviation Hamiltonian
associated to the Landau equation\label{subsec:Asymptotic-expansions-Landau-Hamiltonian}}

In this section, we detail the computation of the large deviation
Hamiltonian for the Landau equation starting from the Hamiltonian
\eqref{eq:H_Boltzmann_Homogene} for the Boltzmann equation and using
the grazing collision limit.

First, let us rewrite this Hamiltonian

\[
H[f,p]=\frac{\Lambda}{2}\int\text{d}\mathbf{r}\text{d\ensuremath{\mathbf{v_{1}}}}\text{d}\mathbf{v}_{2}\text{d}\mathbf{q\,}w\left(\mathbf{v}_{1}+\frac{1}{2}\mathbf{q},\mathbf{v}_{2}-\mathbf{q};\mathbf{q}\right)f(\mathbf{v}_{1})f(\mathbf{v}_{2})\left\{ \text{e}^{\left[-p(\mathbf{v}_{1})-p(\mathbf{v}_{2})+p(\mathbf{\mathbf{v}}_{1}+\mathbf{q})+p(\mathbf{\mathbf{v}}_{2}-\mathbf{q})\right]}-1\right\} .
\]
In order to obtain a Hamiltonian associated with the Landau equation,
we will use the same hypothesis of grazing collisions and a Taylor
expansion in $\mathbf{q}$ to the same order

\[
\begin{cases}
w(\mathbf{v}_{1}+\frac{1}{2}\mathbf{q},\mathbf{v}_{2}-\frac{1}{2}\mathbf{q};\mathbf{q}) & =w(\mathbf{v}_{1},\mathbf{v}_{2};\mathbf{q})+\frac{1}{2}\left(\frac{\partial w}{\partial\mathbf{v}_{1}}-\frac{\partial w}{\partial\mathbf{v}_{2}}\right).\mathbf{q}+\mathcal{O}(q^{2})\\
\text{e}^{\left[-p(\mathbf{v}_{1})-p(\mathbf{v}_{2})+p(\mathbf{\mathbf{v}}_{1}+\mathbf{q})+p(\mathbf{\mathbf{v}}_{2}-\mathbf{q})\right]}-1 & =\left(\frac{\partial p}{\partial\mathbf{v}_{1}}-\frac{\partial p}{\partial\mathbf{v}_{2}}\right).\mathbf{q}+\\
 & +\frac{1}{2}\left\{ \frac{\partial^{2}p}{\partial\mathbf{v}_{1}\partial\mathbf{v}_{1}}+\frac{\partial^{2}p}{\partial\mathbf{v}_{2}\partial\mathbf{v}_{2}}+\left(\frac{\partial p}{\partial\mathbf{v}_{1}}-\frac{\partial p}{\partial\mathbf{v}_{2}}\right)\left(\frac{\partial p}{\partial\mathbf{v}_{1}}-\frac{\partial p}{\partial\mathbf{v}_{2}}\right)\right\} :\mathbf{qq}+\mathcal{O}(q{}^{3}).
\end{cases}
\]
We evaluate the terms of $H$ order by order. There is no term of
order zero. The term of order one in $\mathbf{q}$ is

\[
\frac{\Lambda}{2}\int\text{d}\mathbf{r}\text{d\ensuremath{\mathbf{v_{1}}}}\text{d}\mathbf{v}_{2}\text{d}\mathbf{q}\,w(\mathbf{v}_{1},\mathbf{v}_{2};\mathbf{q})f(\mathbf{v}_{1})f(\mathbf{v}_{2})\left(\frac{\partial p}{\partial\mathbf{v}_{1}}-\frac{\partial p}{\partial\mathbf{v}_{2}}\right).\mathbf{q},
\]
which is zero because $w(\mathbf{v}_{1},\mathbf{v}_{2};\mathbf{q})$
is an even function of $\mathbf{q}$ (see point 3 of appendix \eqref{subsec:The-Landau-collision}).
At second order in $\mathbf{q}$ the Hamiltonian reads

\begin{eqnarray*}
H_{\text{Landau}}[f,p] & = & \frac{\Lambda}{4}\int\text{d}\mathbf{r}\text{d\ensuremath{\mathbf{v}_{1}}}\text{d}\mathbf{v}_{2}\text{d}\mathbf{q}\,f(\mathbf{v}_{1})f(\mathbf{v}_{2})\left\{ w\left[\frac{\partial^{2}p}{\partial\mathbf{v}_{1}\partial\mathbf{v}_{1}}+\frac{\partial^{2}p}{\partial\mathbf{v}_{2}\partial\mathbf{v}_{2}}+\left(\frac{\partial p}{\partial\mathbf{v}_{1}}-\frac{\partial p}{\partial\mathbf{v}_{2}}\right)\left(\frac{\partial p}{\partial\mathbf{v}_{1}}-\frac{\partial p}{\partial\mathbf{v}_{2}}\right)\right]\right.\\
 & + & \left.\left(\frac{\partial p}{\partial\mathbf{v}_{1}}-\frac{\partial p}{\partial\mathbf{v}_{2}}\right)\left(\frac{\partial w}{\partial\mathbf{v}_{1}}-\frac{\partial w}{\partial\mathbf{v}_{2}}\right)\right\} :\mathbf{qq}.
\end{eqnarray*}
In this expression, in order to make appear the tensor $\mathbf{\mathbf{B}}(\mathbf{v},\mathbf{v}_{2})=\Lambda\int\text{d}\mathbf{q}\,w(\mathbf{v},\mathbf{v}_{2};\mathbf{q})\mathbf{q}\mathbf{\mathbf{q}}/2$
(see equation \eqref{eq:tensor_B}), we integrate by parts the terms
involving $\frac{\partial w}{\partial\mathbf{v}_{1}}$ and $\frac{\partial w}{\partial\mathbf{v}_{2}}$,
we develop the derivatives of products generated by partial integration,
we use equation (\ref{eq:tensor_B}) and we obtain

\begin{eqnarray*}
H_{\text{Landau}}[f,p] & = & \frac{1}{2}\int\text{d}\mathbf{r}\text{d\ensuremath{\mathbf{v}_{1}}}\text{d}\mathbf{v}_{2}\,\mathbf{B}(\mathbf{v}_{1},\mathbf{v}_{2})\left\{ f(\mathbf{v}_{1})f(\mathbf{v}_{2})\left(\frac{\partial p}{\partial\mathbf{v}_{1}}-\frac{\partial p}{\partial\mathbf{v}_{2}}\right)\left(\frac{\partial p}{\partial\mathbf{v}_{1}}-\frac{\partial p}{\partial\mathbf{v}_{2}}\right)\right.+\\
 & + & \left.\left(\frac{\partial p}{\partial\mathbf{v}_{1}}-\frac{\partial p}{\partial\mathbf{v}_{2}}\right)\left(\frac{\partial f}{\partial\mathbf{v}_{2}}f(\mathbf{v}_{1})-\frac{\partial f}{\partial\mathbf{v}_{1}}f(\mathbf{v}_{2})\right)\right\} .
\end{eqnarray*}
Using the property that $\mathbf{B}(\mathbf{v}_{1},\mathbf{v}_{2})=\mathbf{B}(\mathbf{v}_{2},\mathbf{v}_{1})$
(see appendix \ref{subsec:The-Landau-collision}), we have for every
function $g$ of $(\mathbf{v}_{1},\mathbf{v}_{2})$: $\int\text{d\ensuremath{\mathbf{v_{1}}}}\text{d}\mathbf{v}_{2}\mathbf{\mathbf{B}}(\mathbf{v}_{1},\mathbf{v}_{2})g(\mathbf{v}_{1},\mathbf{v}_{2})=\int\text{d\ensuremath{\mathbf{v_{1}}}}\text{d}\mathbf{v}_{2}\mathbf{\mathbf{B}}(\mathbf{v}_{1},\mathbf{v}_{2})g(\mathbf{v}_{2},\mathbf{v}_{1})$.
Using this property we have

\[
H_{\text{Landau}}[f,p]=\int\text{d}\mathbf{r}\text{d\ensuremath{\mathbf{v}_{1}}}\text{d}\mathbf{v}_{2}\,\mathbf{\mathbf{B}}(\mathbf{v}_{1},\mathbf{v}_{2})\left\{ f(\mathbf{v}_{1})f(\mathbf{v}_{2})\left(\frac{\partial p}{\partial\mathbf{v}_{1}}\frac{\partial p}{\partial\mathbf{v}_{1}}-\frac{\partial p}{\partial\mathbf{v}_{1}}\frac{\partial p}{\partial\mathbf{v}_{2}}\right)+\frac{\partial p}{\partial\mathbf{v}_{1}}\frac{\partial f}{\partial\mathbf{v}_{2}}f(\mathbf{v}_{1})-\frac{\partial p}{\partial\mathbf{v}_{1}}\frac{\partial f}{\partial\mathbf{v}_{1}}f(\mathbf{v}_{2})\right\} .
\]
We integrate by parts the last term with respect to $\mathbf{v}_{1}$
to obtain

\begin{eqnarray*}
H_{\text{Landau}}[f,p] & = & \int\text{d}\mathbf{r}\text{d\ensuremath{\mathbf{v}_{1}}}\text{d}\mathbf{v}_{2}\,f(\mathbf{v}_{1})\left\{ \frac{\partial p}{\partial\mathbf{v}_{1}}\mathbf{\mathbf{B}}(\mathbf{v}_{1},\mathbf{v}_{2})\frac{\partial f}{\partial\mathbf{v}_{2}}+\frac{\partial p}{\partial\mathbf{v}_{1}}\frac{\partial p}{\partial\mathbf{v}_{1}}\mathbf{\mathbf{B}}(\mathbf{v}_{1},\mathbf{v}_{2})f(\mathbf{v}_{2})+\frac{\partial}{\partial\mathbf{v}_{1}}\left(\mathbf{\mathbf{B}}(\mathbf{v}_{1},\mathbf{v}_{2})f(\mathbf{v}_{2})\frac{\partial p}{\partial\mathbf{v}_{1}}\right)\right\} \\
 &  & -\int\text{d}\mathbf{r}\text{d}\mathbf{v}_{1}\text{d}\mathbf{v}_{2}\,f(\mathbf{v}_{1})f(\mathbf{v}_{2})\frac{\partial p}{\partial\mathbf{v}_{1}}\frac{\partial p}{\partial\mathbf{v}_{2}}\mathbf{\mathbf{B}}(\mathbf{v}_{1},\mathbf{v}_{2}).
\end{eqnarray*}
From here, using equation (\ref{eq:coeffLandau}) we obtain 
\begin{equation}
H_{\text{Landau}}[f,p]=H_{MF}\left[f,p\right]+H_{I}\left[f,p\right],\label{eq:ham_from_boltz-1}
\end{equation}
with 
\[
H_{MF}\left[f,p\right]=\int\text{d}\mathbf{r}\text{d}\mathbf{v}_{1}f\left\{ \mathbf{b}\left[f\right].\frac{\partial p}{\partial\mathbf{v}_{1}}+\frac{\partial}{\partial\mathbf{v}_{1}}\left(\mathbf{D}\left[f\right]\frac{\partial p}{\partial\mathbf{v}_{1}}\right)+\mathbf{D}\left[f\right]:\frac{\partial p}{\partial\mathbf{v}_{1}}\frac{\partial p}{\partial\mathbf{v}_{1}}\right\} ,
\]
and 
\[
H_{I}\left[f,p\right]=-\int\text{d}\mathbf{r}\text{d\ensuremath{\mathbf{v}_{1}}}\text{d}\mathbf{v}_{2}f(\mathbf{v}_{1})f(\mathbf{v}_{2})\frac{\partial p}{\partial\mathbf{v}_{1}}\frac{\partial p}{\partial\mathbf{v}_{2}}:\mathbf{B}\left(\mathbf{v}_{1},\mathbf{v}_{2}\right).
\]

\section{Useful formulas}

In this appendix, we list and prove some formulas used in section
\ref{sec:LD-from-microscopic-dynamics}.

\subsection{Sokhotski--Plemelj formula}

We have 
\begin{equation}
\underset{\tilde{\epsilon}\rightarrow0^{+}}{\lim}\frac{1}{x-i\tilde{\epsilon}}=P\left(\frac{1}{x}\right)+i\pi\delta\left(x\right),\label{eq:Sokhotski=002013Plemelj}
\end{equation}
from which we also have 
\begin{equation}
\Im\left(\frac{1}{x-i\tilde{\epsilon}}\right)=\pi\delta\left(x\right),\label{eq:Imaginary-Sokhotski=002013Plemelj}
\end{equation}
where $\Im(z)$ is the imaginary part of the complex number $z$.

\subsection{Some properties of the dielectric function }

We discuss a few useful properties of the dielectric function. By
definition the dielectric function (\ref{eq:dielectric_function})
is 
\begin{equation}
\varepsilon[f](\mathbf{k},\omega)=1-\hat{W}\left(\mathbf{k}\right)\int\text{d}\mathbf{v}\frac{\mathbf{k}.\frac{\partial f}{\partial\mathbf{v}}}{\mathbf{k}.\mathbf{v}-\omega-i\tilde{\epsilon}},\label{eq:dielectric_function-1}
\end{equation}

Using the definition of the dielectric function (\ref{eq:dielectric_function-1})
and (\ref{eq:Imaginary-Sokhotski=002013Plemelj}), we have 
\begin{equation}
\Im\left[\varepsilon[f](\mathbf{k},\omega)\right]=-\pi\hat{W}\left(\mathbf{k}\right)\int\text{d}\mathbf{v}\,\mathbf{k}.\frac{\partial f}{\partial\mathbf{v}}\delta\left(\omega-\mathbf{k}.\mathbf{v}\right).\label{eq:Imaginary-Dispersion-Relation}
\end{equation}
From (\ref{eq:dielectric_function-1}), we readily see that 
\begin{equation}
\varepsilon^{*}[f](\mathbf{k},\omega)=\varepsilon[f](-\mathbf{k},-\omega)\label{eq:Dielectric_Complex_Conjugated}
\end{equation}
where $\bar{z}$ is the imaginary part of the complex number $z$.

\subsection{\label{subsec:Double-integral-of}Double integral of a homogeneous
kernel}

\subsubsection{Symmetric kernel}

Let $f$ be a function for which $\left|\int_{0}^{\infty}f(t)\text{d}t\right|<\infty$.
Then,
\[
\frac{1}{2T}\int_{0}^{T}\int_{0}^{T}\text{d}t_{1}\text{d}t_{2}\,f\left(\left|t_{1}-t_{2}\right|\right)\underset{T\rightarrow\infty}{\longrightarrow}\int_{0}^{\infty}\text{d}\tau\,f(\tau)=\frac{1}{2}\int_{-\infty}^{\infty}\text{d}\tau\,f(\left|\tau\right|).
\]

\paragraph{Proof}

Using the parity of $f\left(\left|\cdot\right|\right)$, we get
\[
\frac{1}{2T}\int_{0}^{T}\int_{0}^{T}\text{d}t_{1}\text{d}t_{2}\,f\left(\left|t_{1}-t_{2}\right|\right)=\frac{1}{T}\int_{0}^{T}\text{d\ensuremath{t_{1}\int_{0}^{t_{1}}\text{d}t_{2}\,f\left(\left|t_{1}-t_{2}\right|\right)}}.
\]
Rewriting the integrals with the change of variable $\left(t_{1},t_{2}\right)\rightarrow\left(\tau,\tau'\right)=\left(t_{1},t_{1}-t_{2}\right)$
leads to
\[
\frac{1}{2T}\iint_{0}^{T}\text{d}t_{1}\text{d}t_{2}\,f\left(\left|t_{1}-t_{2}\right|\right)=\frac{1}{T}\int_{0}^{T}\text{d\ensuremath{\tau_{1}\int_{0}^{\tau_{1}}\text{d}\tau_{2}\,f\left(\left|\tau_{2}\right|\right)}}.
\]
Defining the function $g$ by $g(\tau)=\int_{0}^{\tau}\text{d}\tau_{2}\,f\left(\left|\tau_{2}\right|\right)$,
noting that $g$ has a finite limit for large $\tau$ and that the
integral on $[0,\infty[$ of $g$ does diverge, we obtain asymptotically,
\[
\int_{0}^{T}g(\tau)\text{d}\tau\underset{T\rightarrow\infty}{\sim}T\int_{0}^{\infty}f(\tau)\text{d}\tau.
\]
Combining the last two equations gives the result we wanted to prove
\[
\frac{1}{2T}\iint_{0}^{T}\text{d}t_{1}\text{d}t_{2}\,f\left(\left|t_{1}-t_{2}\right|\right)\underset{T\rightarrow\infty}{\longrightarrow}\frac{1}{2}\int_{-\infty}^{\infty}\text{d}\tau\,f(\left|\tau\right|).
\]

\subsubsection{General kernel}

Let $f$ be a function for which $\left|\int_{0}^{\infty}f(t)\text{d}t\right|<\infty$
and $\left|\int_{-\infty}^{0}f(t)\text{d}t\right|<\infty$. Then,
\begin{equation}
\frac{1}{T}\iint_{0}^{T}\text{d}t_{1}\text{d}t_{2}\,f\left(t_{1}-t_{2}\right)\underset{T\rightarrow\infty}{\longrightarrow}\frac{1}{T}\int_{-\infty}^{\infty}\text{d}\tau\,f(\tau).\label{eq:general_kernel}
\end{equation}

\paragraph{Proof}

First, let us rewrite the integral on $t_{2}$ using the additivity
of integration on intervals
\begin{equation}
\frac{1}{T}\iint_{0}^{T}\text{d}t_{1}\text{d}t_{2}\,f\left(t_{1}-t_{2}\right)=\frac{1}{T}\int_{0}^{T}\text{d}t_{1}\int_{0}^{t_{1}}\text{d}t_{2}\,f\left(t_{1}-t_{2}\right)+\frac{1}{T}\int_{0}^{T}\text{d}t_{1}\int_{t_{1}}^{T}\text{d}t_{2}\,f\left(t_{1}-t_{2}\right).\label{eq:interannex}
\end{equation}
Rewriting the first term of (\ref{eq:interannex}) with the change
of variable $\left(t_{1},t_{2}\right)\rightarrow\left(\tau,\tau'\right)=\left(t_{1},t_{1}-t_{2}\right)$
leads to
\[
\frac{1}{T}\int_{0}^{T}\text{d}t_{1}\int_{0}^{t_{1}}\text{d}t_{2}\,f\left(t_{1}-t_{2}\right)=\frac{1}{T}\int_{0}^{T}\text{d}\tau\int_{0}^{\tau}\text{d}\tau'\,f\left(\tau'\right),
\]
and using the Fubini theorem and the change of variable $\left(t_{1},t_{2}\right)\rightarrow\left(\tau',\tau\right)=\left(t_{2}-t_{1},t_{2}\right)$
leads to 
\[
\frac{1}{T}\int_{0}^{T}\text{d}t_{1}\int_{t_{1}}^{T}\text{d}t_{2}\,f\left(t_{1}-t_{2}\right)=\frac{1}{T}\int_{0}^{T}\text{d}\tau\int_{0}^{\tau}\text{d}\tau'\,f\left(-\tau'\right).
\]
We noticed during the previous proof that 
\[
\lim_{T\rightarrow\infty}\frac{1}{T}\int_{0}^{T}\text{d}\tau\int_{0}^{\tau}\text{d}\tau'\,f\left(\tau'\right)=\int_{0}^{\infty}f(\tau)\text{d}\tau.
\]
With a similar computation, we can show that
\[
\lim_{T\rightarrow\infty}\frac{1}{T}\int_{0}^{T}\text{d}\tau\int_{0}^{\tau}\text{d}\tau'\,f\left(-\tau'\right)=\int_{0}^{\infty}f(-\tau)\text{d}\tau=-\int_{-\infty}^{0}f(\tau)\text{d}\tau.
\]
Then, gathering the two terms of (\ref{eq:interannex}) and taking
the limit as $T$ goes to infinity, we find that
\[
\frac{1}{T}\iint_{0}^{T}\text{d}t_{1}\text{d}t_{2}\,f\left(t_{1}-t_{2}\right)\underset{T\rightarrow\infty}{\longrightarrow}\frac{1}{T}\int_{-\infty}^{\infty}\text{d}\tau\,f(\tau),
\]
which is what we wanted to prove.

\subsection{Fourier--Laplace representation of a product}

Let $\phi$ and $\psi$ and two functions that admit Fourier--Laplace
transforms $\tilde{\phi}$ and $\tilde{\psi}$ as defined in equation
(\ref{eq:FL_Transf}).  Then,
\begin{equation}
\int\text{d}\mathbf{r}\int_{-\infty}^{\infty}\text{d}t\,\phi\left(\mathbf{r},t\right)\psi\left(\mathbf{r},t\right)=\frac{1}{2\pi}\left(\frac{\lambda_{D}}{L}\right)^{3}\sum_{\mathbf{k}\in\left(2\pi\lambda_{D}/L\right)\mathbb{Z}^{3}}\int_{\Gamma}\text{d}\omega\,\tilde{\phi}\left(\mathbf{k},\omega\right)\tilde{\psi}\left(-\mathbf{k},-\omega\right).\label{eq:FL_Produit}
\end{equation}
\textbf{Proof}

Given the definition (\ref{eq:FL_Transf}) of the Fourier-Laplace
transform, the inversion formula is 
\[
\varphi\left(\mathbf{\mathbf{r}},t\right)=\frac{1}{2\pi}\left(\frac{\lambda_{D}}{L}\right)^{3}\sum_{\mathbf{k}\in\left(2\pi\lambda_{D}/L\right)\mathbb{Z}^{3}}\int_{\Gamma}\text{d}\omega\,\text{e}^{i\left(\mathbf{k}.\mathbf{r}-\omega t\right)}\tilde{\varphi}\left(\mathbf{k},\omega\right),
\]
where $\Gamma$ is a contour to be chosen to insure the convergence.
Using the inversion formula, the proof of the result is straightforward.

\section{Computation of the linear part and the quadratic part of the large
deviation Hamiltonian \label{sec:Computation-Linear-Part-Hamiltonian}}

\subsection{Computation of the first cumulant (linear part)\label{subsec:Comput-linear}}

In this appendix, we explicit the computations of $\mathbf{C}^{(1)}$,
using (\ref{eq:firstcumul}) and (\ref{eq:Autocorrelation_VG_Fourier_Laplace}).
This computation is different, but analogous to the one in \S 51 of
\cite{Lifshitz_Pitaevskii_1981_Physical_Kinetics}. We start from
(\ref{eq:firstcumul}) which leads to 
\[
\mathbf{C}^{(1)}(\mathbf{v})=-\lim_{T\rightarrow\infty}\frac{1}{T}\int_{0}^{T}\text{d}t\,\mathbb{E}\left(\frac{\partial V\left[\delta g_{\Lambda}\right]}{\partial\mathbf{r}}\delta g_{\Lambda}\right).
\]
We notice that over time $t$ long enough to forget the information
about the initial condition, but short enough such that the velocity
distribution has not changed much, $\mathbb{E}\left(\frac{\partial V\left[\delta g_{\Lambda}\right]}{\partial\mathbf{r}}\delta g_{\Lambda}\right)$
reaches a finite limit. In this limit, we simply obtain 
\[
\mathbf{C}^{(1)}(\mathbf{v})=-\mathbb{E}_{S}\left(\frac{\partial V\left[\delta g_{\Lambda}\right]}{\partial\mathbf{r}}\delta g_{\Lambda}\right).
\]
 We express each of the two terms $\frac{\partial V\left[\delta g_{\Lambda}\right]}{\partial\mathbf{r}}\left(\mathbf{r},t\right)$
and $\delta g_{\Lambda}\left(\mathbf{r},t\right)$ through their Fourier--Laplace
transforms, and we apply $\mathbb{E}_{S}$ using (\ref{eq:Autocorrelation_VG})
to get

\[
\mathbf{C}^{(1)}(\mathbf{v})=\frac{-i}{2\pi}\left(\frac{\lambda_{D}}{L}\right)^{3}\sum_{\mathbf{k}}\int_{\Gamma}\text{d}\omega\,\mathbf{k}\widetilde{\mathcal{C}_{VG}}\left(\mathbf{k},\omega,\mathbf{v}\right).
\]
Using (\ref{eq:Autocorrelation_VG_Fourier_Laplace}), we obtain 
\[
\mathbf{C}^{(1)}(\mathbf{v})=\left(\mathbf{b'}\left[f\right]\left(\mathbf{v}\right)f\left(\mathbf{v}\right)-\mathbf{D}'\left[f\right]\left(\mathbf{v}\right).\frac{\partial f}{\partial\mathbf{v}}\right)
\]
 with 
\[
\mathbf{D}'\left[f\right]\left(\mathbf{v}\right)=i\left(\frac{\lambda_{D}}{L}\right)^{3}\sum_{\mathbf{k}}\int_{\Gamma}\text{d}\omega\int\text{d}\mathbf{v}_{2}\,\frac{\mathbf{k}\mathbf{k}}{\mathbf{k}.\mathbf{v}-\omega+i\tilde{\epsilon}}f\left(\mathbf{v}_{2}\right)\delta\left(\omega-\mathbf{k}.\mathbf{v}_{2}\right)\frac{\hat{W}(\mathbf{k})^{2}}{\left|\varepsilon[f]\left(\omega,\mathbf{k}\right)\right|^{2}},
\]
 and 
\[
\mathbf{b'}\left[f\right]\left(\mathbf{v}\right)=-i\left(\frac{\lambda_{D}}{L}\right)^{3}\sum_{\mathbf{k}}\int_{\Gamma}\text{d}\omega\,\frac{\mathbf{k}\hat{W}\left(\mathbf{k}\right)}{\varepsilon\left(\mathbf{k},\omega\right)}\delta\left(\omega-\mathbf{k}.\mathbf{v}\right).
\]
Using (\ref{eq:Imaginary-Sokhotski=002013Plemelj}), we compute the
real part of $\mathbf{\overleftrightarrow{D}}'$and get
\[
\Re\left(\mathbf{D}'\left[f\right]\left(\mathbf{v}\right)\right)=\mathbf{D}\left[f\right]\left(\mathbf{v}\right)=\int\text{d}\mathbf{v}_{2}\mathbf{B}\left[f\right](\mathbf{v},\mathbf{v}_{2})f(\mathbf{v}_{2}),
\]
where 
\[
\mathbf{B}(\mathbf{v},\mathbf{v}_{2})=\pi\left(\frac{\lambda_{D}}{L}\right)^{3}\int_{-\infty}^{+\infty}\text{d}\omega\sum_{\mathbf{k}}\,\delta\left(\omega-\mathbf{k}.\mathbf{v}\right)\delta\left(\omega-\mathbf{k}.\mathbf{v}_{2}\right)\frac{\mathbf{k}\mathbf{k}\hat{W}(\mathbf{k})^{2}}{\left|\varepsilon[f]\left(\omega,\mathbf{k}\right)\right|^{2}}
\]
is the tensor defined in equation (\ref{eq:B_GLB}).

Using (\ref{eq:Imaginary-Dispersion-Relation}), we compute the real
part of $\mathbf{b'}$, and get 

\[
\Re\left(\mathbf{b'}\left[f\right]\left(\mathbf{v}\right)\right)=\mathbf{b}\left[f\right](\mathbf{v})=\int\text{d}\mathbf{v}_{2}\mathbf{B}\left[f\right](\mathbf{v},\mathbf{v}_{2})\frac{\partial f}{\partial\mathbf{v}_{2}}.
\]
It is also easily checked that $\Im\left[\mathbf{C}^{(1)}(\mathbf{v})\right]=0$
. We have thus justified that 
\[
\mathbf{C}^{(1)}(\mathbf{v})=\int\text{d}\mathbf{v}_{2}\,\mathbf{B}\left[f\right](\mathbf{v},\mathbf{v}_{2})\left(\frac{\partial f}{\partial\mathbf{v}_{2}}f(\mathbf{v})-f(\mathbf{v}_{2})\frac{\partial f}{\partial\mathbf{v}}\right)=\left(\mathbf{b}\left[f\right]\left(\mathbf{v}\right)f\left(\mathbf{v}\right)-\mathbf{D}\left[f\right]\left(\mathbf{v}\right).\frac{\partial f}{\partial\mathbf{v}}\right),
\]
where $\mathbf{B}$ is the tensor defined in equation (\ref{eq:B_GLB}).

\subsection{Computation of the second cumulant (quadratic part)\label{subsec:Appendice_Second_Cumulant}}

In this appendix, in order to compute the second order cumulant of
the large deviation Hamiltonian we use Wick's theorem to express the
four-points correlation functions $\mathbb{E}_{S}\left(\frac{\partial V\left[\delta g_{\Lambda}\right]^{(1)}}{\partial\mathbf{r}}\frac{\partial V\left[\delta g_{\Lambda}\right]^{(2)}}{\partial\mathbf{r}}\delta g_{\Lambda}^{(1)}\delta g_{\Lambda}^{(2)}\right)$
as a sum of products of two-point correlation functions. In such formulas,
the superscripts $(1)$ or $(2)$ mean that the quantities are evaluated
at either $\left(\mathbf{r}_{1},t_{1}\right)$ and $\left(\mathbf{r}_{2},t_{2}\right)$,
respectively, or $\left(\mathbf{r}_{1},\mathbf{v}_{1},t_{1}\right)$
and $\left(\mathbf{r}_{2},\mathbf{v}_{2},t_{2}\right)$, respectively.
We obtain 
\begin{multline*}
\mathbb{E}_{S}\left(\frac{\partial V\left[\delta g_{\Lambda}\right]^{(1)}}{\partial\mathbf{r}}\frac{\partial V\left[\delta g_{\Lambda}\right]^{(2)}}{\partial\mathbf{r}}\delta g_{\Lambda}^{(1)}\delta g_{\Lambda}^{(2)}\right)-\mathbb{E}_{S}\left(\frac{\partial V\left[\delta g_{\Lambda}\right]^{(1)}}{\partial\mathbf{r}}\delta g_{\Lambda}^{(1)}\right)\mathbb{E}_{S}\left(\frac{\partial V\left[\delta g_{\Lambda}\right]^{(2)}}{\partial\mathbf{r}}\delta g_{\Lambda}^{(2)}\right)=\\
\mathbb{E}_{S}\left(\frac{\partial V\left[\delta g_{\Lambda}\right]^{(1)}}{\partial\mathbf{r}}\frac{\partial V\left[\delta g_{\Lambda}\right]^{(2)}}{\partial\mathbf{r}}\right)\mathbb{E}_{S}\left(\delta g_{\Lambda}^{(1)}\delta g_{\Lambda}^{(2)}\right)+\mathbb{E}_{S}\left(\frac{\partial V\left[\delta g_{\Lambda}\right]^{(1)}}{\partial\mathbf{r}}\delta g_{\Lambda}^{(2)}\right)\mathbb{E}_{S}\left(\frac{\partial V\left[\delta g_{\Lambda}\right]^{(2)}}{\partial\mathbf{r}}\delta g_{\Lambda}^{(1)}\right)
\end{multline*}
Using (\ref{eq:Definition_C}), we thus obtain 
\[
\mathbf{C}=\mathbf{C}_{\alpha}+\mathbf{C}_{\beta}
\]
with 
\[
\mathbf{C}_{\alpha}=\lim_{N\rightarrow\infty}\frac{1}{2T}\int\text{d}\mathbf{r}_{1}\text{d}\mathbf{r}_{2}\iint_{0}^{T}\text{d}t_{1}\text{d}t_{2}\mathbb{E}_{S}\left(\frac{\partial V\left[\delta g_{\Lambda}\right]^{(1)}}{\partial\mathbf{r}}\frac{\partial V\left[\delta g_{\Lambda}\right]^{(2)}}{\partial\mathbf{r}}\right)\mathbb{E}_{S}\left(\delta g_{\Lambda}^{(1)}\delta g_{\Lambda}^{(2)}\right),
\]
and
\[
\mathbf{C}_{\beta}=\lim_{N\rightarrow\infty}\frac{1}{2T}\int\text{d}\mathbf{r}_{1}\text{d}\mathbf{r}_{2}\iint_{0}^{T}\text{d}t_{1}\text{d}t_{2}\,\mathbb{E}_{S}\left(\frac{\partial V\left[\delta g_{\Lambda}\right]^{(1)}}{\partial\mathbf{r}}\delta g_{\Lambda}^{(2)}\right)\mathbb{E}_{S}\left(\frac{\partial V\left[\delta g_{\Lambda}\right]^{(2)}}{\partial\mathbf{r}}\delta g_{\Lambda}^{(1)}\right).
\]

Due to spatial and temporal homogeneity, the correlations functions
only depend on the difference of the positions and times on which
they are computed: $\mathbb{E}_{S}\left(V\left[\delta g_{\Lambda}\right]\left(\mathbf{r}_{1},t_{1}\right)V\left[\delta g_{\Lambda}\right]\left(\mathbf{r}_{2},t_{2}\right)\right)=\mathcal{C}_{VV}\left(\mathbf{r}_{1}-\mathbf{r}_{2},t_{1}-t_{2}\right)$,
and $\mathbb{E}_{S}\left(\delta g_{\Lambda}\left(\mathbf{r}_{1},\mathbf{v}_{1},t_{1}\right)\delta g_{\Lambda}\left(\mathbf{r}_{2},\mathbf{v}_{2},t_{2}\right)\right)=\mathcal{C}_{GG}\left(\mathbf{r}_{1}-\mathbf{r}_{2},t_{1}-t_{2},\mathbf{v}_{1},\mathbf{v}_{2}\right)$.
We use 
\[
\mathbb{E}_{S}\left(\frac{\partial V\left[\delta g_{\Lambda}\right]^{(1)}}{\partial\mathbf{r}}\frac{\partial V\left[\delta g_{\Lambda}\right]^{(2)}}{\partial\mathbf{r}}\right)=-\frac{\partial}{\partial\mathbf{\mathbf{r}}}\frac{\partial}{\partial\mathbf{\mathbf{r}}}\left[\mathcal{C}_{VV}\right]\left(\mathbf{r}_{1}-\mathbf{r}_{2},t_{1}-t_{2}\right),
\]
and apply the result (\ref{eq:general_kernel}) from annex \ref{subsec:Double-integral-of}
to find
\[
\mathbf{C}_{\alpha}\left(\mathbf{v}_{1},\mathbf{v}_{2}\right)=-\frac{1}{2}\int\text{d}\mathbf{r}_{1}\text{d}\mathbf{r}_{2}\int_{-\infty}^{\infty}\text{d}t\,\frac{\partial}{\partial\mathbf{\mathbf{r}}}\frac{\partial}{\partial\mathbf{\mathbf{r}}}\left[\mathcal{C}_{VV}\right]\left(\mathbf{r}_{1}-\mathbf{r}_{2},t\right)\mathcal{C}_{GG}\left(\mathbf{r}_{1}-\mathbf{r}_{2},t,\mathbf{v}_{1},\mathbf{v}_{2}\right).
\]
\\
Then, we apply the change of variables $\left(\mathbf{r}_{1},\mathbf{r}_{2}\right)\rightarrow\left(\mathbf{r}=\mathbf{r}_{1}-\mathbf{r}_{2},\mathbf{r}'=\mathbf{r}_{2}\right)$,
integrate over $\mathbf{r}'$, apply the result (\ref{eq:FL_Produit})
from annex \ref{subsec:Double-integral-of}, to obtain\\
\[
\mathbf{C}_{\alpha}\left(\mathbf{v}_{1},\mathbf{v}_{2}\right)=\frac{1}{2\left(2\pi\right)}\sum_{\mathbf{k}}\int_{\Gamma}\text{d}\omega\,\mathbf{k}\mathbf{k}\widetilde{\mathcal{C}_{VV}}\left(\mathbf{k},\omega\right)\widetilde{\mathcal{C}_{GG}}\left(-\mathbf{k},-\omega,\mathbf{v}_{1},\mathbf{v}_{2}\right),
\]
Similarly for $\mathbf{C}_{\beta}$, one obtains 
\[
\mathbf{C}_{\beta}\left(\mathbf{v}_{1},\mathbf{v}_{2}\right)=\frac{-1}{2\left(2\pi\right)}\sum_{\mathbf{k}}\int_{\Gamma}\text{d}\omega\,\mathbf{k}\mathbf{k}\widetilde{\mathcal{C}_{VG}}\left(\mathbf{k},\omega,\mathbf{v}_{1}\right)\widetilde{\mathcal{C}_{VG}}\left(\mathbf{k},\omega,\mathbf{v}_{2}\right).
\]
Summing these two terms we obtain

\begin{equation}
\mathbf{C}=\frac{1}{2\left(2\pi\right)}\sum_{\mathbf{k}}\int_{\Gamma}\text{d}\omega\,\mathbf{k}\mathbf{k}\left\{ \widetilde{\mathcal{C}_{VV}}\left(\mathbf{k},\omega\right)\widetilde{\mathcal{C}_{GG}}\left(\mathbf{k},\omega,\mathbf{v}_{1},\mathbf{v}_{2}\right)-\widetilde{\mathcal{C}_{VG}}\left(\mathbf{k},\omega,\mathbf{v}_{1}\right)\widetilde{\mathcal{C}_{VG}}\left(\mathbf{k},\omega,\mathbf{v}_{2}\right)\right\} .\label{eq:C_erratum}
\end{equation}
Let us define $\mathcal{A}$ and $\mathcal{B}$ as 
\[
\mathcal{A}\equiv\widetilde{\mathcal{C}_{VV}}\left(\mathbf{k},\omega\right)\widetilde{\mathcal{C}_{GG}}\left(\mathbf{k},\omega,\mathbf{v}_{1},\mathbf{v}_{2}\right),
\]
and
\[
\mathcal{B}\equiv\widetilde{\mathcal{C}_{VG}}\left(\mathbf{k},\omega,\mathbf{v}_{1}\right)\widetilde{\mathcal{C}_{VG}}\left(\mathbf{k},\omega,\mathbf{v}_{2}\right).
\]
From (\ref{eq:Autocorrelation_Potentiel_Fourier_Laplace}) and (\ref{eq:Autocorrelation_GG_Fourier_Laplace}),
$\mathcal{A}$ reads
\begin{multline*}
\mathcal{A}=2\pi\delta\left(\mathbf{v}_{1}-\mathbf{v}_{2}\right)f\left(\mathbf{v}_{1}\right)\delta\left(\omega-\mathbf{k}.\mathbf{v}_{1}\right)\widetilde{\mathcal{C}_{VV}}\left(\mathbf{k},\omega\right)\\
+\frac{\left(\widetilde{\mathcal{C}_{VV}}\left(\mathbf{k},\omega\right)\right)^{2}}{\left(\omega-\mathbf{k}.\mathbf{v}_{1}+i\tilde{\epsilon}\right)\left(\omega-\mathbf{k}.\mathbf{v}_{2}-i\tilde{\epsilon}\right)}\mathbf{k}.\frac{\partial f}{\partial\mathbf{v}_{1}}\mathbf{k}.\frac{\partial f}{\partial\mathbf{v}_{2}}\\
-2\pi\widetilde{\mathcal{C}_{VV}}\left(\mathbf{k},\omega\right)\hat{W}\left(\mathbf{k}\right)\mathbf{k}.\left\{ \frac{\partial f}{\partial\mathbf{v}_{1}}\frac{f(\mathbf{v}_{2})\delta\left(\omega-\mathbf{k}.\mathbf{v}_{2}\right)}{\varepsilon\left(\mathbf{k},\omega\right)\left(\omega-\mathbf{k}.\mathbf{v}_{1}+i\tilde{\epsilon}\right)}+\frac{\partial f}{\partial\mathbf{v}_{2}}\frac{f(\mathbf{v}_{1})\delta\left(\omega-\mathbf{k}.\mathbf{v}_{1}\right)}{\bar{\varepsilon}\left(\mathbf{k},\omega\right)\left(\omega-\mathbf{k}.\mathbf{v}_{2}-i\tilde{\epsilon}\right)}\right\} .
\end{multline*}
Similarly, from (\ref{eq:Autocorrelation_VG_Fourier_Laplace}), we
can deduce an expression for $\mathcal{B}$
\begin{multline*}
\mathcal{B=}\left(\frac{2\pi\hat{W}\left(\mathbf{k}\right)}{\varepsilon\left(\mathbf{k},\omega\right)}\right)^{2}f\left(\mathbf{v}_{1}\right)f\left(\mathbf{v}_{2}\right)\delta\left(\omega-\mathbf{k}.\mathbf{v}_{1}\right)\delta\left(\omega-\mathbf{k}.\mathbf{v}_{2}\right)\\
+\frac{\left(\widetilde{\mathcal{C}_{VV}}\left(\mathbf{k},\omega\right)\right)^{2}}{\left(\omega-\mathbf{k}.\mathbf{v}_{1}-i\tilde{\epsilon}\right)\left(\omega-\mathbf{k}.\mathbf{v}_{2}-i\tilde{\epsilon}\right)}\mathbf{k}.\frac{\partial f}{\partial\mathbf{v}_{1}}\mathbf{k}.\frac{\partial f}{\partial\mathbf{v}_{2}}\\
-2\pi\widetilde{\mathcal{C}_{VV}}\left(\mathbf{k},\omega\right)\hat{W}\left(\mathbf{k}\right)\mathbf{k}.\left\{ \frac{\partial f}{\partial\mathbf{v}_{1}}\frac{f(\mathbf{v}_{2})\delta\left(\omega-\mathbf{k}.\mathbf{v}_{2}\right)}{\varepsilon\left(\mathbf{k},\omega\right)\left(\omega-\mathbf{k}.\mathbf{v}_{1}-i\tilde{\epsilon}\right)}+\frac{\partial f}{\partial\mathbf{v}_{2}}\frac{f(\mathbf{v}_{1})\delta\left(\omega-\mathbf{k}.\mathbf{v}_{1}\right)}{\varepsilon\left(\mathbf{k},\omega\right)\left(\omega-\mathbf{k}.\mathbf{v}_{2}-i\tilde{\epsilon}\right)}\right\} .
\end{multline*}
To compute the second cumulant from (\ref{eq:C_erratum}), we are
specifically interested in the difference $\mathcal{A-\mathcal{B}}$.
Let us split this difference into five terms, labelled $\Delta_{1}$,...
$\Delta_{5}$ with 
\[
\Delta_{1}=2\pi\delta\left(\mathbf{v}_{1}-\mathbf{v}_{2}\right)f\left(\mathbf{v}_{1}\right)\delta\left(\omega-\mathbf{k}.\mathbf{v}_{1}\right)\widetilde{\mathcal{C}_{VV}}\left(\mathbf{k},\omega\right),
\]
\[
\Delta_{2}=\frac{\left(\widetilde{\mathcal{C}_{VV}}\left(\mathbf{k},\omega\right)\right)^{2}}{\omega-\mathbf{k}.\mathbf{v}_{2}-i\epsilon}\mathbf{k}.\frac{\partial f}{\partial\mathbf{v}_{1}}\mathbf{k}.\frac{\partial f}{\partial\mathbf{v}_{2}}\left\{ \frac{1}{\omega-\mathbf{k}.\mathbf{v}_{1}+i\tilde{\epsilon}}-\frac{1}{\omega-\mathbf{k}.\mathbf{v}_{1}-i\tilde{\epsilon}}\right\} ,
\]
\[
\Delta_{3}=2\pi\widetilde{\mathcal{C}_{VV}}\left(\mathbf{k},\omega\right)\hat{W}\left(\mathbf{k}\right)\mathbf{k}.\frac{\partial f}{\partial\mathbf{v}_{1}}\frac{f(\mathbf{v}_{2})\delta\left(\omega-\mathbf{k}.\mathbf{v}_{2}\right)}{\varepsilon\left(\mathbf{k},\omega\right)}\left\{ \frac{1}{\omega-\mathbf{k}.\mathbf{v}_{1}-i\tilde{\epsilon}}-\frac{1}{\omega-\mathbf{k}.\mathbf{v}_{1}+i\tilde{\epsilon}}\right\} ,
\]
\[
\Delta_{4}=2\pi\widetilde{\mathcal{C}_{VV}}\left(\mathbf{k},\omega\right)\hat{W}\left(\mathbf{k}\right)\mathbf{k}.\frac{\partial f}{\partial\mathbf{v}_{2}}\frac{f(\mathbf{v}_{1})\delta\left(\omega-\mathbf{k}.\mathbf{v}_{1}\right)}{\omega-\mathbf{k}.\mathbf{v}_{2}-i\tilde{\epsilon}}\left\{ \frac{1}{\varepsilon\left(\mathbf{k},\omega\right)}-\frac{1}{\bar{\varepsilon}\left(\mathbf{k},\omega\right)}\right\} ,
\]
and
\[
\Delta_{5}=-\left(\frac{2\pi\hat{W}\left(\mathbf{k}\right)}{\varepsilon\left(\mathbf{k},\omega\right)}\right)^{2}f\left(\mathbf{v}_{1}\right)f\left(\mathbf{v}_{2}\right)\delta\left(\omega-\mathbf{k}.\mathbf{v}_{1}\right)\delta\left(\omega-\mathbf{k}.\mathbf{v}_{2}\right),
\]
such that
\begin{equation}
\mathbf{C}=\frac{1}{2\left(2\pi\right)}\sum_{\mathbf{k}}\int_{\Gamma}\text{d}\omega\,\mathbf{k}\mathbf{k}\left\{ \Delta_{1}+\Delta_{2}+\Delta_{3}+\Delta_{4}+\Delta_{5}\right\} .\label{eq:C_Delta}
\end{equation}
The first term $\Delta_{1}$ is already explicit, there is nothing
more to do. For the other terms, we will use the fact that for any
complex number $z$, we have $z-\bar{z}=2i\Im\left(z\right).$ For
$\Delta_{2}$, using the Sokhotski-Plemelj formula (\ref{eq:Sokhotski=002013Plemelj}),
we have
\[
\frac{1}{\omega-\mathbf{k}.\mathbf{v}_{1}+i\tilde{\epsilon}}-\frac{1}{\omega-\mathbf{k}.\mathbf{v}_{1}-i\tilde{\epsilon}}=-2i\pi\delta\left(\omega-\mathbf{k}.\mathbf{v}_{1}\right),
\]
and then,
\[
\Delta_{2}=-2i\pi\delta\left(\omega-\mathbf{k}.\mathbf{v}_{1}\right)\frac{\left(\widetilde{\mathcal{C}_{VV}}\left(\mathbf{k},\omega\right)\right)^{2}}{\left(\omega-\mathbf{k}.\mathbf{v}_{2}-i\tilde{\epsilon}\right)}\mathbf{k}.\frac{\partial f}{\partial\mathbf{v}_{1}}\mathbf{k}.\frac{\partial f}{\partial\mathbf{v}_{2}}.
\]
We can notice that the imaginary part of $\Delta_{2}$ is odd in $\left(\mathbf{k},\omega\right)$,
whereas its real part is even. Given how $\Delta_{2}$ comes into
play in the expression of the second cumulant (\ref{eq:C_Delta}),
only the real (and even) part of $\Delta_{2}$ will contribute to
the second cumulant, and 
\[
\Re\left(\Delta_{2}\right)=2\pi^{2}\left(\widetilde{\mathcal{C}_{VV}}\left(\mathbf{k},\omega\right)\right)^{2}\mathbf{k}.\frac{\partial f}{\partial\mathbf{v}_{1}}\mathbf{k}.\frac{\partial f}{\partial\mathbf{v}_{2}}\delta\left(\omega-\mathbf{k}.\mathbf{v}_{1}\right)\delta\left(\omega-\mathbf{k}.\mathbf{v}_{2}\right).
\]
In a similar way, we can prove that only the real parts of $\Delta_{3}$
and $\Delta_{4}$ contribute to (\ref{eq:C_Delta}) and that
\[
\Re\left(\Delta_{3}\right)=-4\pi^{3}\widetilde{\mathcal{C}_{VV}}\left(\mathbf{k},\omega\right)\frac{\hat{W}\left(\mathbf{k}\right)^{2}}{\left|\varepsilon\left(\mathbf{k},\omega\right)\right|^{2}}\mathbf{k}.\frac{\partial f}{\partial\mathbf{v}_{1}}f\left(\mathbf{v}_{2}\right)\delta\left(\omega-\mathbf{k}.\mathbf{v}_{1}\right)\delta\left(\omega-\mathbf{k}.\mathbf{v}_{2}\right)\int\text{d}\mathbf{v}'\,\mathbf{k}.\frac{\partial f}{\partial\mathbf{v}'}\delta\left(\omega-\mathbf{k}.\mathbf{v}'\right),
\]
and
\[
\Re\left(\Delta_{4}\right)=-4\pi^{3}\widetilde{\mathcal{C}_{VV}}\left(\mathbf{k},\omega\right)\frac{\hat{W}\left(\mathbf{k}\right)^{2}}{\left|\varepsilon\left(\mathbf{k},\omega\right)\right|^{2}}\mathbf{k}.\frac{\partial f}{\partial\mathbf{v}_{2}}f\left(\mathbf{v}_{1}\right)\delta\left(\omega-\mathbf{k}.\mathbf{v}_{1}\right)\delta\left(\omega-\mathbf{k}.\mathbf{v}_{2}\right)\int\text{d}\mathbf{v}'\,\mathbf{k}.\frac{\partial f}{\partial\mathbf{v}'}\delta\left(\omega-\mathbf{k}.\mathbf{v}'\right).
\]
To compute the contribution of $\Delta_{5}$, let use this simple
identity 
\[
\frac{1}{\varepsilon^{2}}=\frac{1}{\varepsilon^{2}}-\frac{1}{\left|\varepsilon\right|^{2}}+\frac{1}{\left|\varepsilon\right|^{2}}.
\]
And in a similar way that we did for the other terms, we notice that
\[
\frac{1}{\varepsilon^{2}}-\frac{1}{\left|\varepsilon\right|^{2}}=\frac{1}{\varepsilon\left|\varepsilon\right|^{2}}\left\{ \bar{\varepsilon}-\varepsilon\right\} =\frac{-2i\Im\left(\varepsilon\right)}{\varepsilon\left|\varepsilon\right|^{2}}.
\]
Then, the fifth term $\Delta_{5}$ reads
\begin{multline*}
\Delta_{5}=\frac{2i\Im\left(\varepsilon\right)}{\varepsilon\left|\varepsilon\right|^{2}}\left(2\pi\hat{W}\left(\mathbf{k}\right)\right)^{2}f\left(\mathbf{v}_{1}\right)f\left(\mathbf{v}_{2}\right)\delta\left(\omega-\mathbf{k}.\mathbf{v}_{1}\right)\delta\left(\omega-\mathbf{k}.\mathbf{v}_{2}\right)\\
-\left(\frac{2\pi\hat{W}\left(\mathbf{k}\right)}{\left|\varepsilon\left(\mathbf{k},\omega\right)\right|^{2}}\right)^{2}f\left(\mathbf{v}_{1}\right)f\left(\mathbf{v}_{2}\right)\delta\left(\omega-\mathbf{k}.\mathbf{v}_{1}\right)\delta\left(\omega-\mathbf{k}.\mathbf{v}_{2}\right).
\end{multline*}
Once again, with parity arguments we can show that only the real part
of $\Delta_{5}$ contributes to (\ref{eq:C_Delta}), that is to say
\begin{multline*}
\Re\left(\Delta_{5}\right)=\frac{2\Im\left(\varepsilon\right)^{2}}{\left|\varepsilon\right|^{4}}\left(2\pi\hat{W}\left(\mathbf{k}\right)\right)^{2}f\left(\mathbf{v}_{1}\right)f\left(\mathbf{v}_{2}\right)\delta\left(\omega-\mathbf{k}.\mathbf{v}_{1}\right)\delta\left(\omega-\mathbf{k}.\mathbf{v}_{2}\right)\\
-\left(\frac{2\pi\hat{W}\left(\mathbf{k}\right)}{\left|\varepsilon\left(\mathbf{k},\omega\right)\right|^{2}}\right)^{2}f\left(\mathbf{v}_{1}\right)f\left(\mathbf{v}_{2}\right)\delta\left(\omega-\mathbf{k}.\mathbf{v}_{1}\right)\delta\left(\omega-\mathbf{k}.\mathbf{v}_{2}\right).
\end{multline*}
Thanks to this analysis, we can compute $\mathbf{C}$ as following
\begin{equation}
\mathbf{C}=\frac{1}{2\left(2\pi\right)}\sum_{\mathbf{k}}\int_{\Gamma}\text{d}\omega\,\mathbf{k}\mathbf{k}\Re\left\{ \Delta_{1}+\Delta_{2}+\Delta_{3}+\Delta_{4}+\Delta_{5}\right\} .\label{eq:C_Real_delta}
\end{equation}
Furthermore, the quadratic term of the Hamiltonian $H^{(2)}$ is linked
to this cumulant via the following formula
\begin{equation}
H^{(2)}=\int\text{d}\mathbf{r}\text{d}\mathbf{v}_{1}\text{d}\mathbf{v}_{2}\,\frac{\partial p}{\partial\mathbf{v}_1}\frac{\partial p}{\partial\mathbf{v}_2}\:\mathbf{C}(\mathbf{v}_{1},\mathbf{v}_{2}).\label{eq:linkH_2_C}
\end{equation}
Using equations (\ref{eq:Imaginary-Dispersion-Relation}, \ref{eq:Autocorrelation_Potentiel_Fourier_Laplace},
\ref{eq:C_Real_delta}, \ref{eq:linkH_2_C}), we can show that
\begin{multline}
H^{(2)}=\int\text{d}\mathbf{r}\text{d}\mathbf{v}_{1}\,\frac{\partial p}{\partial\mathbf{v}_{1}}\frac{\partial p}{\partial\mathbf{v}_{1}}:\mathbf{D}(\mathbf{v}_{1})f\left(\mathbf{v}_{1}\right)\\
-\int\text{d}\mathbf{r}\text{d}\mathbf{v}_{1}\text{d}\mathbf{v}_{2}\,\frac{\partial p}{\partial\mathbf{v}_{1}}\frac{\partial p}{\partial\mathbf{v}_{2}}:\mathbf{B}[f](\mathbf{v}_{1},\mathbf{v}_{2})f\left(\mathbf{v}_{1}\right)f\left(\mathbf{v}_{2}\right)\\
+\int\text{d}\mathbf{r}\text{d\ensuremath{\mathbf{v}_{1}}}\text{d}\mathbf{v}_{2}\text{d\ensuremath{\mathbf{v}_{3}}}\text{d}\mathbf{v}_{4}\,\frac{\partial p}{\partial\mathbf{v}_{1}}\frac{\partial p}{\partial\mathbf{v}_{2}}\mathbf{B}^{(2)}\left(\mathbf{v}_{1},\mathbf{v}_{2},\mathbf{v}_{3},\mathbf{v}_{4}\right)\left\{ f(\mathbf{v}_{1})f(\mathbf{v}_{2})\frac{\partial f}{\partial\mathbf{v}_{3}}\frac{\partial f}{\partial\mathbf{v}_{4}}\right.\\
\left.-2f(\mathbf{v}_{1})\frac{\partial f}{\partial\mathbf{v}_{2}}f(\mathbf{v}_{3})\frac{\partial f}{\partial\mathbf{v}_{4}}+\frac{\partial f}{\partial\mathbf{v}_{1}}\frac{\partial f}{\partial\mathbf{v}_{2}}f(\mathbf{v}_{3})f(\mathbf{v}_{4})\right\} ,\label{eq:H^2_rec}
\end{multline}
with
\begin{equation}
\mathbf{B}^{(2)}\left(\mathbf{v}_{1},\mathbf{v}_{2},\mathbf{v}_{3},\mathbf{v}_{4}\right)=2\pi^{3}\left(\frac{\lambda_{D}}{L}\right)^{3}\sum_{\mathbf{k}}\int\text{d}\omega\,\mathbf{k}\mathbf{k}\mathbf{k}\mathbf{k}\frac{\hat{W}\left(\mathbf{k}\right)^{4}}{\left|\varepsilon\left(\mathbf{k},\omega\right)\right|^{4}}\prod_{i=1}^{4}\delta\left(\omega-\mathbf{k}.\mathbf{v}_{i}\right),
\label{eq:B^2}
\end{equation}
being a fully symmetric order-4 tensor.

\section{Expression of the fourth cumulant\label{sec:Computation-of-higher}}

In this appendix, we report the result of the computation of $H^{(4)}$
the contribution of the fourth cumulant to the large deviation Hamiltonian:
\begin{multline}
H^{(4)}=\int\text{d}\mathbf{r}\text{d}\mathbf{v}_{1}\cdots\text{d}\mathbf{v}_{4}\,\left\{ \frac{\partial p}{\partial\mathbf{v}_{1}}\frac{\partial p}{\partial\mathbf{v}_{1}}-\frac{\partial p}{\partial\mathbf{v}_{2}}\frac{\partial p}{\partial\mathbf{v}_{1}}\right\} \left\{ \frac{\partial p}{\partial\mathbf{v}_{3}}\frac{\partial p}{\partial\mathbf{v}_{3}}-\frac{\partial p}{\partial\mathbf{v}_{4}}\frac{\partial p}{\partial\mathbf{v}_{3}}\right\} \mathbf{B}^{(2)}f\left(\mathbf{v}_{1}\right)f\left(\mathbf{v}_{2}\right)f\left(\mathbf{v}_{3}\right)f\left(\mathbf{v}_{4}\right)\\
+3\int\text{d}\mathbf{r}\text{d\ensuremath{\mathbf{v}_{1}}}\ldots\text{d}\mathbf{v}_{6}\,\frac{\partial p}{\partial\mathbf{v}_{1}}\frac{\partial p}{\partial\mathbf{v}_{2}}\frac{\partial p}{\partial\mathbf{v}_{3}}\left\{ \frac{\partial p}{\partial\mathbf{v}_{3}}-\frac{\partial p}{\partial\mathbf{v}_{4}}\right\} \mathbf{B}^{(3)} f(\mathbf{v}_{3})f(\mathbf{v}_{4})\left\{ f\left(\mathbf{v}_{1}\right)f(\mathbf{v}_{2})\frac{\partial f}{\partial\mathbf{v}_{5}}\frac{\partial f}{\partial\mathbf{v}_{6}}\right.\\
\left.-2\frac{\partial f}{\partial\mathbf{v}_{1}}f(\mathbf{v}_{2})f\left(\mathbf{v}_{5}\right)\frac{\partial f}{\partial\mathbf{v}_{6}}+\frac{\partial f}{\partial\mathbf{v}_{1}}\frac{\partial f}{\partial\mathbf{v}_{2}}f\left(\mathbf{v}_{5}\right)f\left(\mathbf{v}_{6}\right)\right\} \\
+\int\text{d}\mathbf{r}\text{d\ensuremath{\mathbf{v}_{1}}}\ldots\text{d}\mathbf{v}_{8}\,\frac{\partial p}{\partial\mathbf{v}_{1}}\frac{\partial p}{\partial\mathbf{v}_{2}}\frac{\partial p}{\partial\mathbf{v}_{3}}\frac{\partial p}{\partial\mathbf{v}_{4}}\mathbf{B}^{(4)}\left\{ f(\mathbf{v}_{1})f(\mathbf{v}_{2})f(\mathbf{v}_{3})f\left(\mathbf{v}_{4}\right)\frac{\partial f}{\partial\mathbf{v}_{5}}\frac{\partial f}{\partial\mathbf{v}_{6}}\frac{\partial f}{\partial\mathbf{v}_{7}}\frac{\partial f}{\partial\mathbf{v}_{8}}\right.\\
-4\frac{\partial f}{\partial\mathbf{v}_{1}}f(\mathbf{v}_{2})f(\mathbf{v}_{3})f\left(\mathbf{v}_{4}\right)f\left(\mathbf{v}_{5}\right)\frac{\partial f}{\partial\mathbf{v}_{6}}\frac{\partial f}{\partial\mathbf{v}_{7}}\frac{\partial f}{\partial\mathbf{v}_{8}}
+6\frac{\partial f}{\partial\mathbf{v}_{1}}\frac{\partial f}{\partial\mathbf{v}_{2}}f(\mathbf{v}_{3})f\left(\mathbf{v}_{4}\right)f\left(\mathbf{v}_{5}\right)f\left(\mathbf{v}_{6}\right)\frac{\partial f}{\partial\mathbf{v}_{7}}\frac{\partial f}{\partial\mathbf{v}_{8}}\\
-4\frac{\partial f}{\partial\mathbf{v}_{1}}\frac{\partial f}{\partial\mathbf{v}_{2}}\frac{\partial f}{\partial\mathbf{v}_{3}}f\left(\mathbf{v}_{4}\right)f\left(\mathbf{v}_{5}\right)f\left(\mathbf{v}_{5}\right)f\left(\mathbf{v}_{6}\right)f\left(\mathbf{v}_{7}\right)\frac{\partial f}{\partial\mathbf{v}_{8}}
\left.+\frac{\partial f}{\partial\mathbf{v}_{1}}\frac{\partial f}{\partial\mathbf{v}_{2}}\frac{\partial f}{\partial\mathbf{v}_{3}}\frac{\partial f}{\partial\mathbf{v}_{4}}f\left(\mathbf{v}_{5}\right)f\left(\mathbf{v}_{6}\right)f\left(\mathbf{v}_{7}\right)f\left(\mathbf{v}_{8}\right)\right\} .
\label{eq:H_4}
\end{multline}

\begin{acknowledgements}
 We thank J. Barré and G. Eyink for pointing us to relevant literature and for interesting comments on this manuscript. The research leading to this work was supported by a Subagreement from the Johns Hopkins University with funds provided by Grant No. 663054 from Simons Foundation. Its contents are solely the responsibility of the authors and do not necessarily represent the official views of Simons Foundation or the Johns Hopkins University. We thank the two anonymous reviewers for their useful recommendations, which helped us to improve our first version of our manuscript.
\end{acknowledgements}

\bibliographystyle{spmpsci}      
\bibliography{reflandau}
\newpage{}

\end{document}